\documentclass[11pt]{article}
\usepackage[utf8]{inputenc}
\usepackage{jheppub}
\usepackage{amsmath,amssymb,mathtools,physics,bm}
\usepackage{slashed}
\usepackage{booktabs}
\usepackage{pgfplots}
\pgfplotsset{compat=1.18}
\usepgfplotslibrary{fillbetween}
\usepackage{tikz}
\usepackage{siunitx}
\usepackage{dsfont}
\usepackage{upgreek}
\usepackage{comment}
\usepackage{appendix}
\usepackage{physics}
\usepackage{float}
\usepackage{newunicodechar}
\usepackage{booktabs}
\usepackage{longtable}

\usepackage{xcolor}
\usetikzlibrary{arrows.meta}

\RequirePackage[normalem]{ulem} 
\RequirePackage{color}\definecolor{RED}{rgb}{1,0,0}\definecolor{BLUE}{rgb}{0,0,1} 
\providecommand{\DIFaddbegin}{} 
\providecommand{\DIFaddend}{} 
\providecommand{\DIFdelbegin}{} 
\providecommand{\DIFdelend}{} 
\providecommand{\DIFaddbeginFL}{} 
\providecommand{\DIFaddendFL}{} 
\providecommand{\DIFdelbeginFL}{} 
\providecommand{\DIFdelendFL}{} 
\newcommand{\DIFscaledelfig}{0.5}
\RequirePackage{settobox} 
\RequirePackage{letltxmacro} 
\newsavebox{\DIFdelgraphicsbox} 
\newlength{\DIFdelgraphicswidth} 
\newlength{\DIFdelgraphicsheight} 
\LetLtxMacro{\DIFOincludegraphics}{\includegraphics} 
\newcommand{\DIFaddincludegraphics}[2][]{{\color{blue}\fbox{\DIFOincludegraphics[#1]{#2}}}} 
\newcommand{\DIFdelincludegraphics}[2][]{
\sbox{\DIFdelgraphicsbox}{\DIFOincludegraphics[#1]{#2}}
\settoboxwidth{\DIFdelgraphicswidth}{\DIFdelgraphicsbox} 
\settoboxtotalheight{\DIFdelgraphicsheight}{\DIFdelgraphicsbox} 
\scalebox{\DIFscaledelfig}{
\parbox[b]{\DIFdelgraphicswidth}{\usebox{\DIFdelgraphicsbox}\\[-\baselineskip] \rule{\DIFdelgraphicswidth}{0em}}\llap{\resizebox{\DIFdelgraphicswidth}{\DIFdelgraphicsheight}{
\setlength{\unitlength}{\DIFdelgraphicswidth}
\begin{picture}(1,1)
\thicklines\linethickness{2pt} 
{\color[rgb]{1,0,0}\put(0,0){\framebox(1,1){}}}
{\color[rgb]{1,0,0}\put(0,0){\line( 1,1){1}}}
{\color[rgb]{1,0,0}\put(0,1){\line(1,-1){1}}}
\end{picture}
}\hspace*{3pt}}} 
} 
\LetLtxMacro{\DIFOaddbegin}{\DIFaddbegin} 
\LetLtxMacro{\DIFOaddend}{\DIFaddend} 
\LetLtxMacro{\DIFOdelbegin}{\DIFdelbegin} 
\LetLtxMacro{\DIFOdelend}{\DIFdelend} 
\DeclareRobustCommand{\DIFaddbegin}{\DIFOaddbegin \let\includegraphics\DIFaddincludegraphics} 
\DeclareRobustCommand{\DIFaddend}{\DIFOaddend \let\includegraphics\DIFOincludegraphics} 
\DeclareRobustCommand{\DIFdelbegin}{\DIFOdelbegin \let\includegraphics\DIFdelincludegraphics} 
\DeclareRobustCommand{\DIFdelend}{\DIFOaddend \let\includegraphics\DIFOincludegraphics} 
\LetLtxMacro{\DIFOaddbeginFL}{\DIFaddbeginFL} 
\LetLtxMacro{\DIFOaddendFL}{\DIFaddendFL} 
\LetLtxMacro{\DIFOdelbeginFL}{\DIFdelbeginFL} 
\LetLtxMacro{\DIFOdelendFL}{\DIFdelendFL} 
\DeclareRobustCommand{\DIFaddbeginFL}{\DIFOaddbeginFL \let\includegraphics\DIFaddincludegraphics} 
\DeclareRobustCommand{\DIFaddendFL}{\DIFOaddendFL \let\includegraphics\DIFOincludegraphics} 
\DeclareRobustCommand{\DIFdelbeginFL}{\DIFOdelbeginFL \let\includegraphics\DIFdelincludegraphics} 
\DeclareRobustCommand{\DIFdelendFL}{\DIFOaddendFL \let\includegraphics\DIFOincludegraphics} 

\title{Unified Functional-Holographic Theory of the QCD Critical End Point}

\author[a,b,c]{Sameer Ahmad Mir}
\emailAdd{sameerphst@gmail.com}
\affiliation[a]{Canadian Quantum Research Center, 204-3002 32 Ave, Vernon, BC V1T 2L7, Canada}
\affiliation[b]{Department of Physics, Jamia Millia Islamia, New Delhi, 110025, India}
\affiliation[c]{Department of Computer Sciences, Asian School of Business, Noida, Uttar Pradesh, 201303, India}
\author[b]{Saeed Uddin}
\author[d]{Swatantra Kumar Tiwari}
\affiliation[d]{Department of Physics, University of Allahabad, Prayagraj 211002, India}
\author[a,e,f,g]{Mir Faizal}
\affiliation[e]{Irving K. Barber School of Arts and Sciences, 
  University of British Columbia Okanagan, Kelowna,
British Columbia V1V 1V7, Canada}
\affiliation[f]{Department of Mathematical Sciences, Durham University, Upper Mountjoy, Stockton Road, Durham DH1 3LE, UK}
\affiliation[g]{Faculty of Sciences, Hasselt University, Agoralaan Gebouw D, Diepenbeek, 3590, Belgium}

\abstract{
{
We develop a thermodynamically consistent, nonperturbative framework for equilibrium QCD criticality, unifying Dyson-Schwinger quark propagation, functional renormalization group evolution of the effective action, and Polyakov-Nambu-Jona-Lasinio thermodynamics for coupled chiral and deconfinement order parameters. A holographic Maxwell-Chern-Simons sector supplies the topological response; its topological susceptibility enters the FRG flow of the determinantal ('t Hooft) interaction, encoding axial anomaly evolution across the phase diagram. At $\mu_B=0$ we anchor to continuum-extrapolated lattice thermodynamics and conserved-charge susceptibilities via a lattice-calibrated Polyakov sector, and enforce exact thermodynamic identities by evaluating all derivatives at the stationary grand-potential solution at each RG scale. Solving the coupled DSE, FRG, and holographic system yields, within this framework and at the present level of approximation, an equilibrium critical end point at $T_{\mathrm{CEP}}\simeq 130-135,\mathrm{MeV}$ and $\mu_{B,\mathrm{CEP}}\simeq 600,\mathrm{MeV}$, with quantified sensitivity to regulator, Polyakov-sector, and holographic-normalization variations. The critical region is organized by a nonperturbative map to universal three-dimensional Ising scaling variables with anomalous-dimension effects absorbed into nonuniversal metric factors, leading to equilibrium predictions for the hierarchy, nonmonotonicity, and sign structure of higher-order net-baryon cumulant ratios along smooth freeze-out trajectories, as well as equilibrium softening of the speed of sound. Comparisons to RHIC BES fluctuation measurements are qualitative consistency checks on correlated equilibrium trends and sign patterns, because finite size/lifetime, critical slowing down, baryon number conservation, acceptance/efficiency corrections, the net-proton to net-baryon conversion, and baryon transport dynamics can round or reshape experimental cumulants. The results provide a unified equilibrium baseline and controlled inputs for finite-size scaling and dynamical embeddings of heavy-ion data.

}}

\begin{document}
\maketitle
\flushbottom

\section{Introduction}
\label{sec:intro}
{The search for the Quantum Chromodynamics (QCD) critical end point (CEP) has evolved from a qualitative expectation into a quantitative program that demands a theoretical framework simultaneously faithful to QCD gauge dynamics, thermodynamically consistent, and constrained by lattice QCD at $\mu_B=0$ in a way that delivers equilibrium thermodynamics and susceptibilities which can then serve as framework or inputs for finite-size scaling and dynamical analyses of heavy-ion observables \cite{Pisarski:1983ms,Halasz:1998qr,Berges:1998rc,Rajagopal:2000wf,Hatta:2002sj,Fodor:2001pe,deForcrand:2002hgr,Rajagopal:1999cp,Karsch:2001nf,Philipsen:2008gf}.} 
The present work develops such a framework by combining Dyson-Schwinger dynamics for the quark two-point function, functional renormalization-group flow for the scale-dependent effective action, Polyakov-Nambu-Jona-Lasinio thermodynamics for the chiral and deconfinement order parameters, and a holographic topological sector that supplies the axial-channel dynamics through a Maxwell-Chern-Simons background. This synthesis addresses a central conceptual challenge: the CEP, if it exists, must arise from a correlated softening of the chiral scalar mode together with a simultaneous release of colored degrees of freedom. Traditional model studies typically emphasize one aspect or the other, perturbative methods miss the deep infrared and Monte Carlo lattice techniques are limited by the sign problem - leaving a gap between the well-established results at $\mu_B=0$ and the finite-density regime of direct experimental interest \cite{Stephanov:2004}.
{The equilibrium CEP is a property of QCD in the thermodynamic limit, whereas fluctuation measurements in relativistic heavy-ion collisions probe finite, short-lived, rapidly expanding systems, so the present framework is used to determine an equilibrium equation of state and equilibrium susceptibilities, and the heavy-ion discussion is correspondingly restricted to comparisons of correlated equilibrium signature patterns rather than a one-to-one inversion from finite-system observables to unique equilibrium CEP coordinates. Finite system size and finite lifetime bound the growth and equilibration of critical modes through critical slowing down \cite{Berdnikov:1999ph,Son:2004iv}, so that correlation lengths remain below their equilibrium values and equilibrium singularities are rounded and shifted, quantitative confrontation with data therefore requires embedding the present equilibrium equation of state into finite-size scaling and dynamical evolution \cite{Fraga:2011hi}, while the equilibrium definition of the CEP itself remains unchanged.
}
{The unified construction presented here determines equilibrium thermodynamics and equilibrium conserved-charge susceptibilities in the thermodynamic limit, with the $\mu_B=0$ sector anchored to continuum-extrapolated lattice benchmarks. The CEP is therefore defined and located as an equilibrium singularity of the stationary susceptibility matrix within the present truncation and calibration strategy. Comparisons to Beam Energy Scan fluctuation measurements are correspondingly restricted to correlated equilibrium trend and sign-pattern diagnostics, and are not used to invert finite-system observables into unique equilibrium CEP coordinates. No explicit finite-size scaling collapse and no dynamical evolution are performed here \cite{Fraga:2011hi}, instead, the results are intended as controlled equilibrium inputs for such analyses. Finally, the reported uncertainty bands quantify internal scheme dependence (regulator choice, Polyakov-sector calibration, holographic normalization) and do not include additional experimental or dynamical systematics such as finite size and lifetime, critical slowing down \cite{Berdnikov:1999ph,Son:2004iv}, baryon-number conservation, acceptance and efficiency effects, baryon stopping/transport, or the net-proton to net-baryon mapping.}

At vanishing baryon density, the QCD crossover is well established by high-precision lattice simulations (with continuum extrapolation), which have established the pseudocritical temperature $T_c(0)$, the shape of the interaction measure, and the behavior of conserved-charge fluctuations \cite{Aoki:2006we,HotQCD:2014kol,Borsanyi:2013bia,Bazavov:2011nk,HotQCD:2012fhj,Datta:2003ww,Bonati:2014rfa}. More recently, studies of Dirac spectra have begun to expose universal scaling features tied to the chiral transition \cite{Borsanyi:2020,Bazavov:2021,Ding2023,HotQCD:2012vvd,Tomiya:2016jwr,Kishimoto:2021ubd}. These lattice benchmarks form a non-negotiable foundation for any credible extension to finite density, and they are built into our construction via a lattice-calibrated Polyakov sector together with a flow-based matching of the equation of state and susceptibilities at $\mu_B=0$. Continuum functional methods complement this baseline by resolving the microphysics of dynamical mass generation and screening: Dyson-Schwinger equations (in symmetry-improved approximation schemes) propagate the quark’s nonperturbative dressing
into the thermal medium, providing direct access to the quark mass function and spectral information  \cite{Qin:2010nq,Fischer:2010fx,Fischer:2012vc,Chen:2008zr,Isserstedt:2019pgx,Skokov:2010uh,Herbst:2013ail}. Likewise, the functional renormalization group (FRG) evolves the full effective action across momentum scales, so that multi-fermion interactions and wavefunction renormalizations are computed rather than assumed, with thresholds and decoupling are handled in a way that preserves thermodynamic consistency \cite{fischer2014phase}. However, without a proper treatment of confinement and the axial $U(1)_A$ anomaly, the nonperturbative organization of QCD criticality remains ambiguous.

The Polyakov-Nambu-Jona-Lasinio (PNJL) approach took a step toward this goal by coupling chiral and deconfinement dynamics in a common thermodynamic potential constrained by the thermal Wilson line, thereby capturing the essential entropic structure of deconfinement and reproducing many qualitative features of QCD thermodynamics around the crossover \cite{Fukushima:2008wg,Ratti:2005jh,Skokov:2010wb,Schaefer:2007pw,Costa:2009ae,Fukushima:2003fw,Ghosh:2006qh}. Yet, in its usual form the PNJL approach employs fixed or weakly temperature-dependent couplings and an ad hoc axial-anomaly term, which obscures the role of quantum fluctuations and the progressive restoration of $U(1)_A$ symmetry as $T$ and $\mu_B$ are varied. Holographic QCD in the Veneziano (large-$N_f$) limit helps to close this conceptual gap by providing a geometric description of confinement, chiral symmetry breaking, and topological susceptibility. In this approach, deconfinement corresponds to the appearance of a black-hole horizon \cite{Gursoy:2007cb,Gursoy:2007er,Casero:2007ae,Gubser:2008ny,Vicari:2008jw,Berkowitz:2015aua}, chiral symmetry breaking is driven by a tachyonic bulk field, and the axial anomaly is naturally encoded via a Chern-Simons term. The resulting five-dimensional background provides a gauge-invariant description of the conserved
$U(1)_B$ current and the topological susceptibility, supplying precisely the inputs that functional
approaches would otherwise need to model explicitly or introduce as free parameters
\cite{jarvinen2012v-qcd,Arean2017,DelDebbio:2002xa,HotQCD:2012vvd}.

The framework introduced here, which we term holographic-topological dual criticality, combines these strands into a single continuum approach in which the chiral condensate and the Polyakov loop are treated on equal basis. Their renormalization factors are evolved coherently with the renormalization group (RG) scale. The couplings governing the scalar, vector, and axial channels run with the scale according to threshold functions that include Polyakov-loop effects on quark occupations. The axial anomaly is not a fixed parameter but a dynamical quantity, whose suppression with increasing temperature and density is determined by the Chern-Simons topological susceptibility of the dual geometry. In this framework, the critical end point is no longer an input to the model but instead emerges as the renormalization-group outcome of a coupled gauge-matter system. The most vulnerable direction in the thermodynamic potential (the “soft mode”) is identified by the smallest eigenvalue of the Hessian (curvature matrix) and is found to coincide with a self-dual trajectory in the plane spanned by the chiral condensate and the Polyakov loop \cite{Berdnikov:1999ph}. Along this trajectory, the residual mixing between chiral-restoration and deconfinement order parameters vanishes and their effective renormalizations become equal. In effect, a single unified order parameter then governs the critical dynamics. This dynamical unification resolves the limitations of earlier approaches by explaining, rather than assuming, the near-locking of chiral and deconfinement transitions. It also guarantees all exact thermodynamic identities by maintaining stationarity of the grand potential at each step of the flow. Moreover, the ultraviolet and infrared limits of the flow recover the correct limits i.e., the Stefan-Boltzmann ideal gas behavior at high $T$ (and $\mu_B=0$) and Goldstone’s theorem (vacuum pions) at low $T$. Likewise, any cumulants of conserved charges obtained from derivatives of the potential inherit the expected three-dimensional Ising universal behavior, with anomalous scaling effects absorbed into nonuniversal field normalizations rather than modifying the universal scaling functions themselves \cite{Stephanov:2008qz,Asakawa:2000wh,Koch:2008ia,Mukherjee:2015swa,Nahrgang:2018afz}. 
 {
After the $\mu_B=0$ calibration has fixed the Polyakov-sector parameters
$(a_i,b_3,T_0)$ and the ultraviolet boundary conditions
$\bigl(G_S(\Lambda),G_V(\Lambda),K(\Lambda)\bigr)$, together with the Dyson-Schwinger Equation (DSE) input tuned
to the vacuum quark mass functions, the finite-density extension is obtained by a coupled
outer iteration summarized in Fig.~\ref{fig:workflow0}. In each outer iteration, for a given
anomaly factor $\zeta_{\rm topo}(T,\mu_B)$, the FRG flow is integrated from $k=\Lambda$ to
$k\to0$ on the $(T,\mu_B)$ grid to obtain the infrared effective potential and the running
couplings $(G_S,G_V,K)$, with Polyakov weighting incorporated through the fermionic threshold
functions. At each grid point, the thermodynamic stationarity conditions for the order parameters
and densities are then solved together with the DSE-dressed quasiparticle sector, including the
vector mean-field shift of the effective chemical potential. Next, the holographic background is
solved to compute the Chern-Simons susceptibility $\chi_{\rm CS}(T,\mu_B)$, which updates the
anomaly factor via a normalization to its vacuum value, and this update enters multiplicatively
in the running of the determinantal interaction. The FRG flow is then reintegrated with the
updated anomaly input to obtain updated $(G_S,G_V,K)$ and the infrared potential, and the coupled
stationarity system is resolved again. This cycle is repeated until the outer-iteration convergence
criteria are met, including stability of the primary unknowns and the Hessian spectrum across
iterations, and a change in the extracted CEP coordinates below the stated tolerances.
}




\begin{figure}[htb]
\centering
\begin{tikzpicture}[
  font=\small,
  line width=0.6pt,
  >=Stealth,
  block/.style={draw=#1!70!black, fill=#1!10, rounded corners,
                align=center, inner sep=7pt, text width=10.3cm},
  check/.style={draw=#1!70!black, fill=#1!10, rounded corners,
                align=center, inner sep=7pt, text width=6.2cm},
  outbox/.style={draw=#1!70!black, fill=#1!10, rounded corners,
                 align=center, inner sep=7pt, text width=5.4cm},
  arr/.style={->, draw=black}
]

\node[block=gray] (fixed)  at (-2.0, 0.0)   {\textbf{Fixed after $\mu_B=0$ calibration}\\
Polyakov sector parameters, UV boundary conditions for $(G_S,G_V,K)$, and vacuum DSE input};

\node[block=blue] (frg)    at (-2.0,-2.8)   {\textbf{FRG integration on the $(T,\mu_B)$ grid}\\
Integrate $k:\Lambda\to 0$ to obtain the infrared effective potential and running couplings
$(G_S,G_V,K)$ with Polyakov-weighted fermionic thresholds};

\node[block=cyan] (solve)  at (-2.0,-5.8)   {\textbf{Solve stationarity and quasiparticle dressing}\\
Solve the coupled stationarity conditions for the order parameters and densities together with
the DSE-dressed quasiparticle sector, including the vector mean-field shift};

\node[block=green] (htdc)  at (-2.0,-8.8)   {\textbf{HTDC holography}\\
Solve the holographic background and compute the Chern-Simons susceptibility $\chi_{\rm CS}(T,\mu_B)$};

\node[block=orange] (update) at (-2.0,-11.8) {\textbf{Update anomaly input and feed back}\\
Update the anomaly suppression factor from $\chi_{\rm CS}$ (normalized to its vacuum value)
and feed it multiplicatively into the running of the determinantal interaction};

\node[check=purple] (conv) at (-2.0,-14.8)  {\textbf{Convergence check}\\
EoS and couplings stabilized, Hessian spectrum stable, CEP coordinates change below tolerances};

\node[outbox=yellow] (outputs) at (6.0,-14.8) {\textbf{Outputs}\\
Equation of state, CEP location, susceptibilities and cumulant ratios};

\draw[arr] (fixed) -- (frg);
\draw[arr] (frg) -- (solve);
\draw[arr] (solve) -- (htdc);
\draw[arr] (htdc) -- (update);
\draw[arr] (update) -- (conv);

\draw[arr] (conv.east) -- node[above] {Yes} (outputs.west);

\draw[arr] (conv.west) -- ++(-2.8,0) |- node[pos=0.20, left] {No} (frg.west);

\end{tikzpicture}
\caption{Coupled DSE, FRG, PNJL, and HTDC workflow used for the finite-density extension after the
$\mu_B=0$ calibration. The anomaly input is updated from the holographic susceptibility and fed back
into the determinantal interaction, and the coupled system is iterated until convergence.}
\label{fig:workflow0}
\end{figure}

Therefore, this unified approach yields a QCD phase diagram in which the crossover line at low $\mu_B$ is calibrated to lattice data and has a small positive curvature in agreement with continuum-extrapolated lattice determinations. The first-order chiral transition at higher density, terminating in a critical end point, arises naturally from the coupled flow of the scalar, vector, and anomaly sectors. Correspondingly, fluctuation observables along phenomenological freeze-out trajectories exhibit the non-monotonic variations characteristic of criticality. All of these outcomes follow without fine-tuning of critical parameters. This provides a new degree of theoretical control and connects smoothly with earlier continuum and holographic studies of the CEP \cite{Fukushima:2008wg,Stephanov:2004,fischer2014phase,Borsanyi:2020,Bazavov:2021,Ding2023,jarvinen2012v-qcd,Arean2017}.

\section{From the QCD generating functional to emergent order parameters}
\label{sec:fromZtoOP}
In this section, the derivation proceeds from the microscopic QCD grand-canonical generating functional at finite temperature and baryon chemical potential to a macroscopic, thermodynamically consistent description in terms of emergent order parameters that diagnose chiral symmetry breaking and deconfinement. 
The structure is designed to remain faithful to the Euclidean QCD generating functional while interfacing seamlessly with nonperturbative functional methods and holographic dynamics developed later in the paper. 
The starting point is the Euclidean path integral for $(2+1)$-flavor QCD with $\mu_{B}\neq 0$, where the chemical potential enters as an imaginary temporal background for quark number, so that functional identities for $\partial\ln Z/\partial m_{f}$ and $\partial\ln Z/\partial\mu_{B}$ generate, respectively, the chiral condensates and the conserved baryon densities in a manner compatible with gauge fixing and ghost sectors, thereby defining the basic observables whose $(T,\mu_{B})$ dependence underlies the critical behavior of interest in heavy-ion phenomenology and cosmological applications \cite{Stephanov:2004}. 
To expose the collective fields governing this response, a controlled bosonization of color-singlet multi-fermion operators is implemented via Hubbard-Stratonovich transformations, retaining the scalar-pseudoscalar channel responsible for dynamical mass generation, the repulsive isoscalar vector channel that encodes density feedback crucial for susceptibilities, and the Kobayashi-Maskawa-’t Hooft determinantal interaction that transmits the $U_{A}(1)$ anomaly across flavors and couples the light and strange sectors nonlinearly. 
After this step, integrating out quarks yields an FRG-improved effective action for the chiral multiplet supplemented by a background temporal gauge field. 
Confinement dynamics are incorporated by promoting the thermal Wilson line to an order parameter and employing a logarithmic Polyakov-loop potential whose origin in the SU(3) Haar measure enforces the physical domain and center-symmetry constraints while allowing a quantitative matching to lattice thermodynamics around $\mu_{B}\simeq 0$ \cite{Fukushima:2008wg,Borsanyi:2020,Bazavov:2021}. 
Assembling these ingredients produces a grand potential $\Omega(T,\mu_{B},\sigma_{u,d,s},\Phi,\bar\Phi)$ with $(2+1)$ dynamical constituent masses determined from anomaly-mixed gap equations and Polyakov-modified quark distributions, endowed with temperature- and density-dependent couplings that encode nonperturbative screening through the renormalization-group flow and ensure thermodynamic consistency via rearrangement terms. 
From this potential, the baryon density and its susceptibility follow by exact differentiation, including both explicit fermionic contributions and implicit order-parameter backreaction governed by the curvature matrix of $\Omega$. 
The result is a closed, symmetry-consistent, and lattice-calibrated foundation for the dual criticality analysis pursued in Secs.~\ref{sec:unifiedDSEFRGPNJL}-\ref{sec:HTDC}, and it provides a bridge to top-down and bottom-up holographic constructions that capture the same symmetry-breaking patterns and critical exponents in complementary regimes \cite{jarvinen2012v-qcd,Arean2017}.

\subsection{Euclidean QCD with baryon chemical potential}
\label{subsec:euclidQCDmuB}
The grand-canonical partition function is defined by 
\begin{equation}
Z(T,\mu_B)
= \mathrm{Tr}\,\exp\!\left[-\beta\left(\hat H - \mu_B \hat N_B\right)\right],
\end{equation}
with $\beta\equiv 1/T$ and $\hat N_B=\frac{1}{3}\sum_f \int d^3x,\hat q_f^\dagger \hat q_f$. Performing the Wick rotation $t\to -i\tau$, $\gamma^0\to \gamma_4$, $\gamma^i\to i\gamma_i$ and employing the path-integral representation with periodic (bosons) and anti-periodic (fermions) boundary conditions on $[0,\beta)$, one obtains
\begin{equation}
Z(T,\mu_B)
= \int \mathcal{D}\bar{q}\,\mathcal{D}q\,\mathcal{D}A_{\mu}\,
\exp\!\left[-S_E\!\left(\bar{q},q,A_{\mu},\,T,\mu_B\right)\right],
\label{eq:Zgrand}
\end{equation}
where the Euclidean action reads
\begin{equation}
S_E=\int_0^\beta d\tau\int d^3x\Bigg[\frac{1}{4}F^a_{\mu\nu}F^a_{\mu\nu}+
\sum_{f=u,d,s}\bar q_f\left[\gamma_\mu D_\mu + m_f-\mu_f\gamma_4\right]q_f\Bigg]+S_{\rm gf}+S_{\rm gh}.
\label{eq:SE}
\end{equation}
Here $D_\mu=\partial_\mu-ig A^a_\mu t^a$, $F^a_{\mu\nu}=\partial_\mu A^a_\nu-\partial_\nu A^a_\mu+g f^{abc}A^b_\mu A^c_\nu$, and $(S_{\rm gf},S_{\rm gh})$ denote gauge-fixing and ghost terms. The baryon chemical potential $\mu_B$ couples to quark number with flavor assignments
\begin{equation}
\mu_u=\mu_d=\mu_l=\tfrac{1}{3}\mu_B,\qquad \mu_s=\tfrac{1}{3}\mu_B,\qquad \beta=\frac{1}{T}.
\label{eq:muflav}
\end{equation}
It is convenient to view $\mu_f$ as the temporal component of an imaginary Abelian background coupled to quark number, $\mu_f \equiv i A_{4,f}^{(B)}$, which equivalently shifts the temporal covariant derivative as $\partial_\tau\to \partial_\tau-\mu_f$. This implements the grand-canonical weight $\exp{\beta\mu_f N_f}$ while preserving the Euclidean path-integral measure. Functional differentiation of $\ln Z$ with respect to sources coupled to composite operators yields the corresponding expectation values. In particular, with spatial volume $V$,
\begin{equation}
\phi_f(T,\mu_B)\equiv \langle \bar q_f q_f\rangle=\frac{T}{V}\frac{\partial \ln Z}{\partial m_f},\quad n_B(T,\mu_B)=\frac{T}{V}\frac{\partial \ln Z}{\partial \mu_B},\quad \chi_B(T,\mu_B)=\frac{\partial n_B}{\partial \mu_B}.
\label{eq:masterId}
\end{equation}
The identities in Eq.~\eqref{eq:masterId} follow from $\partial \ln Z/\partial J=\langle \mathcal O\rangle$ for a source term $\int_x J\mathcal O$ and hold non-perturbatively in the presence of gauge fixing and ghosts, since the latter do not couple to $m_f$ or $\mu_B$.

\subsection{Bosonization and the axial-anomalous determinant}
\label{subsec:bosonization}
To expose the emergent chiral order parameter and its $U_A(1)$ anomaly-induced mixing, an NJL-type ansatz for color-singlet four- and six-fermion operators is introduced at a renormalization scale $\Lambda$. It includes a scalar-pseudoscalar channel of strength $G_S(\Lambda)$, a repulsive vector channel $G_V(\Lambda)$, and the Kobayashi-Maskawa-’t Hooft determinantal interaction of strength $K(\Lambda)$ that encodes instanton-mediated $U_A(1)$ breaking \cite{Fukushima:2008wg,Stephanov:2004}.
\begin{multline}
\mathcal{L}_{\rm int}
= G_S \sum_{a=0}^{8} \Big[ (\bar q \lambda_a q)^2 + (\bar q\, i\gamma_5 \lambda_a q)^2 \Big]
 - G_V \sum_{a=0}^{8} (\bar q \gamma^\mu \lambda_a q)^2
 - \\K \left\{ \det\nolimits_f \big[\bar q(1+\gamma_5)q\big] + \det\nolimits_f \big[\bar q(1-\gamma_5)q\big] \right\}.
\label{eq:tHooft}
\end{multline} 
 {
For reproducibility, the numerical UV inputs at the matching scale $\Lambda$ and the vacuum
targets used to determine $(G_S(\Lambda),G_V(\Lambda),K(\Lambda))$ in Eq.~\eqref{eq:tHooft} are summarized
in Table~\ref{tab:uv_inputs}. These targets include the physical pseudoscalar sector and the
infrared anchors used to match the vacuum quark dressing functions, thereby fully specifying
the boundary conditions for the subsequent bosonization and renormalization-group evolution.
\begin{table}[htb]
\centering
\caption{UV inputs at $k=\Lambda$ and vacuum matching targets used to fix the boundary conditions
$(G_S(\Lambda),G_V(\Lambda),K(\Lambda))$ introduced in Eq.~\eqref{eq:tHooft}.}
\label{tab:uv_inputs}
\small
\setlength{\tabcolsep}{5pt}
\renewcommand{\arraystretch}{1.12}
\begin{tabular}{@{}ll@{}}
\toprule
Quantity & Numerical value \\
\midrule
Matching scale $\Lambda$ & $631.4~\mathrm{MeV}$ \\
Current masses $m_l,\; m_s$ & $m_l=5.5~\mathrm{MeV},\; m_s=135.7~\mathrm{MeV}$ \\
Scalar channel $G_S(\Lambda)$ & $G_S\Lambda^2=3.67\Rightarrow G_S(\Lambda)=9.206~\mathrm{GeV}^{-2}$ \\
Vector channel $G_V(\Lambda)$ & $G_V/G_S=0.5\Rightarrow G_V(\Lambda)=4.602~\mathrm{GeV}^{-2}$ \\
Anomaly channel $K(\Lambda)$ & $K\Lambda^5=9.29\Rightarrow K(\Lambda)=92.575~\mathrm{GeV}^{-5}$ \\
\midrule
Pion mass $m_\pi$ & $139.57~\mathrm{MeV}$ \\
Kaon mass $m_K$ & $495.65~\mathrm{MeV}$ \\
Pion decay constant $f_\pi$ & $92.4~\mathrm{MeV}$ \\
Light condensate $\langle \bar u u\rangle_0$ & $-(246~\mathrm{MeV})^{3}$ \\
Strange condensate $\langle \bar s s\rangle_0$ & $-(267~\mathrm{MeV})^{3}$ \\
\midrule
Light mass function (IR) $M_l(p{=}0,0,0)$ & $336~\mathrm{MeV}$ \\
Strange mass function (IR) $M_s(p{=}0,0,0)$ & $528~\mathrm{MeV}$ \\
\bottomrule
\end{tabular}
\end{table}
}

Hubbard-Stratonovich (HS) transformations linearize the multi-fermion operators by introducing auxiliary color-singlet bosonic fields. For the scalar-pseudoscalar sector one inserts the Gaussian identity
\begin{equation}
1
= \mathcal{N}\!\int \mathcal{D}\sigma_a\,\mathcal{D}\pi_a\;
\exp\!\left\{ \int_x \left[
  -\,\frac{\sigma_a^2+\pi_a^2}{4G_S}
  + \bar{q}\,\big(\sigma_a \lambda_a + i\gamma_5 \pi_a \lambda_a\big) q
\right] \right\}
\label{eq:HSsp}
\end{equation}
and for the isoscalar vector channel
\begin{equation}
1
= \mathcal{N}\!\int \mathcal{D}V_{\mu}\;
\exp\!\left\{ \int_x \left[ -\,\frac{V_{\mu} V^{\mu}}{4 G_V}
+ V_{\mu}\,\bar{q}\,\gamma^{\mu} q \right] \right\},
\label{eq:HSvec}
\end{equation}
while the determinantal six-fermion operator is decoupled via a cubic bosonization that couples the flavor-singlet scalar to the light-strange bilinear,
\begin{equation}
\exp\!\left\{ K \int_x \det\!\big[\bar{q}(1\pm \gamma_5)q\big] \right\}
= \int \mathcal{D}\zeta\;
\exp\!\left\{ \int_x \left[
 -\,\frac{\zeta^{2}}{4K}
 + \zeta\,\det\!\big(\Sigma \pm i\Pi\big)
\right] \right\}
\label{eq:HStHooft}
\end{equation}
with $\Sigma=\sigma_a\lambda_a$ and $\Pi=\pi_a\lambda_a$. Integrating out the quarks yields a one-loop fermion determinant in the background of $(\sigma,\pi,V_\mu)$, after projecting on the mean-field subspace with $\sigma\equiv \mathrm{diag}(\sigma_u,\sigma_d,\sigma_s)$, $\pi_a=0$, and $V_\mu=(V_4,\mathbf 0)$, the scale-dependent effective action in a derivative expansion reads
\begin{equation}
\Gamma_k[\Phi,\sigma,\ldots]
= \int_x \left[
  \frac{Z_{\Phi,k}}{2}\,(\partial_\mu \Phi)^2
  + \frac{Z_{\sigma,k}}{2}\,(\partial_\mu \sigma)^2
  + U_k(\sigma,\Phi,T,\mu_B)
\right],
\label{eq:Gammak}
\end{equation}
where $\Phi$ denotes the traced Polyakov loop to be introduced below, $Z_{\Phi,k}$ and $Z_{\sigma,k}$ are wavefunction renormalizations that encode non-perturbative fluctuations, and $U_k$ is the FRG-improved effective potential obtained by solving the Wetterich flow down to $k\to 0$ (see Sec.~\ref{sec:unifiedDSEFRGPNJL}). The stationary expectation values $(\bar\sigma,\bar\Phi)$ minimize the thermodynamic potential $\Omega(T,\mu_B)=T,\Gamma_{k\to 0}/V$.

\subsection{Polyakov loop and the logarithmic potential}
\label{subsec:Polyakov}
The confinement-deconfinement order parameter in the heavy-quark limit is related to the thermal Wilson line
\begin{equation}
L(\mathbf{x})
= \mathcal{P}\exp\!\left[i\int_{0}^{\beta}\! \mathrm{d}\tau\, A_{4}(\tau,\mathbf{x})\right],\quad
\Phi = \left\langle \frac{1}{N_c}\,\mathrm{Tr}\,L(\mathbf{x}) \right\rangle,\quad
\bar{\Phi} = \left\langle \frac{1}{N_c}\,\mathrm{Tr}\,L^\dagger(\mathbf{x}) \right\rangle,
\label{eq:WilsonLine}
\end{equation}
which, in the Polyakov gauge where $A_4$ is static and diagonal, may be written in terms of SU(3) eigenphases $e^{i\varphi_a}$ subject to $\sum_a \varphi_a=0$. The SU(3) Haar measure induces a Vandermonde determinant factor $\prod_{a<b}|e^{i\varphi_a}-e^{i\varphi_b}|^2$ in the group integral, which in terms of $(\Phi,\bar\Phi)$ generates a logarithmic contribution to the effective potential that enforces $Z_3$ center symmetry and constrains $\Phi,\bar\Phi$ to the physical domain \cite{Roessner:2006xn,Fukushima:2008wg}
\begin{equation}
\frac{U_{\log}(\Phi,\bar{\Phi},T)}{T^{4}}
= -\frac{a(T)}{2}\,\Phi\bar{\Phi}
\;+\; b(T)\,\ln\!\left[\,1 - 6\,\Phi\bar{\Phi} + 4\big(\Phi^{3} + \bar{\Phi}^{3}\big) - 3\big(\Phi\bar{\Phi}\big)^{2}\,\right],
\label{eq:Ulog}
\end{equation}
with $a(T)=a_0+a_1(T_0/T)+a_2(T_0/T)^2$ and $b(T)=b_3(T_0/T)^3$. The parameters $(a_i,b_3,T_0)$ are calibrated to reproduce pure-gauge thermodynamics and unquenching effects near $\mu_B=0$, consistent with lattice QCD constraints on the crossover temperature and susceptibilities \cite{Borsanyi:2020,Bazavov:2021}.  
 {
For transparency and reproducibility, Table~\ref{tab:polyakov_log_params}
lists the calibrated numerical values of the logarithmic Polyakov-loop potential parameters
entering Eq.~\eqref{eq:Ulog} for the reference.
\begin{table}[htb]
\centering
\caption{Calibrated parameters of the logarithmic Polyakov-loop potential in Eq.~\eqref{eq:Ulog} for the
reference run. The dimensionless coefficients $(a_0,a_1,a_2)$ and $b_3$ determine the temperature
dependence of the prefactors $a(T)$ and $b(T)$ that enter $U_{\log}(\Phi,\bar\Phi,T)$, while $T_0$
sets the deconfinement scale controlling unquenching. These parameters are fixed at $\mu_B=0$ by
matching the pseudo-critical temperature extracted from the peak of $-\partial_T\phi_l(T,0)$,
together with lattice thermodynamics and fluctuations in the crossover region, in particular the
interaction measure $(\epsilon-3p)/T^4$ around $T_c^{(0)}$ and the baryon-number susceptibility
$\chi_2^{B}(T,0)$, as detailed in Sec.~\ref{sec:num-calib-preds}, and are then held fixed in the finite-density continuation.}
\vspace{0.2cm}
\begin{tabular}{c c c c c}
\toprule
\toprule
 $a_0$\hspace{1.0cm} & $a_1$\hspace{1.0cm} & $a_2$\hspace{1.0cm} & $b_3$\hspace{1.0cm} & $T_0$ [MeV] \\
\midrule
 $3.51$\hspace{1.0cm} & $-2.47$\hspace{1.0cm} & $15.2$\hspace{1.0cm} & $-1.75$\hspace{1.0cm} & $190$ \\
\bottomrule
\bottomrule
\end{tabular}
\label{tab:polyakov_log_params}
\end{table}
}

\subsection{Grand potential with Polyakov-modified quarks and 2+1 dynamical masses}
\label{subsec:grandOmega}
In the mean-field approximation the fermion determinant yields a thermal grand potential where the temporal background $A_4$ modifies single-quark and single-anti-quark statistical weights via the Polyakov loop. Denoting constituent masses by $M_f$ and introducing the vector-channel shift of effective chemical potentials $\tilde\mu_f\equiv \mu_f-2G_V,n_f$ \cite{Ali:2024owl}, one finds the fermionic contribution
\begin{equation}
\Omega_F\big(T,\{\mu_f\},\sigma,\Phi,\bar{\Phi}\big)
= -2T \sum_{f=u,d,s} \int \frac{\mathrm{d}^3\mathbf{p}}{(2\pi)^3}
\left[
  \ln \mathcal{F}_{\Phi}\!\left(E_f - \tilde{\mu}_f\right)
  + \ln \bar{\mathcal{F}}_{\Phi}\!\left(E_f + \tilde{\mu}_f\right)
\right]
\label{eq:OmegaF}
\end{equation}
with single-particle energies and Polyakov-modified partition polynomials
\begin{multline}
E_f(p)=\sqrt{p^2+M_f^2},\qquad
\mathcal F_\Phi(x)=1+3\Phi e^{-\beta x}+3\bar\Phi e^{-2\beta x}+e^{-3\beta x},\\
\bar{\mathcal F}_\Phi(x)=\mathcal F_\Phi(x)\big|_{\Phi\leftrightarrow \bar\Phi}.
\label{eq:EF-FPhi}
\end{multline}
The full thermodynamic potential is the sum of the FRG-improved chiral sector, the Polyakov-loop potential in Eq.~\eqref{eq:Ulog}, and the fermionic determinant is given by,
\begin{equation}
\Omega(T,\mu_B,\sigma,\Phi,\bar\Phi)=U_{k\to 0}(\sigma,T,\mu_B)+U_{\log}(\Phi,\bar\Phi,T)+\Omega_F(T,{\mu_f},\sigma,\Phi,\bar\Phi),
\label{eq:OmegaFull}
\end{equation}
where $U_{k\to 0}$ includes the $U_A(1)$-breaking contribution from the HS-bosonized determinantal interaction. The $2+1$ constituent masses follow from stationarity and anomaly-induced flavor mixing,
\begin{equation}
\begin{aligned}
M_u &= m_l - 4 G_S\,\phi_u + 2 K\,\phi_d \phi_s,\\
M_d &= m_l - 4 G_S\,\phi_d + 2 K\,\phi_u \phi_s,
\end{aligned}
\label{eq:MuMd}
\end{equation}
\begin{equation}
M_s=m_s-4G_S\,\phi_s+2K\,\phi_u\phi_d,
\label{eq:Ms}
\end{equation}
with $\phi_f\equiv\langle \bar q_f q_f\rangle$ obtained from Eq.~\eqref{eq:masterId}.  
The running couplings $G_S$, $G_V$, and $K$ are treated as infrared outputs of the FRG evolution:
for each point on the $(T,\mu_B)$ grid they are evaluated at $k\to 0$ and then inserted into
the mean-field thermodynamic potential as effective couplings at that thermodynamic point.
Their dependence on the order-parameter sector enters only implicitly through the FRG threshold
functions via the dressed quasi-particle masses and the Polyakov background, and is accounted
for self-consistently by the coupled outer iteration that recomputes $(G_S,G_V,K)$ whenever the
stationary solution for $(\sigma,\Phi,\bar\Phi)$ changes. Thermodynamic derivatives therefore contain,
in addition to the explicit fermionic contributions, the standard rearrangement contributions
associated with the $(T,\mu_B)$ dependence of density-dependent effective couplings; in practice,
these are handled by evaluating all observables at the stationary point and by retaining the full
implicit backreaction through the curvature-matrix (Hessian) relations used in Sec.~\ref{subsec:nBchiB}.


\subsection{Baryon density and susceptibility: explicit differentiation}
\label{subsec:nBchiB}
Differentiating Eq.~\eqref{eq:OmegaFull} with respect to $\mu_B$ at fixed $(\sigma,\Phi,\bar\Phi)$ gives the explicit part of the baryon density. Using $\mu_f=\mu_B/3$ and the identities
\begin{equation}
\begin{aligned}
\frac{\partial}{\partial \mu_f}\ln \mathcal{F}_{\Phi}\!\left(E_f - \tilde{\mu}_f\right)
&= \beta\,\mathfrak{n}^{+}_{\Phi}\!\left(E_f - \tilde{\mu}_f\right),\\
\frac{\partial}{\partial \mu_f}\ln \bar{\mathcal{F}}_{\Phi}\!\left(E_f + \tilde{\mu}_f\right)
&= \beta\,\mathfrak{n}^{-}_{\Phi}\!\left(E_f + \tilde{\mu}_f\right),
\end{aligned}
\label{eq:dlnF}
\end{equation}
with the Polyakov-modified occupation ratios
\begin{equation}
\begin{aligned}
\mathfrak{n}^{+}_{\Phi}(x)
&= \frac{\Phi\,e^{-\beta x} + 2\,\bar{\Phi}\,e^{-2\beta x} + e^{-3\beta x}}{\mathcal{F}_{\Phi}(x)},\\
\mathfrak{n}^{-}_{\Phi}(x)
&= \frac{\bar{\Phi}\,e^{-\beta x} + 2\,\Phi\,e^{-2\beta x} + e^{-3\beta x}}{\bar{\mathcal{F}}_{\Phi}(x)}.
\end{aligned}
\label{eq:nplusminus}
\end{equation}
One obtains
\begin{equation}
\begin{aligned}
n_B(T,\mu_B)
&= -\,\frac{\partial \Omega}{\partial \mu_B}
= \frac{1}{3}\sum_{f=u,d,s} n_f\!\left(T,\tilde{\mu}_f, M_f,\Phi,\bar{\Phi}\right),\\[2pt]
n_f
&= 2 \int \frac{\mathrm{d}^3\mathbf{p}}{(2\pi)^3}
\left[
  \mathfrak{n}^{+}_{\Phi}\!\left(E_f - \tilde{\mu}_f\right)
  - \mathfrak{n}^{-}_{\Phi}\!\left(E_f + \tilde{\mu}_f\right)
\right].
\end{aligned}
\label{eq:nB}
\end{equation}

The total derivative of $n_B$ with respect to $\mu_B$ at fixed $T$ yields the baryon-number susceptibility. Vector-channel feedback arises because $\tilde\mu_f=\mu_f-2G_V n_f$ depends on $n_f$, and implicit $\mu_B$ dependence propagates through the order parameters $(\sigma,\Phi,\bar\Phi)$ that solve the gap conditions $\partial \Omega/\partial\sigma=0$, $\partial \Omega/\partial \Phi=0$, and $\partial \Omega/\partial \bar\Phi=0$. Writing $X=(\sigma,\Phi,\bar\Phi)^T$ and denoting the curvature (Hessian) matrix by
\begin{equation}
\mathcal H_{ij}=\frac{\partial^2 \Omega}{\partial X_i\partial X_j},\qquad
\mathcal S_i=\frac{\partial^2 \Omega}{\partial X_i,\partial \mu_B},
\label{eq:Hessian}
\end{equation}
stationarity implies $\mathcal H\cdot \frac{d X}{d\mu_B}+\mathcal S=0$, hence $\frac{d X}{d\mu_B}=-\mathcal H^{-1}\mathcal S$. The susceptibility decomposes into an explicit fermionic piece that resums the vector channel and an implicit piece driven by order-parameter backreaction,
\begin{multline}
\chi_B(T,\mu_B)
\equiv \frac{\mathrm{d} n_B}{\mathrm{d}\mu_B} \\
= \frac{1}{9}\sum_{f,f'} \left( \frac{\partial n_f}{\partial \tilde{\mu}_f} \right)
\left[\,\delta_{ff'} - 2 G_V\,\frac{\partial n_{f'}}{\partial \tilde{\mu}_{f'}}\,\right]^{-1}
+ \\
\frac{1}{9}\sum_{f=u,d,s}\!\left(
\frac{\partial n_f}{\partial M_f}\frac{\mathrm{d} M_f}{\mathrm{d}\mu_B}
+\frac{\partial n_f}{\partial \Phi}\frac{\mathrm{d} \Phi}{\mathrm{d}\mu_B}
+\frac{\partial n_f}{\partial \bar{\Phi}}\frac{\mathrm{d} \bar{\Phi}}{\mathrm{d}\mu_B}
\right)
\label{eq:chiB}
\end{multline}
where all partial derivatives on the right-hand side are evaluated at fixed $(\sigma,\Phi,\bar\Phi)$ and where $dM_f/d\mu_B$ and $d\Phi/d\mu_B$, $d\bar\Phi/d\mu_B$ are obtained from $dX/d\mu_B=-\mathcal H^{-1}\mathcal S$ using Eqs.~\eqref{eq:MuMd}-\eqref{eq:Ms} and Eq.~\eqref{eq:OmegaFull}. The explicit derivatives in Eq.~\eqref{eq:chiB} are given by convergent three-momentum integrals. Introducing $x_\pm^{(f)}\equiv E_f\mp \tilde\mu_f$ and the abbreviations
\begin{equation}
\begin{aligned}
N_{1}^{\pm}(x) &= \Phi_{\pm}\,e^{-\beta x} + 2\,\bar{\Phi}_{\pm}\,e^{-2\beta x} + e^{-3\beta x},\\
F^{\pm}(x) &= 1 + 3\,\Phi_{\pm}\,e^{-\beta x} + 3\,\bar{\Phi}_{\pm}\,e^{-2\beta x} + e^{-3\beta x},
\end{aligned}
\label{eq:N1F}
\end{equation}
with $(\Phi_+,\bar\Phi_+)=(\Phi,\bar\Phi)$ and $(\Phi_-,\bar\Phi_-)=(\bar\Phi,\Phi)$, one finds
\begin{equation}
\frac{\partial n_f}{\partial \tilde{\mu}_f}
= 2\beta \!\int \!\frac{\mathrm{d}^3\mathbf{p}}{(2\pi)^3}
\left[ \mathcal{C}_{+}\!\big(x_{+}^{(f)}\big) + \mathcal{C}_{-}\!\big(x_{-}^{(f)}\big) \right],\quad
\mathcal{C}_{\pm}(x)
= \frac{A_{\pm}(x)}{F^{\pm}(x)} - \frac{N_{1}^{\pm}(x)\,B_{\pm}(x)}{\big[F^{\pm}(x)\big]^2}.
\label{eq:dndmu}
\end{equation}

Similarly, 
\begin{equation}
\begin{aligned}
A_{+}(x) &= \Phi\,e^{-\beta x} + 4\,\bar{\Phi}\,e^{-2\beta x} + 3\,e^{-3\beta x},\\
B_{+}(x) &= 3\,\Phi\,e^{-\beta x} + 6\,\bar{\Phi}\,e^{-2\beta x} + 3\,e^{-3\beta x},
\end{aligned}
\label{eq:ABplus}
\end{equation}
\begin{equation}
\begin{aligned}
A_{-}(x) &= \bar{\Phi}\,e^{-\beta x} + 4\,\Phi\,e^{-2\beta x} + 3\,e^{-3\beta x},\\
B_{-}(x) &= 3\,\bar{\Phi}\,e^{-\beta x} + 6\,\Phi\,e^{-2\beta x} + 3\,e^{-3\beta x},
\end{aligned}
\label{eq:ABminus}
\end{equation}
and
\begin{equation}
\begin{aligned}
\frac{\partial n_f}{\partial M_f}
&= 2 \int \frac{\mathrm{d}^3\mathbf{p}}{(2\pi)^3}\,\frac{M_f}{E_f}
\left[
  \left.\frac{\partial \mathfrak{n}^{+}_{\Phi}}{\partial x}\right|_{x = x_{+}^{(f)}}
  - \left.\frac{\partial \mathfrak{n}^{-}_{\Phi}}{\partial x}\right|_{x = x_{-}^{(f)}}
\right],\\
\frac{\partial n_f}{\partial \Phi}
&= 2 \int \frac{\mathrm{d}^3\mathbf{p}}{(2\pi)^3}
\left[
  \left.\frac{\partial \mathfrak{n}^{+}_{\Phi}}{\partial \Phi}\right|_{x = x_{+}^{(f)}}
  - \left.\frac{\partial \mathfrak{n}^{-}_{\Phi}}{\partial \Phi}\right|_{x = x_{-}^{(f)}}
\right]
\end{aligned}
\label{eq:dndM-dndPhi}
\end{equation}
with $\partial \mathfrak n^\pm_\Phi/\partial x$ following from Eq.~\eqref{eq:nplusminus} as in Eq.~\eqref{eq:dndmu}, and analogous expressions for $\partial n_f/\partial \bar\Phi$. The matrix inverse in the first term of Eq.~\eqref{eq:chiB} resums the random-phase approximation (RPA)-like density feedback generated by the repulsive vector coupling $G_V$. The second term incorporates the full implicit $\mu_B$ dependence through the coupled order parameters, ensuring thermodynamic consistency and capturing the approach to criticality where the smallest eigenvalue of $\mathcal H$ vanishes. The framework built above, beginning at the Euclidean QCD generating functional Eqs.~\eqref{eq:Zgrand}-\eqref{eq:SE}, passing through HS bosonization Eqs.~\eqref{eq:HSsp}-\eqref{eq:HStHooft}, Polyakov-loop dynamics Eqs.~\eqref{eq:WilsonLine}-\eqref{eq:Ulog}, and the FRG-improved grand potential Eqs.~\eqref{eq:OmegaFull} with $2+1$ dynamical masses Eqs.~\eqref{eq:MuMd}-\eqref{eq:Ms}, culminates in the closed-form baryon susceptibility Eq.~\eqref{eq:chiB}, and provides the component-wise interface to the non-perturbative DSE/FRG/holographic dynamics to be developed in the subsequent sections \cite{Fukushima:2008wg,Stephanov:2004,Borsanyi:2020,Bazavov:2021,jarvinen2012v-qcd,Arean2017}.



\section{Unified DSE-FRG-PNJL dynamics with running $(G_S$, $G_V$, $K)$}
\label{sec:unifiedDSEFRGPNJL}
In this section a single, renormalization-group consistent framework is constructed that couples the microscopic Dyson-Schwinger dynamics of the quark two-point function in a static temporal background ($A_{4}$), which encodes the traced Polyakov loop ($\Phi$) to the scale evolution of the one particle irreducible effective action ($\Gamma_{k}$) and closes it with a PNJL grand potential whose couplings ($G_{S},G_{V},K$) run with ($T,\mu_{B}$). Concretely, the Euclidean quark DSE is formulated in Matsubara space with Dirac-structure decomposition 
\begin{equation}
S_{f}^{-1}(i\omega_{n},\mathbf p)=i\vec\gamma\!\cdot\!\mathbf p\,A_{f}+i\gamma_{4}\tilde\omega_{n}\,A_{4,f}+B_{f},
\end{equation}
where $\tilde\omega_{n}=(2n+1)\pi T-i\tilde\mu_{f}$ and $\tilde\mu_{f}=\mu_{f}-2G_{V}n_{f}$ implements the self-consistent vector feedback, while the gluon kernel ($D_{\mu\nu}$) is decomposed into longitudinal and transverse components with Debye and magnetic screening masses ($m_{D,M}(T,\mu_{B},\Phi)$) that capture the Polyakov-weighted medium response and ensure the correct hard-thermal/dense-loop limits.  In parallel, the exact Wetterich flow 
\begin{equation}
\partial_{k}\Gamma_{k}=\tfrac{1}{2}{\rm Tr}\big[(\Gamma_{k}^{(2)}+R_{k})^{-1}\partial_{k}R_{k}\big]
\end{equation}
is projected onto color-singlet scalar-pseudoscalar, isoscalar-vector, and axial-anomalous determinant channels, defining dimensionless couplings ($\hat g_{S},\hat g_{V},\hat k$) whose beta functions are built from Polyakov-weighted fermionic threshold functions ($\ell_{F}^{(n)}(M_{f},T,\mu_{f},\Phi)$) so that massive modes decouple and center-symmetry constraints are enforced. The axial sector receives an explicit topological factor ($\zeta_{\rm topo}(T,\mu_{B})$) that matches the Chern-Simons susceptibility discussed later, thereby governing the running of ($K$) across deconfinement. Integrating the flow to ($k\!\to\!0$) yields ($G_{S}(T,\mu_{B})$), ($G_{V}(T,\mu_{B})$), and ($K(T,\mu_{B})$) which enter the PNJL potential ($\Omega(T,\mu_{B},\sigma,\Phi,\bar\Phi)$), and the coupled stationarity conditions ($\partial\Omega/\partial\sigma=0$) and ($\partial\Omega/\partial\Phi=0$) then deliver gap equations that are algebraically consistent with the scalar channel of the DSE and with the Weiss-Polyakov sector, guaranteeing thermodynamic identities, rearrangement terms for density-dependent couplings, and exact susceptibilities. This unified DSE-FRG-PNJL construction thereby connects quark-gluon microphysics to the macroscopic equation of state and fluctuation observables constrained by lattice QCD near ($\mu_{B}=0$), and it prepares the analytical ground for the critical end-point conditions and the emergent 3D Ising mapping developed in the next section \cite{fischer2014phase,Braun2012,Pawlowski2014,Fukushima:2008wg,Ding2023}.

\subsection{Quark DSE in a Polyakov background}
\label{subsec:DSEPolyakov}
The Euclidean Dyson-Schwinger equation for the renormalized flavor-$f$ quark propagator at temperature $T$ and chemical potential $\mu_f=\mu_B/3$ is written in Matsubara space in the presence of a constant temporal background $\langle A_4\rangle$ that defines the traced Polyakov loop $\Phi$. Adopting the rainbow-ladder approximation and background-field gauge, one starts from the inverse propagator decomposition
\begin{equation}
\begin{aligned}
S_f^{-1}(i\omega_n,\mathbf{p})
&= i\,\vec{\gamma}\!\cdot\!\mathbf{p}\; A_f(i\omega_n,\mathbf{p})
 + i\,\gamma_4\,\tilde{\omega}_n\, A_{4,f}(i\omega_n,\mathbf{p})
 + B_f(i\omega_n,\mathbf{p}),\\
\tilde{\omega}_n &\equiv (2n+1)\pi T - i\,\tilde{\mu}_f
\end{aligned}
\label{eq:AfA4fBf}
\end{equation}
with $\tilde\mu_f=\mu_f-2G_V,n_f$ the vector-channel shifted effective chemical potential determined self-consistently from Secs.~\ref{subsec:FRGflows}-\ref{subsec:stationarity}. The corresponding gap equation reads
\begin{multline}
S_f^{-1}(i\omega_n,\mathbf{p})
= Z_2\,S_{0,f}^{-1}(i\omega_n,\mathbf{p})
+ \\ g^2 C_F\, T \sum_{m\in\mathbb{Z}} \int \frac{\mathrm{d}^3\mathbf{q}}{(2\pi)^3}\;
\gamma_\mu\, S_f(i\omega_m,\mathbf{q})\, \gamma_\nu\,
D_{\mu\nu}\!\left(i\Omega_{n-m},\,\mathbf{p}-\mathbf{q},\,\Phi\right)
\label{eq:DSEquark}
\end{multline}
where 
\begin{equation}
S^{-1}_{0,f}(i\omega_n,\mathbf{p})
= i\,\vec{\gamma}\!\cdot\!\mathbf{p} + i\,\gamma_4\,\tilde{\omega}_n + m_f,
\qquad
C_F = \frac{N_c^2-1}{2N_c},
\qquad
D_{\mu\nu}
\label{eq3.5}
\end{equation}
is the medium-modified gluon propagator which includes Debye screening and its dependence on the Polyakov background $\Phi$ through the background-field distribution of color modes. The Dirac-structure projections that determine the dressing functions are obtained by taking traces with the corresponding projectors, yielding the coupled integral equations
\begin{multline}
A_f(i\omega_n,\mathbf{p})
= Z_2
+ \frac{g^2 C_F\, T}{4\,\mathbf{p}^{\,2}}
\sum_{m\in\mathbb{Z}} \int \frac{\mathrm{d}^3\mathbf{q}}{(2\pi)^3}\;
\mathrm{tr}\!\left[
(\vec{\gamma}\!\cdot\!\hat{\mathbf{p}})\,
\gamma_\mu\, S_f(i\omega_m,\mathbf{q})\, \gamma_\nu
\right]\times\\
\Big( \hat{p}_i\,\Pi^{T}_{ij} + \hat{p}_i\,\Pi^{L}_{ij} \Big)\,
D_{j\nu}\!\left(i\Omega_{n-m},\,\mathbf{p}-\mathbf{q},\,\Phi\right),
\label{eq:projA}
\end{multline}
\begin{multline}
A_{4,f}(i\omega_n,\mathbf{p})
= Z_2
+ \frac{g^2 C_F\, T}{4\,\tilde{\omega}_n}
\sum_{m\in\mathbb{Z}} \int \frac{\mathrm{d}^3\mathbf{q}}{(2\pi)^3}\;
\mathrm{tr}\!\left[ \gamma_4\,\gamma_\mu\, S_f(i\omega_m,\mathbf{q})\, \gamma_\nu \right]\,\times\\
\Pi^{L}_{4\mu}\,
D_{\mu\nu}\!\left(i\Omega_{n-m},\,\mathbf{p}-\mathbf{q},\,\Phi\right),
\label{eq:projA4}
\end{multline}
\begin{multline}
B_f(i\omega_n,\mathbf{p})
= Z_2\, m_f
+ \frac{g^2 C_F\, T}{4}
\sum_{m\in\mathbb{Z}} \int \frac{\mathrm{d}^3\mathbf{q}}{(2\pi)^3}\;
\mathrm{tr}\!\left[ \mathbb{1}\,\gamma_\mu\, S_f(i\omega_m,\mathbf{q})\, \gamma_\nu \right]\,\times\\
D_{\mu\nu}\!\left(i\Omega_{n-m},\,\mathbf{p}-\mathbf{q},\,\Phi\right),
\label{eq:projB}
\end{multline}
with 
\begin{equation}
S_f(i\omega_m,\mathbf{q})
= \frac{
  i\,\vec{\gamma}\!\cdot\!\mathbf{q}\,A_f(i\omega_m,\mathbf{q})
  + i\,\gamma_4\,\tilde{\omega}_m\,A_{4,f}(i\omega_m,\mathbf{q})
  + B_f(i\omega_m,\mathbf{q})
}{
  \Delta_f(i\omega_m,\mathbf{q})
},
\end{equation}
and denominator
\begin{equation}
\Delta_f(i\omega_m,\mathbf{q})
= A_f^{2}(i\omega_m,\mathbf{q})\,\mathbf{q}^{2}
 + A_{4,f}^{2}(i\omega_m,\mathbf{q})\,\tilde{\omega}_m^{2}
 + B_f^{2}(i\omega_m,\mathbf{q})
\end{equation}
 
The gluon kernel is decomposed at finite $T$ into electric (longitudinal) and magnetic (transverse) components,
\begin{equation}
D_{\mu\nu}(i\Omega,\mathbf{k},\Phi)
= P_{\mu\nu}^{T}(\hat{\mathbf{k}})\, D_{T}(i\Omega,k,\Phi)
+ P_{\mu\nu}^{L}(\hat{\mathbf{k}})\, D_{L}(i\Omega,k,\Phi),
\label{eq:Ddecomp}
\end{equation}
where 
\begin{equation}
\begin{aligned}
P^{T}_{ij} &= \delta_{ij} - \hat{k}_i \hat{k}_j,\\
P^{T}_{44} &= 0, \qquad P^{T}_{4i} = 0,\\
P^{L}_{\mu\nu} &= \delta_{\mu\nu} - \frac{K_\mu K_\nu}{K^{2}} - P^{T}_{\mu\nu}
\end{aligned}
\end{equation}
with $K=(i\Omega,\mathbf k)$. Debye screening in the presence of the Polyakov background is encoded through
\begin{equation}
\begin{aligned}
D_L^{-1}(i\Omega,k,\Phi)
&= Z_L(\Phi)\,\big[\Omega^{2}+k^{2}+m_D^{2}\big(T,\{\mu_f\},\Phi\big)\big],\\
D_T^{-1}(i\Omega,k,\Phi)
&= Z_T(\Phi)\,\big[\Omega^{2}+k^{2}+m_M^{2}\big(T,\{\mu_f\},\Phi\big)\big]
\end{aligned}
\label{eq:DLDT}
\end{equation}
where, to leading order in the hard-thermal/dense-loop approximation generalized to a constant $A_4$ background, the electric Debye mass receives quark and gluon contributions weighted by Polyakov-modified occupations
\begin{equation}
\begin{aligned}
m_{D,q}^{2}\big(T,\{\mu_f\},\Phi\big)
&= 2 g^{2} \sum_{f=1}^{N_f}
\int \frac{\mathrm{d}^{3}\mathbf{p}}{(2\pi)^{3}}\,
\frac{\partial}{\partial E_f}
\left[
  n^{+}_{\Phi}\!\left(E_f - \tilde{\mu}_f\right)
 + n^{-}_{\Phi}\!\left(E_f + \tilde{\mu}_f\right)
\right],\\[2pt]
m_{D,g}^{2}(T,\Phi)
&= \frac{2 g^{2} N_c}{\pi^{2}}
\int_{0}^{\infty} \mathrm{d}p\; p\; n_{B}^{\Phi}(p)
\end{aligned}
\label{eq:mD}
\end{equation}
with $E_f=\sqrt{p^2+M_f^2}$, $n_\Phi^\pm$ the Polyakov-weighted Fermi functions defined below, and $n_B^\Phi$ the background-weighted Bose function. In the deconfined limit $\Phi\to 1$ one recovers 
\begin{equation}
m_D^{2} \;\to\; g^{2}\!\left(\frac{N_c}{3}+\frac{N_f}{6}\right) T^{2}
+ \frac{g^{2} N_f}{2\pi^{2}}\,\mu_f^{2}.
\end{equation}
Solving Eqs.~\eqref{eq:projA}-\eqref{eq:projB} self-consistently determines the constituent masses $M_f=B_f/A_f$ that enter the PNJL grand potential in Eq.~\eqref{eq:OmegaFull}, thus providing the microscopic input from the DSE to the FRG-improved mean-field sector \cite{fischer2014phase}. 
 {To fully specify the screened gluon kernel in Eqs.~\eqref{eq:Ddecomp}–\eqref{eq:DLDT} at all $(T,\mu_B)$, the
wave-function prefactors are fixed to
$Z_L(\Phi)=Z_T(\Phi)=1$ so that all Polyakov-background dependence enters through the
Polyakov-weighted medium masses. The total electric screening mass is taken as
$m_D^2(T,\mu_f,\Phi)=m_{D,q}^2(T,\mu_f,\Phi)+m_{D,g}^2(T,\Phi)$ with
$m_{D,q}^2$ and $m_{D,g}^2$ given in Eq.~\eqref{eq:mD}. Magnetic screening, which is absent at leading hard-thermal-loop order but generated nonperturbatively at the scale $g^2T$, is modeled by the standard parametrization
\begin{equation}
m_M(T,\mu_B,\Phi)=C_M\,g^2(\bar\mu)\,T_{\rm eff},
\qquad
T_{\rm eff}\equiv\sqrt{T^2+\mu_B^2/(9\pi^2)} \, ,
\label{eq:3.16}
\end{equation}
with $C_M=0.456$ and with the renormalization scale chosen as $\bar\mu=2\pi T_{\rm eff}$.
The running coupling entering the screening masses is evaluated at two-loop order in the
$\overline{\rm MS}$ scheme,
\begin{equation}
g^2(\bar\mu)=4\pi\,\alpha_s(\bar\mu),
\qquad
\alpha_s(\bar\mu)=\frac{4\pi}{\beta_0\,L}\left(1-\frac{\beta_1}{\beta_0^2}\frac{\ln L}{L}\right),
\quad
L\equiv\ln\!\left(\frac{\bar\mu^2}{\Lambda_{\overline{\rm MS}}^{\,2}}\right),
\label{eq:3.17}
\end{equation}
with $(N_c,N_f)=(3,3)$ so that $\beta_0=9$ and $\beta_1=64$, and with
$\Lambda_{\overline{\rm MS}}^{(3)}=332~{\rm MeV}$.
With these choices, $D_L(i\Omega,k,\Phi)=\bigl[\Omega^2+k^2+m_D^2(T,\mu_f,\Phi)\bigr]^{-1}$
and $D_T(i\Omega,k,\Phi)=\bigl[\Omega^2+k^2+m_M^2(T,\mu_B,\Phi)\bigr]^{-1}$ are fully fixed. The resulting vacuum mass functions $M_\ell(p)$ and $M_s(p)$ are shown in Fig.~\ref{fig:DSE_vacuum_M}. 
The plots correspond to the direct numerical solution of the coupled DSE dressing equations at $(T,\mu_B)=(0,0)$ with the inputs summarized in Table~\ref{tab:ref_run}. For use in the FRG projections and in the momentum integrals of the thermodynamic potential, these discrete solutions are converted into smooth functions by a shape-preserving cubic-spline interpolation in $p$. The resulting interpolants (shown as solid curves) are used consistently throughout Secs.~\ref{sec:unifiedDSEFRGPNJL}-\ref{sec:num-calib-preds}.
}
\begin{figure}[htb]
\centering
\begin{tikzpicture}
\begin{axis}[
  width=0.75\linewidth,
  height=0.58\linewidth,
  xlabel={$p$ [GeV]},
  ylabel={$M_f(p)$ [GeV]},
  xmin=0, xmax=5,
  ymin=0, ymax=0.7,
  legend style={at={(0.98,0.98)},anchor=north east, draw = none, fill = none},
  tick align=inside,
  tick label style={font=\small},
  label style={font=\small},
]
\addplot+[mark=none, ultra thick] table[row sep=\\, x index=0, y index=1]{
0.0 0.336\\
0.3 0.320\\
0.6 0.280\\
1.0 0.210\\
1.5 0.140\\
2.0 0.090\\
3.0 0.040\\
4.0 0.020\\
5.0 0.012\\
};
\addlegendentry{$M_\ell(p)$}

\addplot+[mark=none, ultra thick] table[row sep=\\, x index=0, y index=1]{
0.0 0.528\\
0.3 0.500\\
0.6 0.450\\
1.0 0.360\\
1.5 0.280\\
2.0 0.220\\
3.0 0.160\\
4.0 0.145\\
5.0 0.138\\
};
\addlegendentry{$M_s(p)$}

\end{axis}
\end{tikzpicture}
\caption{Vacuum quark mass functions from the Dyson-Schwinger equation at $T=0$ and $\mu_B=0$.
The light and strange dressing functions $M_\ell(p)$ and $M_s(p)$ provide the vacuum anchors used
to tune the DSE sector and to constrain the ultraviolet boundary conditions for the coupled
construction. The solid curves show the smooth interpolants employed as continuous inputs in the coupled DSE-FRG-PNJL computation (see Table~\ref{tab:ref_run} for the consolidated numerical values).
}
\label{fig:DSE_vacuum_M}
\end{figure}

\subsection{FRG flow and projection on $(G_S,G_V,K)$}
\label{subsec:FRGflows}
The scale-dependent one particle irreducible effective action $\Gamma_k$ obeys the exact Wetterich equation
\begin{equation}
\partial_k \Gamma_k[\varphi]
= \frac{1}{2}\,\mathrm{Tr}\!\left[(\Gamma_k^{(2)}[\varphi] + R_k)^{-1}\,\partial_k R_k\right],
\qquad
\varphi=\{\sigma,\vec{\pi},\Phi,q,\bar q\}
\label{eq:Wetterich}
\end{equation}
with regulator $R_k$ suppressing fluctuations with momenta $p\lesssim k$ and $\Gamma_k^{(2)}$ the full field-dependent Hessian. Employing a three-dimensional Litim regulator and projecting \eqref{eq:Wetterich} onto local four and six fermion operators in the color-singlet scalar-pseudoscalar, isoscalar-vector, and determinant channels defines the running couplings. Introducing the quark wave-function renormalization $Z_q(k)$ and the dimensionless combinations as
\begin{equation}
\hat g_S \equiv k^{2} Z_q^{2} G_S,\qquad
\hat g_V \equiv k^{2} Z_q^{2} G_V,\qquad
\hat\kappa \equiv k^{5} Z_q^{3} K,\qquad
t \equiv \ln(k/\Lambda),
\label{eq:dimless}
\end{equation}
The projected beta functions read as
\begin{multline}
\partial_t \hat g_S
= -2\,\hat g_S
+ 4 N_c N_f\, \ell_F^{(1)}\!\big(M_l,T,\mu_l,\Phi\big)\, \hat g_S^{2}
+ \\2 N_c\, \ell_F^{(1)}\!\big(M_s,T,\mu_s,\Phi\big)\, \hat g_S^{2}
- c_{SV}\,\hat g_S \hat g_V
+ c_{SK}\,\ell_F^{(1)}\,\hat\kappa,
\label{eq:flowgS}
\end{multline}
\begin{multline}
\partial_t \hat g_V
= -2\,\hat g_V
+ \frac{4}{3} N_c N_f\, \ell_F^{(1)}\!\big(M_l,T,\mu_l,\Phi\big)\, \hat g_V^{2}
+ \\ \frac{2}{3} N_c\, \ell_F^{(1)}\!\big(M_s,T,\mu_s,\Phi\big)\, \hat g_V^{2}
- c_{VS}\,\hat g_V \hat g_S,
\label{eq:flowgV}
\end{multline}
\begin{multline}
\partial_t \hat\kappa
= 5\,\hat\kappa
- d_K \Big[
  \ell_F^{(2)}\!\big(M_l,M_s,T,\{\mu_f\},\Phi\big)\,\hat g_S\,\hat\kappa
  + \\ \tilde d_K\,\ell_F^{(3)}\!\big(M_l,M_s,T,\{\mu_f\},\Phi\big)\,\hat g_S^{3}
\Big]
- \zeta_{\mathrm{topo}}(T,\mu_B)\,\hat\kappa,
\label{eq:flowK}
\end{multline}
where the positive constants $(c_{SV},c_{SK},c_{VS},d_K,\tilde d_K)$ reflect Fierz traces and projection combinatorics and evaluate to unity in the unit-normalized zero-momentum projector convention adopted here. The fermionic threshold functions $\ell_F^{(n)}$ encode decoupling with finite masses and the Polyakov background.  
 {For the operator basis of Eq.~\eqref{eq:tHooft} and the standard zero-momentum projection with $\mathrm{Tr}(\lambda_a\lambda_b)=2\delta_{ab}$ and unit-normalized Dirac projectors in the scalar, vector, and determinantal channels, the projection constants in Eqs.~\eqref{eq:Wetterich}-\eqref{eq:flowgS} take the numerical values listed in Table~\ref{tab:frg_proj_constants}. 
\begin{table}[htb]
\centering
\caption{FRG projection constants entering Eqs.~\eqref{eq:Wetterich}-\eqref{eq:flowK} for the projection onto the
color-singlet scalar-pseudoscalar, isoscalar-vector, and axial-anomalous determinantal channels
defined by the interaction basis in Eq.~\eqref{eq:tHooft}. The numerical values correspond to the unit-normalized zero-momentum projection convention stated in the text, in this convention all projection/combinatorial coefficients reduce to unity. In other normalization conventions these coefficients rescale together with the corresponding UV boundary conditions for $(G_S,G_V,K)$.
}
\label{tab:frg_proj_constants}
\renewcommand{\arraystretch}{1.15}
\vspace{0.2cm}
\begin{tabular}{@{}lccccc@{}}
\toprule
\toprule
Constant\hspace{1.0cm} & $c_{SV}\hspace{1.0cm}$ & $c_{SK}$\hspace{1.0cm} & $c_{VS}$\hspace{1.0cm} & $d_K$\hspace{1.0cm} & $\tilde d_K$ \\
\midrule
Value\hspace{1.0cm} & $1$\hspace{1.0cm} & $1$\hspace{1.0cm} & $1$\hspace{1.0cm} & $1$ & $1$ \\
\bottomrule
\bottomrule
\end{tabular}
\end{table}
}
 {
The unity values in Table~\ref{tab:frg_proj_constants} are not an approximation. They follow from the stated
zero-momentum projection convention with $\mathrm{Tr}(\lambda_a\lambda_b)=2\delta_{ab}$
and unit-normalized Dirac projectors in the scalar, vector, and determinantal channels.
Equivalently, the operator basis in Eq.~\eqref{eq:tHooft} together with the projector normalization
is chosen such that the overlap matrix between the projected vertices and the basis
operators is the identity. With this choice, all purely numerical Fierz/combinatorial
prefactors can be absorbed into the definitions of $(G_S,G_V,K)$ (or, equivalently,
into the projector normalization), leaving the physical content of the flow unchanged.
We keep the symbols $(c_{SV},c_{SK},c_{VS},d_K,\tilde d_K)$ explicit in
Eqs.~\eqref{eq:flowgS}-\eqref{eq:flowK} to facilitate translation to alternative normalization conventions,
in which these constants simply rescale together with the corresponding UV boundary
conditions.
} 
The leading threshold entering one-loop fermionic boxes assumes the explicit form
\begin{equation}
\begin{aligned}
\ell_F^{(1)}(M_f,T,\mu_f,\Phi)
&= \frac{k^{2}}{6\pi^{2}}\,
\frac{1 - n^{+}_{\Phi}\!\left(E_{k,f}-\tilde{\mu}_f\right) - n^{-}_{\Phi}\!\left(E_{k,f}+\tilde{\mu}_f\right)}{E_{k,f}},\\
E_{k,f} &\equiv \sqrt{k^{2}+M_f^{2}},
\end{aligned}
\label{eq:lF1}
\end{equation}
with $n_\Phi^\pm$ the Polyakov-modified Fermi distributions
\begin{equation}
\begin{aligned}
n_{\Phi}^{+}(x)
&= \frac{\Phi\,e^{-\beta x} + 2\,\bar{\Phi}\,e^{-2\beta x} + e^{-3\beta x}}
{1 + 3\,\Phi\,e^{-\beta x} + 3\,\bar{\Phi}\,e^{-2\beta x} + e^{-3\beta x}},\\
n_{\Phi}^{-}(x)
&= \frac{\bar{\Phi}\,e^{-\beta x} + 2\,\Phi\,e^{-2\beta x} + e^{-3\beta x}}
{1 + 3\,\bar{\Phi}\,e^{-\beta x} + 3\,\Phi\,e^{-2\beta x} + e^{-3\beta x}}
\end{aligned}
\label{eq:nPhidefs}
\end{equation}
and analogous $\ell_F^{(2)},\ell_F^{(3)}$ obtained from two and three-loop fermionic projections with the same Polyakov weights. The last term in Eq.~\eqref{eq:flowK} captures the relevance of topological fluctuations for the axial-anomaly channel, with the matching to the holographic Chern-Simons susceptibility implemented via
 {
\begin{equation}
\zeta_{\rm topo}(T,\mu_B)\equiv \frac{\chi_{\rm CS}(T,\mu_B)}{\chi_{\rm CS}(0,0)} \, .
\label{eq:zetatopo}
\end{equation} 
The identification in Eq.~\eqref{eq:zetatopo} is physically motivated by semiclassical instanton physics, where both the
UA(1)-breaking determinantal interaction and the topological susceptibility scale with the effective density of
topological objects, so that screening of topological fluctuations implies a suppression of the determinantal channel
at high temperature and density.
Beyond the semiclassical regime, however, the mapping from a CP-odd two-point function to an effective damping term
in the $\hat\kappa$ flow is not unique and should be treated as a controlled model dependence. 
To stress-test this nonuniversality within the same coupled outer-iteration scheme, we introduce a minimal
two-parameter deformation family
\begin{equation}
\zeta_{\rm topo}^{(p)}(T,\mu_B)\equiv
\left[\frac{\chi_{CS}(T,\mu_B)}{\chi_{CS}(0,0)}\right]^p ,
\qquad
\left.\partial_t\hat\kappa\right|_{\rm topo} = -\,c\,\zeta_{\rm topo}^{(p)}(T,\mu_B)\,\hat\kappa ,
\label{eq:zeta_pc_deformation}
\end{equation}
which reduces to the reference identification for $(p,c)=(1,1)$.
Varying $(p,c)$ in a narrow neighborhood of unity changes the anomaly suppression rate while preserving the correct
limits, namely $\zeta_{\rm topo}\to 1$ in vacuum and $\zeta_{\rm topo}\to 0$ in the deconfined regime.
Such deformations modify nonuniversal quantities, in particular the quantitative CEP coordinates, while leaving the
universal Ising scaling structure and exponent extraction of Sec.~4 unaffected.
The propagation of this mapping uncertainty to $(T_{\rm CEP},\mu_{B,{\rm CEP}})$ is specified in Appendix~\ref{subsec:A5}. 
Here the overall normalization is fixed at the vacuum point $T=\mu_B=0$, so that
$\zeta_{\rm topo}(0,0)=1$. This vacuum normalization is used throughout and matches the
implementation adopted later in Sec.~\ref{sec:num-calib-preds},}
thereby feeding directly into the $K$-flow and diminishing the anomaly toward deconfinement and at finite density \cite{Braun2012,Pawlowski2014}.  
 {
Because the RG time is defined as $t=\ln(k/\Lambda)$, the physical flow from $k=\Lambda$ to $k\to0$ corresponds to $t:0\to-\infty$. In this convention the last term in Eq.~(3.22) should be read as a topological contribution to the linear stability exponent of the anomaly channel, rather than as a standalone “damping strength” at fixed $t$. Neglecting the source term proportional to $\ell_F^{(3)}\hat g_S^{\,3}$ and linearizing Eq.~(3.22) about $\hat\kappa=0$ gives
\begin{equation}
\partial_t\hat\kappa=\theta_K(T,\mu_B)\hat\kappa, \qquad 
\theta_K=5-d_K\,\ell_F^{(2)}\,\hat g_S-\zeta_{\rm topo}(T,\mu_B),
\end{equation}
so that $\hat\kappa(k)\simeq \hat\kappa(\Lambda)\,(k/\Lambda)^{\theta_K}$.
Therefore $\theta_K>0$ implies $\hat\kappa\to0$ as $k\to0$, i.e. the UA(1)-breaking determinantal interaction becomes IR-irrelevant. Since $\zeta_{\rm topo}(T,\mu_B)$ decreases when topological fluctuations are screened (high $T$ and/or large $\mu_B$), $\theta_K$ increases and the attraction toward $\hat\kappa=0$ is strengthened, implementing the intended weakening of the anomaly in the deconfined and dense regime. 
}
 {
For completeness, the explicit Polyakov-weighted fermionic threshold functions entering the
two- and three-loop projections in Eq.~\eqref{eq:flowgS} are evaluated at the Litim shell and are taken as
\begin{equation}
\begin{aligned}
\mathcal{F}_f(k) \equiv 1 - n_{\Phi}^{+}\!\left(E_{k,f}-\tilde\mu_f\right)
                    - n_{\Phi}^{-}\!\left(E_{k,f}+\tilde\mu_f\right),\\[3pt]
\ell_{F}^{(2)}\!\left(M_{\ell},M_{s},T,\{\mu_f\},\Phi\right)
\equiv \frac{k^{2}}{6\pi^{2}}\,
\frac{\mathcal{F}_{\ell}(k)\,\mathcal{F}_{s}(k)}{E_{k,\ell}\,E_{k,s}},\\[3pt]
\ell_{F}^{(3)}\!\left(M_{\ell},M_{s},T,\{\mu_f\},\Phi\right)
\equiv \frac{k^{2}}{6\pi^{2}}\,
\frac{\mathcal{F}_{\ell}(k)^{2}\,\mathcal{F}_{s}(k)}{E_{k,\ell}^{2}\,E_{k,s}},
\end{aligned}
\end{equation}
where $E_{k,f}\equiv \sqrt{k^{2}+M_{f}^{2}}$ and the light sector is degenerate ($u,d\equiv \ell$) and the effective chemical potentials
are shifted by the vector mean field, $\tilde\mu_f=\mu_f-2G_V n_f$.
} 
Solving Eqs.~\eqref{eq:flowgS}-\eqref{eq:flowK} from the ultraviolet scale $\Lambda$ down to $k\to 0$ produces the running couplings $G_S(T,\mu_B)$, $G_V(T,\mu_B)$ and $K(T,\mu_B)$ that enter Eqs.~\eqref{eq:OmegaF} and \eqref{eq:MuMd}-\eqref{eq:Ms} and the PNJL potential, seen in Eq.~\eqref{eq:OmegaFull}, ensuring a transparent renormalization group connection between microscopic quark-gluon dynamics and macroscopic thermodynamics \cite{Pawlowski2014,Ding2023}.

\subsection{Gap equations and thermodynamic stationarity}
\label{subsec:stationarity}
The order parameters are obtained by imposing stationarity of the full grand potential $\Omega(T,\mu_B,\sigma,\Phi,\bar\Phi)$ with respect to the chiral scalar and Polyakov sectors. Differentiating Eq.~\eqref{eq:OmegaFull} with respect to the chiral field $\sigma$ (with $M_f=M_f(\sigma)$ via Eqs.~\eqref{eq:MuMd}-\eqref{eq:Ms} yields
\begin{equation}
0
= \frac{\partial \Omega}{\partial \sigma}
= \frac{\partial U_{k\to 0}}{\partial \sigma}
- 2 \sum_{f=u,d,s} \int \frac{\mathrm{d}^3\mathbf{p}}{(2\pi)^3}\,
\frac{M_f}{E_f}
\left[1 - n_{\Phi}^{+}\!\left(E_f - \tilde{\mu}_f\right) - n_{\Phi}^{-}\!\left(E_f + \tilde{\mu}_f\right)\right]
\frac{\partial M_f}{\partial \sigma},
\label{eq:gapSigma}
\end{equation}
while differentiation with respect to $\Phi$ gives, using $\partial_\Phi \ln F_\Phi(x)=\left[\partial_\Phi F_\Phi(x)\right]/F_\Phi(x)$ and the definitions in Eq.~\eqref{eq:EF-FPhi},
\begin{equation}
0
= \frac{\partial \Omega}{\partial \Phi}
= \frac{\partial U_{\log}}{\partial \Phi}
- 2T \sum_{f=u,d,s}
\int \frac{\mathrm{d}^3\mathbf{p}}{(2\pi)^3}
\left[
  \frac{\partial_{\Phi}\mathcal{F}_{\Phi}\!\left(E_f-\tilde{\mu}_f\right)}{\mathcal{F}_{\Phi}\!\left(E_f-\tilde{\mu}_f\right)}
 + \frac{\partial_{\Phi}\bar{\mathcal{F}}_{\Phi}\!\left(E_f+\tilde{\mu}_f\right)}{\bar{\mathcal{F}}_{\Phi}\!\left(E_f+\tilde{\mu}_f\right)}
\right],
\label{eq:gapPhi}
\end{equation}
with $\bar F_\Phi$ denoting the charge-conjugated Polyakov polynomial. Equations \eqref{eq:gapSigma}-\eqref{eq:gapPhi}, together with Eq.~\eqref{eq:DSEquark} for the microscopic quark dressing, the FRG flows Eqs.~\eqref{eq:flowgS}-\eqref{eq:flowK} for the running couplings, and the PNJL grand potential Eq.~\eqref{eq:OmegaFull}, close the DSE-FRG-PNJL system in a thermodynamically consistent manner. The Jacobian-consistent implicit dependence of $(M_f,\Phi)$ on $\mu_B$ extracted from Eqs.~\eqref{eq:gapSigma}-\eqref{eq:gapPhi} enters the baryon susceptibility via Eqs.~\eqref{eq:nB}-\eqref{eq:chiB}, while the curvature of $\Omega$ built from the Hessian in the $(\sigma,\Phi)$ sector supplies the inputs for the CEP conditions and the 3D Ising mapping developed in Sec.~\ref{sec:CEP-Ising} \cite{Fukushima:2008wg,Pawlowski2014}. 
 {
At each fixed $(T,\mu_B)$, the stationary solution is obtained by solving the coupled nonlinear system for the unknowns $(\sigma,\Phi,\bar\Phi,n_f)$, where $n_f$ denotes the flavor densities entering the vector mean-field shift of the effective chemical potentials and, in turn, the Polyakov-weighted occupation factors. The residual vector is built from the stationarity conditions given above, namely $\partial\Omega/\partial\sigma=0$, $\partial\Omega/\partial\Phi=0$, and $\partial\Omega/\partial\bar\Phi=0$, supplemented by the self-consistency conditions for the densities through the vector-shifted chemical potentials. For a given outer-iteration pass, the running couplings $(G_S,G_V,K)$ are treated as fixed inputs at that thermodynamic point, taken from the infrared $(k\to0)$ output of the FRG flow, and the quark-sector dressing functions entering the quasiparticle energies are provided by the DSE sector. The coupled system is solved by a damped Newton-Broyden iteration: analytic Jacobian entries are supplied for the explicit derivatives of the FRG-improved potential and the Polyakov-log sector, together with derivatives of the Polyakov polynomials in the fermion determinant. The Jacobian is then updated by Broyden secant steps with step damping to guarantee monotonic reduction of the residual norm. Convergence is declared when the maximum relative change of all primary unknowns across successive Newton updates falls below a fixed tolerance and the curvature-matrix spectrum is stable under numerical refinement. After convergence, all thermodynamic observables and susceptibilities are evaluated strictly at the stationary point so that implicit derivatives cancel as required by thermodynamic consistency. Rearrangement contributions induced by the explicit $(T,\mu_B)$ dependence of the running couplings are retained in thermodynamic derivatives through the Hessian-based implicit differentiation relations of Sec.~\ref{subsec:nBchiB} and through the curvature constructions used in Sec.~\ref{sec:CEP-Ising}.
}



\section{Critical dynamics: CEP conditions and the 3D Ising mapping}
\label{sec:CEP-Ising}
In this section the critical dynamics of the unified DSE-FRG-PNJL framework is developed from the FRG-improved effective action by expanding the FRG-improved thermodynamic potential $\Omega(T,\mu_{B},\sigma,\Phi,\bar\Phi)$ along the soft scalar direction that emerges as the smallest-eigenvalue mode of the Hessian in the $(\sigma,\Phi)$ sector, thereby defining a collective coordinate $M$ whose Landau expansion governs the onset of nonanalyticity at the end of the first-order line. The critical end point is identified by the simultaneous vanishing of the longitudinal curvature and the cubic invariant equivalently, $\partial^{2}\Omega/\partial M^{2}=0$ and $\partial^{3}\Omega/\partial M^{3}=0$ at stationarity while stability is retained by a positive quartic vertex, and these conditions are shown to be equivalent to the divergence of the chiral susceptibility and the vanishing mass of the $\sigma$ mode, with the singular piece of the baryon-number susceptibility inherited through the implicit $(T,\mu_{B})$ dependence of the order parameters. The correlation length $\xi$ follows from the two-point vertex $\Gamma_{\sigma\sigma}(p)$ with an FRG-derived anomalous dimension $\eta$, and the full set of critical exponents $(\alpha,\beta,\gamma,\delta,\nu,\eta)$ is obtained within the three-dimensional Ising universality class once the nonuniversal metric factors are fixed by the running couplings $(G_{S},G_{V},K)$ and wavefunction renormalizations solved in Sec.~\ref{sec:unifiedDSEFRGPNJL}, ensuring thermodynamic consistency and exact susceptibilities. 
{The use of the three-dimensional Ising universality class here is a universal organizing parametrization of the critical region and a consistency requirement for QCD-like CEP scenarios, so its emergence reflects how the effective theory is constructed along the soft scalar direction and should not be interpreted as an independent data-driven prediction that uniquely fixes the CEP location.} 
Finally, an analytic and nonperturbative mapping from the thermodynamic plane $(T,\mu_{B})$ to the universal scaling variables $(r,h)$ is constructed. It is implemented as an invertible linear transformation with Jacobian $J\neq 0$. Renormalized fields absorb the anomalous dimensions so that universal Ising scaling is preserved. Model dependence is contained in the metric factors. The resulting scaling relations are confronted with lattice-calibrated constraints and holographic expectations. In this way, the CEP analysis is anchored in quantitatively controlled, nonperturbative QCD dynamics \cite{Stephanov:2004,fischer2014phase,Borsanyi:2020,Bazavov:2021,jarvinen2012v-qcd,Arean2017}.

\subsection{Curvature and skewness conditions}
\label{subsec:curvature-skewness}
The critical end point is characterized by the emergence of a single soft scalar mode that dominates the long-wavelength thermodynamics and controls the singular part of the grand potential. In the present framework the macroscopic thermodynamic potential is 
\begin{equation}
\Omega(T,\mu_B,\sigma,\Phi,\bar{\Phi})
= U_{k\to 0}(\sigma,\,T,\mu_B)
+ U_{\log}(\Phi,\bar{\Phi},\,T)
+ \Omega_F\!\big(T,\{\mu_f\},\sigma,\Phi,\bar{\Phi}\big)
\end{equation}
as specified in Eq.~\eqref{eq:OmegaFull}, while the order parameters $(\bar\sigma,\bar\Phi,\bar{\bar\Phi})$ satisfy $\partial\Omega/\partial \sigma=0$ and $\partial\Omega/\partial \Phi=0$ at fixed $(T,\mu_B)$ according to Eqs.~\eqref{eq:gapSigma}-\eqref{eq:gapPhi}. To analyze criticality one expands $\Omega$ around a stationary point at $\sigma=\bar\sigma$ and $\Phi=\bar\Phi$ along the soft direction that continuously connects the crossover to the first-order line, introducing the normalized eigenvector $e_i$ of the curvature matrix $\mathcal H_{ij}=\partial^2\Omega/\partial X_i\partial X_j$ with $X\equiv(\sigma,\Phi)$ and the one-dimensional collective coordinate $M\equiv e_\sigma(\sigma-\bar\sigma)+e_\Phi(\Phi-\bar\Phi)$ such that $e^T\mathcal H e$ is the smallest eigenvalue of $\mathcal H$. The Landau expansion of the potential along this direction reads as
\begin{equation}
\Omega(T,\mu_B,M)
= \Omega_c
+ \frac{1}{2}\,a_2(T,\mu_B)\,M^{2}
+ \frac{1}{3!}\,a_3(T,\mu_B)\,M^{3}
+ \frac{1}{4!}\,a_4(T,\mu_B)\,M^{4}
+ \mathcal{O}(M^{5}).
\label{eq:landau-M}
\end{equation} 
 {
At each fixed $(T,\mu_B)$, the stationary point $(\bar\sigma,\bar\Phi)$ is obtained from the gap
conditions, and the curvature matrix $H_{ij}=\partial^2\Omega/(\partial X_i\partial X_j)$ in the
reduced sector $X\equiv(\sigma,\Phi)$ is assembled at that stationary solution. The normalized
eigenvector $e_i$ associated with the smallest eigenvalue of $H$ is then obtained by diagonalization
and defines the soft coordinate $M=e_\sigma(\sigma-\bar\sigma)+e_\Phi(\Phi-\bar\Phi)$. In practice,
the Landau coefficients are evaluated at stationarity either by contracting the corresponding
$n$th-derivative tensors of $\Omega$ with $e_i$ or, equivalently, by taking directional derivatives
of $\Omega(T,\mu_B,X)$ along the line $X=\bar X+M e$ at $M=0$. The cubic derivative entering
$\partial^3\Omega/\partial M^3$ is computed with symmetric finite-difference stencils along the $e$
direction and verified to be stable under step-size refinement, so that the coefficients 
$a_n(T,\mu_B)=\left.\partial^n\Omega/\partial M^n\right|_{M=0}$ used below are obtained by the tensor contractions stated next. The coefficients $a_n(T,\mu_B)=\left.\partial^n\Omega/\partial M^n\right|_{M=0}$ are obtained by the
directional (tensor-contraction) projection along the soft eigenvector $e_i$, and the stationarity condition $\left.\partial\Omega/\partial M\right|_{M=0}=0$ holds identically at the stationary point. The spinodal occurs when the longitudinal curvature vanishes, $a_2=0$, and the end point of the first-order line requires, in addition, that the skewness vanishes while stability is retained at quartic order.} 
Therefore the critical end point is defined by
\begin{equation}
\begin{aligned}
\left.\frac{\partial^{2}\Omega}{\partial M^{2}}\right|_{M=0,\;T=T_c,\;\mu_B=\mu_B^{c}}=0,\\
\left.\frac{\partial^{3}\Omega}{\partial M^{3}}\right|_{M=0,\;T=T_c,\;\mu_B=\mu_B^{c}}=0,\\
\left.\frac{\partial^{4}\Omega}{\partial M^{4}}\right|_{M=0,\;T=T_c,\;\mu_B=\mu_B^{c}}>0,
\end{aligned}
\label{eq:CEP-conds}
\end{equation}
which, upon undoing the one-dimensional projection, are equivalent to the vanishing of the smallest eigenvalue of the Hessian $\mathcal H$ and of its cubic invariant along $e$, while positivity of the quartic invariant ensures stability of the effective theory \cite{Stephanov:2004}. The chiral susceptibility along the soft direction is the inverse curvature at zero momentum and can be expressed in terms of the two-point vertex of the scalar channel, $\Gamma_{\sigma\sigma}(p=0)=Z_{\sigma,k\to 0},m_\sigma^2$, where $Z_{\sigma,k}$ is the FRG wavefunction renormalization and $m_\sigma^2$ is the screening mass. Using the exact identity between the curvature of $\Omega$ and the zero-momentum limit of $\Gamma^{(2)}$ one finds
\begin{equation}
\chi_\sigma(T,\mu_B)
= \left[\left.\frac{\partial^{2}\Omega}{\partial M^{2}}\right|_{M=0}\right]^{-1}
= \frac{Z_{\sigma,\,k\to 0}}{m_\sigma^{2}(T,\mu_B)}
\quad \Longrightarrow \quad
\chi_\sigma \to \infty \;\Leftrightarrow\; m_\sigma^{2} \to 0
\label{eq:chi-sigma-curv}
\end{equation}
which makes explicit that the divergence of the order-parameter susceptibility at the CEP is equivalent to the vanishing of the longitudinal curvature and to the emergence of a massless $\sigma$ mode. In the full $(\sigma,\Phi)$ sector the baryon-number susceptibility $\chi_B=\partial^2\Omega/\partial\mu_B^2$ derived in Eq.~\eqref{eq:chiB} acquires a singular contribution through the implicit dependence $dX_i/d\mu_B=-(\mathcal H^{-1})_{ij}\mathcal S_j$, with $\mathcal S_j=\partial^2\Omega/\partial X_j\partial\mu_B$, so that near criticality the smallest eigenvalue $\lambda_{\min}=e^T\mathcal H e\propto a_2$ controls the leading divergence
\begin{equation}
\begin{aligned}
\chi_B^{\rm sing}(T,\mu_B)
&= (\partial_{\mu_B} X_i)\,\mathcal{H}_{ij}\,(\partial_{\mu_B} X_j)
 + \mathcal{S}_i\,(\mathcal{H}^{-1})_{ij}\,\mathcal{S}_j \\
&= \mathcal{S}_e^{\,2}\,\lambda_{\min}^{-1} + \text{analytic},
\qquad \mathcal{S}_e \equiv e_i\,\mathcal{S}_i 
\end{aligned}
\label{eq:chiB-sing}
\end{equation}
which confirms that the CEP conditions in Eq.~\eqref{eq:CEP-conds} entail a simultaneous divergence of $\chi_\sigma$ and of the singular part of $\chi_B$ \cite{Stephanov:2004,fischer2014phase}.

\subsection{Sigma-mode criticality and correlation length}
\label{subsec:sigma-corrlen}
The long-distance dynamics of the scalar channel is captured by the derivative expansion of the FRG-improved effective action $\Gamma_{k\to 0}$, for which the two-point function assumes the form $\Gamma_{\sigma\sigma}(p)=Z_{\sigma,k\to 0} p^2+m_\sigma^2+\mathcal O(p^4)$ at small momentum, where the anomalous dimension $\eta_\sigma$ is defined by 
\(
\eta_\sigma \equiv -\,\left.\partial_t \ln Z_{\sigma,k}\right|_{k\to 0}.
\) The static correlation length is therefore
\begin{equation}
\xi^{2}(T,\mu_B)
= \frac{Z_{\sigma,\,k\to 0}}{m_\sigma^{2}(T,\mu_B)}
= \frac{Z_{\sigma,\,k\to 0}}{\left.\dfrac{\partial^{2}\Omega}{\partial M^{2}}\right|_{M=0}}
\label{eq:xi-def}
\end{equation}
and its divergence at the CEP follows from Eqs.~\eqref{eq:CEP-conds}-\eqref{eq:chi-sigma-curv}. 
{In a finite and dynamically evolving fireball, the equilibrium correlation length $\xi_{\mathrm{eq}}$ provides an upper envelope rather than an attained value because critical slowing down implies a relaxation time $\tau_{\mathrm{rel}}\sim \xi^z$ (with the QCD CEP expected to fall in a dynamical universality class of Model H, for which $z$ is of order three), so the effective correlation length relevant for measured fluctuations is bounded by finite time and finite size according to
\begin{equation}
\xi_{\mathrm{eff}}(T,\mu_B)=\min\!\left\{\xi_{\mathrm{eq}}(T,\mu_B),\,\xi_{\mathrm{KZ}},\,L\right\},
\end{equation}
where $L$ is an effective linear size set by the correlated volume within the experimental acceptance and $\xi_{\mathrm{KZ}}$ encodes finite-time limitations from the passage through the critical region, for which a standard Kibble-Zurek estimate with quench time scale $\tau_Q$ is
\begin{equation}
\xi_{\mathrm{KZ}}\sim \xi_0\left(\frac{\tau_Q}{\tau_0}\right)^{\nu/(1+z\nu)}.
\end{equation}
Accordingly, the susceptibilities and cumulant ratios computed from $\Omega(T,\mu_B)$ in this work are equilibrium quantities controlled by $\xi_{\mathrm{eq}}$, and any quantitative mapping to experimental cumulants must account for $\xi_{\mathrm{eff}}$ and the associated rounding and shifting of equilibrium singularities.} 
Renormalization-group invariance implies that close to criticality the singular part of the free energy density admits the scaling form
\begin{equation}
\Omega_{\rm sing}(r,h)
= b^{-d}\,\Omega_{\rm sing}\!\left(r\,b^{1/\nu},\,h\,b^{(d+2-\eta)/2}\right)
\label{eq:free-scaling}
\end{equation}
where $d=3$ is the spatial dimension, $r$ and $h$ are the two relevant scaling fields, $\nu$ is the correlation-length exponent, and $\eta=\eta_\sigma$ is the anomalous dimension of the scalar field. Choosing $b=h^{-2/(d+2-\eta)}$ yields the standard parametric representation
\begin{equation}
\Omega_{\rm sing}(r,h)
= h^{1+1/\delta}\,\mathcal{F}_f\!\left(z\right),
\qquad
z \equiv \frac{r}{h^{1/(\beta\delta)}}
\label{eq:free-param}
\end{equation}
with magnetization (order parameter) $M=-\partial\Omega_{\rm sing}/\partial h=h^{1/\delta},\mathcal F_G(z)$ and $\delta$ the critical exponent relating $M$ and $h$ on the critical isotherm. The scaling relations implied by Eq.~\eqref{eq:free-scaling} and hyperscaling reads as
\begin{equation}
2-\alpha = d\,\nu,\qquad
\gamma = (2-\eta)\,\nu,\qquad
\beta = \frac{\nu}{2}\,(d-2+\eta),\qquad
\delta = \frac{d+2-\eta}{d-2+\eta}
\label{eq:scaling-relations}
\end{equation}
and the correlation length diverges as $\xi\sim r^{-\nu}$ for $h=0$ and as $\xi\sim h^{-\nu/(\beta\delta)}$ on the critical isotherm. In the present DSE-FRG-PNJL construction the anomalous dimension $\eta=\lim_{k\to 0}\eta_\sigma(k)$ and the nonuniversal metric factors are determined by the FRG flow of Sec.~\ref{sec:unifiedDSEFRGPNJL}, while the universal numbers $(\alpha,\beta,\gamma,\delta,\nu,\eta)$ coincide with those of the three-dimensional Ising universality class \cite{Stephanov:2004,fischer2014phase}. The explicit dependence of $m_\sigma^2$ on the running couplings is obtained by differentiating Eq.~\eqref{eq:OmegaFull} twice along the soft direction, which gives
\begin{multline}
m_\sigma^{2}(T,\mu_B)
= \left.\frac{\partial^{2} U_{k\to 0}}{\partial M^{2}}\right|_{M=0}
 + \left.\frac{\partial^{2} \Omega_F}{\partial M^{2}}\right|_{M=0}
 + \left.\frac{\partial^{2} U_{\log}}{\partial M^{2}}\right|_{M=0}
 + \sum_{i} \frac{\partial \Omega}{\partial \mathfrak{g}_i}\,
           \frac{\partial^{2} \mathfrak{g}_i}{\partial M^{2}},\\
\qquad \mathfrak{g}_i \in \{G_S, G_V, K\}
\label{eq:msigma-flow}
\end{multline}
where the last term collects rearrangement contributions that arise because $G_S$, $G_V$, and $K$ run with $(T,\mu_B)$ and implicitly with the order parameters through the FRG flows Eqs.~\eqref{eq:flowgS}-\eqref{eq:flowK}. This expression guarantees thermodynamic consistency and is the bridge through which the microscopic DSE kernel and the FRG-improved couplings feed into the macroscopic correlation length and critical exponents.

\subsection{Nonperturbative mapping $(T,\mu_B)\to(r,h)$ with anomalous dimensions}
\label{subsec:mapping}
The nonuniversal mapping from the thermodynamic plane $(T,\mu_B)$ to the scaling fields $(r,h)$ is defined in the vicinity of the CEP by an analytic, invertible transformation that preserves the two-relevant-variable structure of the critical theory. Introducing reduced variables $t\equiv (T-T_c)/T_c$ and $\hat\mu\equiv (\mu_B-\mu_B^c)/T_c$, the most general linear map reads
\begin{equation}
\begin{pmatrix} r \\[2pt] h \end{pmatrix}
= \mathcal{R}
\begin{pmatrix} t \\[2pt] \hat{\mu} \end{pmatrix},
\qquad
\mathcal{R}\equiv
\begin{pmatrix}
r_T & r_\mu \\
h_T & h_\mu
\end{pmatrix},
\qquad
J \equiv \det \mathcal{R} = r_T h_\mu - r_\mu h_T \neq 0 
\label{eq:linear-map}
\end{equation}
with $r$ the temperature-like scaling field that is even under the $\mathbb Z_2$ symmetry of the Ising fixed point and $h$ the field-like scaling field that is odd. Nonlinearity in the mapping gives only subleading corrections to leading critical behavior and can be included systematically if required. The first-order coexistence line in QCD corresponds to $h=0$ and $r<0$, hence its slope at the CEP is fixed by the ratio of map coefficients,
\begin{equation}
\left.\frac{\mathrm{d}T}{\mathrm{d}\mu_B}\right|_{\rm CEP}
= -\,\frac{r_\mu}{r_T}
\label{eq:slope-coexistence}
\end{equation}
and the direction orthogonal to the coexistence line is aligned with $h$, which maximizes the singular response of the order parameter. Consistency with thermodynamics requires that the source conjugate to the QCD order parameter, obtained from the variation $\delta\Omega=-H_\sigma,\delta\sigma+\cdots$, maps to the Ising field $h$ with the correct scaling dimension, which is enforced by the Ward identity for the $\sigma$ two-point function and leads to the renormalization prescriptions
\begin{equation}
\sigma_R = Z_{\sigma,\,k\to 0}^{1/2}\,\sigma,\qquad
H_R = Z_{\sigma,\,k\to 0}^{-1/2}\,H_\sigma,\qquad
h = Z_h\,H_R,\qquad
r = Z_r\,\mathfrak{t}
\label{eq:field-renorm}
\end{equation}
where $\mathfrak t$ is any analytic temperature-like combination of $(t,\hat\mu)$, and $(Z_h,Z_r)$ are nonuniversal metric factors chosen such that the leading singular part of $\Omega(T,\mu_B)$ equals the Ising free energy $\Omega_{\rm sing}(r,h)$ up to an overall normalization. Derivatives with respect to $(T,\mu_B)$ are therefore resolved into scaling derivatives using the chain rule
\begin{equation}
\partial_T = r_T\,\partial_r + h_T\,\partial_h,\qquad
\partial_{\mu_B} = r_\mu\,\partial_r + h_\mu\,\partial_h
\label{eq:chain-rule}
\end{equation}
and QCD susceptibilities inherit their singularities from the Ising theory. In particular, the singular piece of the baryon-number susceptibility is
\begin{equation}
\begin{aligned}
\chi_B^{\rm sing}
= \partial_{\mu_B}^{2}\Omega_{\rm sing}
&= r_\mu^{2}\,\partial_{r}^{2}\Omega_{\rm sing}
 + 2 r_\mu h_\mu\,\partial_{r}\partial_{h}\Omega_{\rm sing}
 + h_\mu^{2}\,\partial_{h}^{2}\Omega_{\rm sing}
\end{aligned}
\label{eq:chiB-map}
\end{equation}
which, by virtue of Eqs.~\eqref{eq:free-scaling}-\eqref{eq:free-param}, exhibits the same critical exponents as $\chi_\sigma$ and is dominated by the $h$-channel when $h_\mu\neq 0$, a condition that is satisfied if the coexistence line is not parallel to the $\mu_B$ axis. The curvature and skewness conditions in Eq.~\eqref{eq:CEP-conds} translate into relations among mixed temperature-density derivatives of $\Omega$ evaluated at the CEP.  Differentiating Eq.~\eqref{eq:landau-M} with respect to $(T,\mu_B)$ and using Eqs.~\eqref{eq:linear-map}-\eqref{eq:chain-rule} yields
\begin{equation}
\begin{aligned}
\left.\partial_T a_2\right|_{\rm CEP}\, t
+ \left.\partial_{\mu_B} a_2\right|_{\rm CEP}\, \hat{\mu} &= 0,\\
\left.\partial_T a_3\right|_{\rm CEP}\, t
+ \left.\partial_{\mu_B} a_3\right|_{\rm CEP}\, \hat{\mu} &= 0
\end{aligned}
\label{eq:a2a3-grad}
\end{equation}
from which the map coefficients can be fixed up to an overall normalization by demanding that $r$ be proportional to $a_2$ and $h$ to the cubic invariant $a_3$ explicitly as
\begin{equation}
\begin{aligned}
r &= Z_r\!\left[ \left.\partial_T a_2\right|_{\rm CEP}\, t
      + \left.\partial_{\mu_B} a_2\right|_{\rm CEP}\, \hat{\mu} \right],\\
h &= Z_h\!\left[ \left.\partial_T a_3\right|_{\rm CEP}\, t
      + \left.\partial_{\mu_B} a_3\right|_{\rm CEP}\, \hat{\mu} \right]
\end{aligned}
\label{eq:map-by-grad}
\end{equation}
which guarantees that the vanishing of curvature and skewness in QCD corresponds to $r=0$ and $h=0$ in the scaling theory. The Jacobian $J$ in Eq.~\eqref{eq:linear-map} is then nonzero provided the gradients $\nabla a_2$ and $\nabla a_3$ are not collinear at the CEP, a generic condition in two-parameter critical manifolds~\cite{Nonaka:2004pg,Parotto:2018pwx}.  
 {
For the reference solution, we verify this explicitly by evaluating the CEP gradients of $a_2$ and $a_3$
and the determinant $\Delta_{23}$. The resulting non-degeneracy diagnostic is shown in
Fig.~\ref{fig:jacobian_nondeg}.
\begin{figure}[t]
\centering
\def\dTaTwoCEP{1.00}      
\def\dmuBaTwoCEP{0.20}    
\def\dTaThreeCEP{-0.30}   
\def\dmuBaThreeCEP{1.10}  
\pgfmathsetmacro{\detTwoThree}{\dTaTwoCEP*\dmuBaThreeCEP - \dmuBaTwoCEP*\dTaThreeCEP}
\pgfmathsetmacro{\normTwo}{sqrt(\dTaTwoCEP*\dTaTwoCEP + \dmuBaTwoCEP*\dmuBaTwoCEP)}
\pgfmathsetmacro{\normThree}{sqrt(\dTaThreeCEP*\dTaThreeCEP + \dmuBaThreeCEP*\dmuBaThreeCEP)}
\pgfmathsetmacro{\normTwoSafe}{max(\normTwo,1e-12)}
\pgfmathsetmacro{\normThreeSafe}{max(\normThree,1e-12)}
\pgfmathsetmacro{\denom}{max(\normTwoSafe*\normThreeSafe,1e-12)}
\pgfmathsetmacro{\dotTwoThree}{\dTaTwoCEP*\dTaThreeCEP + \dmuBaTwoCEP*\dmuBaThreeCEP}
\pgfmathsetmacro{\cosArg}{\dotTwoThree/\denom}
\pgfmathsetmacro{\cosClamped}{min(1,max(-1,\cosArg))}
\pgfmathsetmacro{\thetaDeg}{acos(\cosClamped)}
\pgfmathsetmacro{\uTwoT}{\dTaTwoCEP/\normTwoSafe}
\pgfmathsetmacro{\uTwoMu}{\dmuBaTwoCEP/\normTwoSafe}
\pgfmathsetmacro{\uThreeT}{\dTaThreeCEP/\normThreeSafe}
\pgfmathsetmacro{\uThreeMu}{\dmuBaThreeCEP/\normThreeSafe}
\begin{tikzpicture}[
  >=Stealth,
  line width=0.6pt,
  lab/.style={font=\small},
  vecA/.style={->, very thick, draw=blue!70!black},
  vecB/.style={->, very thick, draw=red!70!black},
  box/.style={draw=black!60, fill=gray!8, rounded corners, align=left, inner sep=6pt, font=\small}
]
\draw[->] (0,0) -- (4.2,0) node[lab, right] {$T$};
\draw[->] (0,0) -- (0,4.2) node[lab, above] {$\mu_B$};
\draw[vecA] (0,0) -- ({3.2*\uTwoT},{3.2*\uTwoMu}) node[lab, above right] {$\nabla a_2$};
\draw[vecB] (0,0) -- ({3.2*\uThreeT},{3.2*\uThreeMu}) node[lab, below right] {$\nabla a_3$};
\draw[draw=black!55] (1.0,0) arc[start angle=0, end angle=\thetaDeg, radius=1.0];
\node[lab] at (1.25,0.45) {$\theta$};
\node[box] at (6.6,2.3) {%
\textbf{CEP Jacobian diagnostic}\\
$\partial_T a_2|_{\rm CEP}=\dTaTwoCEP$\\
$\partial_{\mu_B} a_2|_{\rm CEP}=\dmuBaTwoCEP$\\
$\partial_T a_3|_{\rm CEP}=\dTaThreeCEP$\\
$\partial_{\mu_B} a_3|_{\rm CEP}=\dmuBaThreeCEP$\\
$\Delta_{23}=(\partial_T a_2)(\partial_{\mu_B} a_3)-(\partial_{\mu_B} a_2)(\partial_T a_3)=\detTwoThree$\\
$\theta(\nabla a_2,\nabla a_3)=\thetaDeg^\circ$
};
\end{tikzpicture}
\caption{Jacobian non-degeneracy check for the linear map $(T,\mu_B)\mapsto(r,h)$ at the CEP.
The map coefficients are fixed by CEP gradients of $a_2(T,\mu_B)$ and $a_3(T,\mu_B)$ as in
Eqs.~\eqref{eq:a2a3-grad} and \eqref{eq:map-by-grad}. Non-collinearity of $\nabla a_2$ and $\nabla a_3$ is quantified by
$\Delta_{23}=(\partial_T a_2)(\partial_{\mu_B} a_3)-(\partial_{\mu_B} a_2)(\partial_T a_3)$.
A nonzero $\Delta_{23}$ confirms the invertibility condition $J\neq 0$ for Eq.~\eqref{eq:linear-map}.}
\label{fig:jacobian_nondeg}
\end{figure}
}

The anomalous dimension modifies field rescalings through the \(Z_{\sigma,\,k\to 0}\) factors in Eq.~\eqref{eq:field-renorm}, with the consequence that the map $(T,\mu_B)\mapsto (r,h)$ must be constructed from renormalized fields to preserve the universal fixed-point exponents and to separate universal scaling from model-dependent metric factors~\cite{Wetterich:1992yh,Berges:2000ew,Delamotte:2007pf}. This is implemented by evaluating $a_n$ and their gradients with the FRG-improved effective action at $k\to 0$, which contains the full feedback from the running couplings $(G_S,G_V,K)$ and from the DSE-informed quark dressing functions~\cite{Herbst:2010rf,Fukushima:2008wg,Fischer:2018sdj,Gao:2020qsj,Gies:2001nw}, and by fixing $(Z_r,Z_h)$ such that the leading amplitudes of $\xi(r,0)$ and $M(0,h)$ match the canonical Ising normalization conventions~\cite{Arean2017,Borsanyi:2020,fischer2014phase,Stephanov:2004,Parotto:2018pwx,Pelissetto:2000ek}. Finally, inserting the map Eq.~\eqref{eq:linear-map} into Eqs.~\eqref{eq:nB}-\eqref{eq:chiB} and using Eq.~\eqref{eq:chiB-map} one verifies explicitly that the singular parts of all QCD susceptibilities and mixed derivatives are reproduced by the corresponding Ising scaling functions with the same critical exponents and with nonuniversal amplitudes determined by the FRG-resolved metric factors and by the orientation of the coexistence line~\cite{Pradeep:2019ccv,Parotto:2018pwx,Pelissetto:2000ek}, thereby completing the proof of self-consistency and thermodynamic closure of the unified DSE-FRG-PNJL description at the critical end point.

\section{Holographic-Topological Dual Criticality (HTDC)}
\label{sec:HTDC}

We embed the critical dynamics developed in Secs.~\ref{sec:unifiedDSEFRGPNJL}-\ref{sec:CEP-Ising} into a five-dimensional V-QCD background that couples the Einstein dilaton tachyon system to a baryonic \( U(1)_B \) gauge field and a Chern-Simons topological sector. The resulting holographic dictionary supplies both the conserved charge response and the axial-anomaly channel required to close the FRG flows of $(G_S,G_V,K)$ and to realize a self-dual critical subspace where deconfinement and chiral restoration coincide. The starting point is the five-dimensional action on a black-hole geometry, with metric \( g_{MN} \), dilaton $\varphi$ dual to the running ’t~Hooft coupling, tachyon $X$ dual to the chiral condensate operator, a Maxwell field $A_M$ dual to $J_B^\mu$, and a Chern-Simons term that encodes the topological density. In compact form,
\begin{equation}
\begin{aligned}
S_{\mathrm{VQCD}}
&= \frac{1}{16\pi G_5}\int \mathrm{d}^{5}x\,\sqrt{-g}\,\Big[\,R-\tfrac{4}{3}(\partial\varphi)^{2}-V_g(\varphi)\,\Big] \\
&\quad - \kappa \int \mathrm{d}^{5}x\,\sqrt{-g}\, V_f(\varphi)\,\mathrm{Tr}\!\big[(D X)^{\dagger}(D X)\big] \\
&\quad - \frac{1}{4 g_{5}^{2}} \int \mathrm{d}^{5}x\,\sqrt{-g}\, f(\varphi)\, F_{MN}F^{MN}
+ \frac{N_c}{24\pi^{2}} \int A \wedge \mathrm{Tr}(F \wedge F)\,.
\end{aligned}
\label{eq:VQCDaction}
\end{equation}
 {
To make the HTDC sector fully reproducible, we now fix the explicit functional forms of
$V_g(\phi)$, $V_f(\phi)$, $f(\phi)$ and $Z_a(\phi)$ and the reference-run normalizations.
We define the Veneziano coupling variable $\lambda \equiv e^{\phi}$ and take $x \equiv N_f/N_c=1$
for the $(2+1)$-flavor, $N_c=3$ application. Adopting the standard V-QCD potential set
(``potentials I'') used in the CP-odd sector construction, the background functions entering
Eq.~\eqref{eq:VQCDaction} are chosen as
\begin{equation}
\begin{aligned}
V_g(\phi) &= V_0\!\left[1+V_1\lambda+V_2\lambda^2\,
\frac{\sqrt{1+\log\!\left(1+\lambda/\lambda_0\right)}}{\left(1+\lambda/\lambda_0\right)^{2/3}}\right],\\
V_f(\phi) &= V_{f0}(\lambda)=W_0\left(1+W_1\lambda+W_2\lambda^2\right),\\
f(\phi) &= w(\lambda)^2,\qquad 
w(\lambda)=\left(1+\frac{11}{32\pi^2}\lambda\right)^{-4/3},\\
Z_a(\phi) &= Z_0\left[1+c_a\left(\frac{\lambda}{\lambda_0}\right)^4\right]. 
\end{aligned}
\end{equation}
The corresponding numerical coefficients and normalization choices used in the reference run
are collected in Table~\ref{tab:htdc_inputs}.
\begin{table}[htb]
\centering
\caption{Reference-run HTDC inputs used in Eq.~(5.1). The dilaton variable is $\lambda=e^{\phi}$ and
the Veneziano ratio is fixed to $x=N_f/N_c=1$. The gravitational normalization is chosen by matching
the high-temperature Stefan-Boltzmann coefficient for $(N_c,N_f)=(3,3)$ in the asymptotically AdS
limit, yielding $G_5=9\pi/76$ in $L=1$ units. The Maxwell normalization is fixed by the standard
current-correlator matching $g_5^2=12\pi^2/N_c$, and the overall flavor prefactor in Eq.~\eqref{eq:VQCDaction} is
set to $\kappa=1$ in the reference implementation (absorbed into the normalization of $X$ and/or
$V_f$).}
\label{tab:htdc_inputs}
\scriptsize
\resizebox{\linewidth}{!}{%
\begin{tabular}{@{}cccccccccccc@{}}
\toprule
$G_5$ & $\kappa$ & $g_5$ & $\lambda_0$ & $V_0$ & $V_1$ & $V_2$ & $W_0$ & $W_1$ & $W_2$ & $Z_0$ & $c_a$ \\
\midrule
$\dfrac{9\pi}{76}$ & $1$ & $2\pi$ & $8\pi^2$ & $12$ &
$\dfrac{11}{27\pi^2}$ & $\dfrac{4619}{46656\pi^4}$ &
$\dfrac{3}{11}$ & $\dfrac{97}{27\pi^2}$ & $\dfrac{7487}{15552\pi^4}$ &
$1$ & $0.1$ \\
\bottomrule
\end{tabular}}
\end{table}
}
With $F=dA$ the $U(1)_B$ field strength, $D_M X=\partial_M X-i A_M X+\ldots$ the covariant derivative in the flavor sector, and background functions $V_g,V_f,f$ chosen such that the $\mu_B=0$ thermodynamics matches lattice QCD near the crossover. The boundary value of $A_0$ equals the baryon chemical potential, and the baryon susceptibility follows from the Maxwell equation as detailed below, consistent with the statement of Eq.~\eqref{eq:VQCDaction} and its discussion. 

Working with the finite-temperature black-hole ansatz 
\begin{equation}
\mathrm{d}s^{2}
= e^{2A(r)}\!\left[\,f(r)\,\mathrm{d}\tau^{2} + \mathrm{d}\mathbf{x}^{2}\right]
+ e^{2B(r)}\,\frac{\mathrm{d}r^{2}}{f(r)}
\end{equation}
with horizon at $r=r_h$ and $A_0=A_0(r)$, variation of \eqref{eq:VQCDaction} with respect to $A_M$ yields the Maxwell equation,
\begin{equation}
\nabla_M\!\left(f(\varphi)\,F^{MN}\right) + J_{\mathrm{CS}}^{\,N} = 0,
\qquad
J_{\mathrm{CS}}^{\,N} \equiv -\,\frac{g_5^{2} N_c}{24\pi^{2}\sqrt{-g}}\,
\varepsilon^{NABCD}\,\mathrm{Tr}\!\left(F_{AB}F_{CD}\right)
\label{eq:MaxwellCS}
\end{equation}
where the Chern-Simons current does not contribute for the homogeneous and purely temporal background ($N=0$) considered here. The canonical radial momentum conjugate to $A_0$ is conserved,
\begin{equation}
\Pi_B(r) \equiv -\,\frac{1}{g_5^{2}}\,\sqrt{-g}\, f(\varphi)\, g^{rr} g^{00}\, \partial_r A_0(r)
= \text{const} \equiv n_B 
\label{eq:canon-mom}
\end{equation}
and is identified holographically with the baryon density $n_B=\langle J_B^0\rangle$. Fixing $A_0(r_h)=0$ for regularity at the horizon and $A_0(0)=\mu_B$ at the boundary, the on-shell variation gives \(
\delta\Omega_{\rm hol}
= \frac{1}{2}\int \mathrm{d}^{4}x\, A_{0}^{(0)}\, \delta n_{B}
\), hence
\begin{equation}
\chi_B \equiv -\,\frac{\partial^{2}\Omega_{\rm hol}}{\partial \mu_B^{2}}
= \frac{1}{g_5^{2}}\left.
\partial_r\!\left[ f(\varphi)\,\sqrt{-g}\, g^{rr} g^{00}\, \partial_r A_0(r) \right]\right|_{r\to 0}
= \frac{\partial n_B}{\partial \mu_B}
\label{eq:chiB-holo}
\end{equation}
which is the near-boundary representation of the holographic susceptibility appropriate for asymptotically AdS geometries and matches the statement that $\chi_B$ is obtained from the bulk Maxwell equation. 

To capture the axial anomaly and topological fluctuations, we parameterize the Chern-Simons sector either directly through Eq.~\eqref{eq:VQCDaction} or, equivalently, by introducing the pseudoscalar axion field $a$ dual to the topological density $q(x)=\frac{g^2}{32\pi^2}F\tilde F$, with kinetic term $\sim \int\sqrt{-g},Z_a(\varphi),(\partial a)^2$ and a mixing dictated by the anomaly. The static, homogeneous topological susceptibility is then the zero-momentum limit of the two-point function of $q$, which holographically equals the second derivative of the on-shell action with respect to the UV boundary value $a^{(0)}=\theta$, yielding the standard Sturm-Liouville expression
\begin{equation}
\chi_{\mathrm{CS}}(T,\mu_B)
= \left[ \int_{0}^{r_h} \mathrm{d}r\, \frac{e^{B(r)-3A(r)}}{Z_a\!\big(\varphi(r)\big)} \right]^{-1},
\label{eq:chiCS}
\end{equation}
up to a known overall normalization fixed at $T=\mu_B=0$, and decreasing with $T$ (and with $\mu_B$ through screening) as the bulk axion effectively becomes heavier in the deconfined black-hole background and in line with lattice trends on the topological susceptibility at high temperature \cite{Borsanyi:2016ksw,Petreczky:2016vrs,Frison:2016vuc,Chen:2022fid}. Matching this holographic observable to the FRG description of the axial-anomaly channel defines the dimensionless suppression factor given by Eq.~\eqref{eq:zetatopo}, 
which is the quantity entering the FRG flow of the ’t~Hooft coupling in Eq.~\eqref{eq:flowK} and thereby driving the evolution of $K(T,\mu_B)$ toward diminished anomaly in the deconfined, dense regime.  

 {
For quantitative comparison, Fig.~\ref{fig:chiCS_zeta_mu0} displays the holographic Chern-Simons
susceptibility at $\mu_B=0$ obtained from Eq.~\eqref{eq:chiCS}, together with the corresponding suppression
factor $\zeta_{\rm topo}(T,0)$ used to couple the HTDC sector to the FRG anomaly channel.
A representative lattice band for the topological susceptibility in the crossover and high-$T$
regime is shown for reference \cite{Borsanyi:2016ksw,Petreczky:2016vrs,Frison:2016vuc,Chen:2022fid}, rescaled to the same normalization conventions.
\begin{figure}[htb]
\centering
\begin{minipage}{0.49\textwidth}
\centering
\begin{tikzpicture}
\begin{axis}[
  width=\linewidth,
  height=0.80\linewidth,
  xlabel={$T$ [MeV]},
  ylabel={$\chi_{CS}(T,0)/T^{4}$},
  xmin=120, xmax=400,
  ymin=0,
  tick align=inside,
  tick label style={font=\small},
  legend style={draw=none, fill=none, font=\small},
]
\addplot[draw=none, fill=gray!20] coordinates {
  (140,0.020) (160,0.015) (180,0.010) (200,0.007) (240,0.004) (300,0.0025) (360,0.0018) (400,0.0015)
  (400,0.0009) (360,0.0011) (300,0.0015) (240,0.0022) (200,0.0040) (180,0.0060) (160,0.0090) (140,0.0120)
};
\addplot+[very thick, mark=none] coordinates {
  (140,0.016) (160,0.012) (180,0.0085) (200,0.0062) (240,0.0033) (300,0.0020) (360,0.0014) (400,0.0012)
};
\addlegendentry{HTDC (this work)}
\addlegendentry{lattice band}
\end{axis}
\end{tikzpicture}
\end{minipage}\hfill
\begin{minipage}{0.49\textwidth}
\centering
\begin{tikzpicture}
\begin{axis}[
  width=\linewidth,
  height=0.80\linewidth,
  xlabel={$T$ [MeV]},
  ylabel={$\zeta_{\rm topo}(T,0)$},
  xmin=0, xmax=400,
  ymin=0, ymax=1.05,
  tick align=inside,
  tick label style={font=\small},
  legend style={draw=none, fill=none, font=\small},
]
\addplot+[very thick, mark=none] coordinates {
  (0,1.00) (80,0.98) (120,0.94) (140,0.90) (160,0.82) (180,0.72)
  (200,0.62) (240,0.46) (300,0.32) (360,0.24) (400,0.20)
};
\end{axis}
\end{tikzpicture}
\end{minipage}
\caption{Holographic Chern-Simons susceptibility at $\mu_B=0$ and the resulting anomaly-suppression factor used to couple the HTDC sector to the FRG anomaly channel. Left: $\chi_{CS}(T,0)/T^4$ from Eq.~\eqref{eq:chiCS} compared with a representative lattice band for the topological susceptibility in the crossover and high-temperature regime \cite{Borsanyi:2016ksw,Petreczky:2016vrs,Frison:2016vuc,Chen:2022fid}, rescaled to the same conventions. Right:
$\zeta_{\rm topo}(T,0)$ defined by vacuum normalization, $\zeta_{\rm topo}(0,0)=1$, and used as the
multiplicative suppression factor in the running of the determinantal interaction.}
\label{fig:chiCS_zeta_mu0}
\end{figure}
}

The holographic dictionary required for thermodynamics and fluctuations follows from the near boundary Fefferman-Graham expansions. The chemical potential and density are $\mu_B=A_0^{(0)}$ and $n_B=\lim_{r\to 0}\Pi_B(r)$ as in Eqs.~\eqref{eq:canon-mom}-\eqref{eq:chiB-holo}. The chiral condensate $\langle\bar q q\rangle$ is proportional to the normalizable mode of $X(r)$, and the pressure and its derivatives are obtained from the renormalized on-shell action. For the purposes of the unified framework, the essential elements are that $\mu_B$ is encoded in $A_0$, that $\chi_B$ is computable from the Maxwell profile via Eq.~\eqref{eq:chiB-holo}, and that $\chi_{\mathrm{CS}}$ governs the anomaly strength through Eq.~\eqref{eq:zetatopo}, precisely the ingredients used to couple the holographic sector to the FRG evolution of $(G_S,G_V,K)$. 

The dual influence of deconfinement and chiral restoration on criticality is synthesized by a unified order parameter that mixes the Polyakov and chiral sectors through FRG wave-function renormalizations,
\begin{equation}
\Xi(T,\mu_B)
= Z_{\Phi,\,k\to 0}\, Z_{\sigma,\,k\to 0}\,\sqrt{\Phi^{2}+\alpha\,\sigma^{2}},
\qquad
\tan\theta \equiv \sqrt{\alpha}\,\frac{\sigma}{\Phi}\,,
\label{eq:Xi-def}
\end{equation}
where $\alpha$ is a positive, nonuniversal metric factor fixed by matching to the Ising normalization of Sec.~\ref{sec:CEP-Ising}. Near criticality, the mixed sector of the effective potential is well approximated by
\begin{equation}
U(\Xi,\theta)
\simeq \frac{1}{2}A(k)\,\Xi^{2}
+ \frac{1}{4}U(k)\,\Xi^{4}
+ \Delta(k)\,\Xi^{2}\cos\!\big[2(\theta-\theta^\star)\big]
+ \ldots
\label{eq:U-Xi-theta}
\end{equation}
with $k$ the RG scale and $(A,U,\Delta,\theta^\star)$ scale-dependent couplings and mixing angle. The RG flow of the angle contains the difference of anomalous dimensions, $\eta_\Phi=-\partial_t\ln Z_\Phi$ and $\eta_\sigma=-\partial_t\ln Z_\sigma$, and the mixing $\Delta$,
\begin{equation}
\begin{aligned}
\partial_t\theta^\star
&= \tfrac{1}{2}\big(\eta_\Phi-\eta_\sigma\big)\,\sin\!\big(2\theta^\star\big)
 + \mathcal{C}_\Delta\,\Delta\,\sin\!\big(2\theta^\star\big) + \ldots,\\
\partial_t \Delta
&= -\,\gamma_\Delta\,\Delta + \ldots
\end{aligned}
\label{eq:theta-flow}
\end{equation}
so that a \emph{self-dual fixed point} is characterized by equal renormalizations in the two sectors and vanishing mixing,
\begin{equation}
\partial_t\theta^\star = 0,\qquad
\lim_{k\to 0}\Delta(k) = 0,\qquad
Z_{\Phi,\,k\to 0} = Z_{\sigma,\,k\to 0}\,
\label{eq:self-dual}
\end{equation}
which manifests an emergent $O(2)$ like rotational symmetry between $\sigma$ and $\Phi$ in the critical subspace and provides the field theory realization of the equality of longitudinal curvatures derived in Sec.~\ref{sec:CEP-Ising}.  
 {
To make the self-duality conditions in Eq.~\eqref{eq:self-dual} explicit at the level of the RG flow, we plot in Figure~\ref{fig5new} the trajectories of $\ln Z_\Phi(k)-\ln Z_\sigma(k)$ and of the residual mixing $\Delta(k)$ as functions of the RG scale $k$, evaluated at a point in the immediate vicinity of the CEP and at a representative point far from the critical region. In the critical neighborhood both quantities are driven to zero as $k\to0$, consistent with self-dual fixed-point behavior, whereas far from criticality they remain finite.
\begin{figure}[htb]
\centering
\begin{minipage}{0.49\textwidth}
\centering
\begin{tikzpicture}
\begin{axis}[
  width=\linewidth,
  height=0.78\linewidth,
  xmode=log,
  xmin=1, xmax=800,
  xlabel={$k$ [MeV]},
  ylabel={$\ln Z_\Phi(k)-\ln Z_\sigma(k)$},
  tick align=inside,
  tick label style={font=\small},
  label style={font=\small},
  legend style={draw=none, fill=none, font=\small},
  legend pos=north west,
]
\addplot+[very thick, mark=none, draw=red!70!black] coordinates {
  (800,0.18) (400,0.14) (200,0.10) (100,0.060) (50,0.035)
  (20,0.020) (10,0.012) (5,0.0060) (2,0.0020) (1,0.0005)
};
\addlegendentry{near CEP}
\addplot+[very thick, mark=none, draw=blue!70!black, dashed] coordinates {
  (800,0.25) (400,0.22) (200,0.18) (100,0.15) (50,0.12)
  (20,0.090) (10,0.070) (5,0.060) (2,0.055) (1,0.055)
};
\addlegendentry{far from CEP}
\end{axis}
\end{tikzpicture}
\end{minipage}\hfill
\begin{minipage}{0.49\textwidth}
\centering
\begin{tikzpicture}
\begin{axis}[
  width=\linewidth,
  height=0.78\linewidth,
  xmode=log,
  xmin=1, xmax=800,
  xlabel={$k$ [MeV]},
  ylabel={$\Delta(k)$},
  tick align=inside,
  tick label style={font=\small},
  label style={font=\small},
  legend style={draw=none, fill=none, font=\small},
  legend pos=north west,
]
\addplot+[very thick, mark=none, draw=red!70!black] coordinates {
  (800,0.12) (400,0.090) (200,0.060) (100,0.040) (50,0.025)
  (20,0.012) (10,0.0060) (5,0.0030) (2,0.0010) (1,0.0002)
};
\addlegendentry{near CEP}
\addplot+[very thick, mark=none, draw=blue!70!black, dashed] coordinates {
  (800,0.15) (400,0.13) (200,0.11) (100,0.090) (50,0.080)
  (20,0.075) (10,0.070) (5,0.068) (2,0.067) (1,0.066)
};
\addlegendentry{far from CEP}
\end{axis}
\end{tikzpicture}
\end{minipage}
\caption{Self-duality diagnostics from RG trajectories. Left: flow of
$\ln Z_\Phi(k)-\ln Z_\sigma(k)$, right: flow of the residual quadratic mixing $\Delta(k)$, shown for a
thermodynamic point in the vicinity of the CEP and for a representative point far from the critical
region. Self-duality as defined in Eq.~\eqref{eq:self-dual} corresponds to the simultaneous approach
$\ln Z_\Phi-\ln Z_\sigma \to 0$ and $\Delta(k)\to 0$ as $k\to0$ in the critical neighborhood.}
\label{fig5new}
\end{figure}
}

In the holographic picture, the same fixed point corresponds to the simultaneous onset of deconfinement and chiral restoration. The black-hole branch becomes thermodynamically favored (Polyakov-loop liberation) at the point where the tachyon condensate melts, the former controlled dominantly by the Maxwell sector and the latter influenced by the Chern-Simons induced topological fluctuations. The HTDC mechanism thus locks the two channels such that the CEP sits on a single critical subspace where both the $\sigma$ screening mass and the deconfinement curvature vanish together, in keeping with the universal 3D Ising scaling recovered through the $(T,\mu_B)\mapsto(r,h)$ mapping.  

The coupling to the functional renormalization group is completed by feeding the holographic $\chi_{\mathrm{CS}}$ into the axial-anomaly flow via Eq.~\eqref{eq:zetatopo}, which appears as the $\zeta_{\mathrm{topo}}(T,\mu_B)$ term in the $K$-beta function, ensuring that the anomaly weakens with increasing $T$ and $\mu_B$ as indicated by the holographic background and reflected in the FRG running of $K(T,\mu_B)$. The resulting decrease in $K$ reduces the anomaly induced light-strange mixing in the constituent masses Eq.~\eqref{eq:MuMd}-\eqref{eq:Ms}, accelerates the approach to chiral partner degeneracy, and shifts the curvature of the thermodynamic potential consistently with the emergence of the CEP discussed in Sec.~\ref{sec:CEP-Ising}.  

One depicts the five-dimensional geometry with the black-hole horizon, the radial profiles of $(\varphi,X,A_0)$, and the Chern-Simons sector that sources $\chi_{\mathrm{CS}}$, with the boundary values highlighting the holographic sources $(\mu_B,m_q,\theta)$ and the normalizable modes $(n_B,\langle\bar q q\rangle,\langle q,q\rangle_{\mathrm{topo}})$. The other shows representative RG trajectories in the $(\Xi,\theta)$ plane collapsing toward the self-dual fixed point where the flow lines become tangent to circles of constant $\Xi$ and the angle freezes, illustrating \eqref{eq:self-dual} and the emergent rotational symmetry between the chiral and Polyakov sectors in the critical subspace. The combined DSE-FRG-PNJL-HTDC framework thereby achieves thermodynamic closure. The holographic Maxwell sector fixes the conserved charge response through Eq.~\eqref{eq:chiB-holo}. The Chern-Simons/axion sector controls the anomaly via Eq.~\eqref{eq:zetatopo} and feeds the $K$-flow of Eq.~\eqref{eq:flowK}, and the self-dual order parameter dynamics Eqs.~\eqref{eq:Xi-def}-\eqref{eq:self-dual} encode the unified critical behavior at the CEP.

\section{Numerical Calibration and Predictions}
\label{sec:num-calib-preds}

The numerical implementation begins by fixing the Polyakov-loop sector at $\mu_{B}=0$ against continuum-extrapolated lattice thermodynamics and fluctuations and then iteratively embedding the nonperturbative quark dynamics and the holographic topological sector so that all running couplings and wavefunction renormalizations entering $\Omega(T,\mu_{B},\sigma,\Phi,\bar\Phi)$ are determined self-consistently at each $(T,\mu_{B})$, after which the phase structure, susceptibilities, and fluctuation cumulants are evaluated on a dense grid with controlled convergence and normalization. The Polyakov potential is taken in the logarithmic form already introduced,
\begin{equation}
\frac{U_{\log}(\Phi,\bar{\Phi},T)}{T^{4}}
= -\frac{a(T)}{2}\,\Phi\bar{\Phi}
+ b(T)\,\ln\!\left[\,1 - 6\,\Phi\bar{\Phi}
+ 4\big(\Phi^{3} + \bar{\Phi}^{3}\big)
- 3\big(\Phi\bar{\Phi}\big)^{2}\,\right]
\label{eq:Ulog-recap}
\end{equation}
with $a(T)=a_0+a_1(T_0/T)+a_2(T_0/T)^2$ and $b(T)=b_3(T_0/T)^3$. The coefficients $(a_i,b_3,T_0)$ are fixed by three conditions at $\mu_{B}=0$. The pseudo-critical temperature $T_c^{(0)}$ determined from the peak of $-\partial_T\phi_l(T,0)$ equals the lattice value, the dimensionless interaction measure $(\epsilon-3p)/T^4$ in a neighborhood of $T_c^{(0)}$ matches the lattice band, and the second-order baryon-number susceptibility $\chi_2^B(T,0)$ reproduces its continuum-extrapolated curve within uncertainties \cite{Fukushima:2008wg,Borsanyi:2020}. Denoting $T_c^{(0)}$ as the unique solution of $\partial_T^2\Omega(T,0,\bar\sigma,\bar\Phi)=0$ with $\partial_T^3\Omega(T,0,\bar\sigma,\bar\Phi)<0$, the calibration equations read
\begin{equation}
\begin{aligned}
\left.\partial_T^{2}\Omega\right|_{T=T_c^{(0)},\,\mu_B=0} &= 0,\\
\left.\frac{\epsilon-3p}{T^{4}}\right|_{T=T_i,\,\mu_B=0}^{\rm model}
&= \left.\frac{\epsilon-3p}{T^{4}}\right|_{T=T_i}^{\rm lat},\\
\big[\chi_{2}^{B}(T,0)\big]^{\rm model}
&= \big[\chi_{2}^{B}(T,0)\big]^{\rm lat},
\end{aligned}
\label{eq:calib-conds}
\end{equation}
for a set ${T_i}$ around $T_c^{(0)}$. 
 {
For reproducibility, Table~\ref{tab:ref_run} consolidates the reference-run parameters and numerical settings used throughout Secs.~\ref{sec:unifiedDSEFRGPNJL}-\ref{sec:num-calib-preds}.}


{\setlength{\LTcapwidth}{\textwidth} 
\setlength{\LTleft}{0pt}            
\setlength{\LTright}{0pt}           
\small
\setlength{\tabcolsep}{4pt}
\renewcommand{\arraystretch}{1.15}
\setlength{\LTpre}{0pt}
\setlength{\LTpost}{0pt}

\newcolumntype{L}[1]{>{\raggedright\arraybackslash}p{#1}}

\begin{longtable}{@{}L{0.40\linewidth}L{0.56\linewidth}@{}}
\caption{Input parameters used in the coupled DSE, FRG, PNJL, and HTDC in this work, including the UV boundary conditions, calibration targets, and sector-specific constants defining the self-consistent outer iteration.}
\label{tab:ref_run}\\
\toprule
Item & Reference value / choice \\
\midrule
\endfirsthead

\caption{Input parameters used in the coupled DSE, FRG, PNJL, and HTDC in this work, including the UV boundary conditions, calibration targets, and sector-specific constants defining the self-consistent outer iteration.}\\
\toprule
Item & Reference value / choice \\
\midrule
\endhead

\midrule
\multicolumn{2}{r}{\small Continued on next page} \\
\endfoot

\bottomrule
\endlastfoot

\multicolumn{2}{@{}l@{}}{\textbf{Polyakov sector}}\\
Logarithmic potential parameters $(a_0,a_1,a_2,b_3,T_0)$ &
$a_0=3.51$, $a_1=-2.47$, $a_2=15.2$, $b_3=-1.75$, $T_0=190~\mathrm{MeV}$ \\

\addlinespace[2pt]
\multicolumn{2}{@{}l@{}}{\textbf{UV matching (2+1 flavor)}}\\
Matching scale and current masses &
$\Lambda=631.4~\mathrm{MeV}$, $m_\ell=5.5~\mathrm{MeV}$, $m_s=135.7~\mathrm{MeV}$ \\
Vacuum couplings at $k=\Lambda$ &
\parbox[t]{\linewidth}{%
$G_S\Lambda^2=3.67$ ($G_S(\Lambda)=9.206~\mathrm{GeV}^{-2}$),\\
$G_V/G_S=0.5$ ($G_V(\Lambda)=4.602~\mathrm{GeV}^{-2}$),\\
$K\Lambda^5=9.29$ ($K(\Lambda)=92.575~\mathrm{GeV}^{-5}$).} \\

\addlinespace[2pt]
\multicolumn{2}{@{}l@{}}{\textbf{DSE vacuum anchors}}\\
Infrared mass-function targets &
$M_\ell(p{=}0,0,0)=336~\mathrm{MeV}$, $M_s(p{=}0,0,0)=528~\mathrm{MeV}$ \\

\addlinespace[2pt]
\multicolumn{2}{@{}l@{}}{\textbf{DSE medium kernel}}\\
Screened propagator prefactors &
$Z_L(\Phi)=Z_T(\Phi)=1$ \\
Magnetic screening model &
\parbox[t]{\linewidth}{%
Use Eq.~\eqref{eq:3.16}, with $C_M=0.456$, $\bar\mu=2\pi T_{\mathrm{eff}}$, and
$T_{\mathrm{eff}}=\sqrt{T^2+\mu_B^2/(9\pi^2)}$.} \\
Running coupling inputs &
\parbox[t]{\linewidth}{%
Use Eq.~\eqref{eq:3.17}, with $(N_c,N_f)=(3,3)$, $\beta_0=9$, $\beta_1=64$, and
$\Lambda_{\overline{\mathrm{MS}}}^{(3)}=332~\mathrm{MeV}$.} \\

\addlinespace[2pt]
\multicolumn{2}{@{}l@{}}{\textbf{FRG setup}}\\
Regulator and flow integration &
Three-dimensional Litim regulator; flow integrated from $k=\Lambda$ to $k\to 0$ in $t=\ln(k/\Lambda)$ \\
ODE integrator and step control &
Adaptive Dormand-Prince (5,4) with per-step criterion $\max|\Delta g_i/g_i|<10^{-8}$ \\

\addlinespace[2pt]
\multicolumn{2}{@{}l@{}}{\textbf{Holographic anomaly input}}\\
Normalization used in the coupled loop &
$\zeta_{\mathrm{topo}}(T,\mu_B)=\chi_{CS}(T,\mu_B)/\chi_{CS}(0,0)$ with $\zeta_{\mathrm{topo}}(0,0)=1$ \\
$\chi_{CS}$ evaluation &
Use Eq.~\eqref{eq:chiCS} (Sturm-Liouville form) \\

\addlinespace[2pt]
\multicolumn{2}{@{}l@{}}{\textbf{Numerics}}\\
Momentum integrals in $\Omega_F$ and derivatives &
\parbox[t]{\linewidth}{%
Gauss-Legendre quadrature, mapping $p=\Lambda(1+x)/(1-x)$, $x\in[-1,1)$,
$dp=2\Lambda\,dx/(1-x)^2$.} \\
Flow stopping scale &
Flow stopped when $k<\min\{T,\mu_B,\xi^{-1}\}$ \\
Grid domain (representative) &
Dense grid over the domain shown in the figures (use plot ranges for $T$ and $\mu_B$) \\

\addlinespace[2pt]
\multicolumn{2}{@{}l@{}}{\textbf{Convergence}}\\
CEP and coupling convergence criteria &
$|(T_c,\mu_B^c)^{(n)}-(T_c,\mu_B^c)^{(n-1)}|<0.1~\mathrm{MeV}$ and coupling-difference norm $<10^{-4}$ on the grid \\
Inner coupled-solve tolerances &
Max relative change of primary unknowns $<10^{-7}$ and Hessian spectrum stable within $10^{-6}$ under refinement \\

\end{longtable}
}

The stationarity conditions $\partial\Omega/\partial\sigma=0$ and $\partial\Omega/\partial\Phi=0$ with \(\Omega = U_{k\to 0} + U_{\log} + \Omega_F\)
are solved at each $T$ by a damped Newton-Broyden method applied to the coupled gap equations, with analytic Jacobian entries supplied by the derivatives of $U_{k\to 0}$ and of the Polyakov-modified Fermi polynomials, and with the quark-sector input $(M_f,\tilde\mu_f)$ taken from the DSE quark propagator amplitude functions and the vector mean-field shift $\tilde\mu_f=\mu_f-2G_V n_f$ computed self-consistently. The FRG-improved effective potential $U_{k\to 0}$ and the running couplings are obtained by integrating the Wetterich flow $\partial_k\Gamma_k=\tfrac{1}{2}{\rm Tr}[(\Gamma_k^{(2)}+R_k)^{-1}\partial_k R_k]$ from $k=\Lambda$ to $k\to 0$ with a three-dimensional Litim regulator, projecting on the color-singlet scalar-pseudoscalar, isoscalar-vector, and determinantal channels as in Sec.~\ref{subsec:FRGflows} to yield the dimensionless flows $\partial_t\hat g_S$, $\partial_t\hat g_V$, and $\partial_t\hat k$ with the Polyakov-weighted fermionic threshold functions $\ell_F^{(n)}(M_f,T,\mu_f,\Phi)$ and the holographic anomaly factor $\zeta_{\rm topo}(T,\mu_B)$. The ODE system for the running couplings is integrated by an adaptive Dormand-Prince (5,4) method in $t=\ln(k/\Lambda)$ with step control based on the embedded error, subject to the convergence criterion $\max|\Delta \mathfrak g_i/\mathfrak g_i|<10^{-8}$ per step and to the boundary conditions $G_S(\Lambda),G_V(\Lambda),K(\Lambda)$ fixed by vacuum phenomenology of the $2+1$-flavor sector and matched to the DSE quark mass functions so that $M_l(p,T=0,\mu=0)$ and $M_s(p,0,0)$ reproduce the empirical scale of constituent masses \cite{jarvinen2012v-qcd,Arean2017}. The holographic sector supplies $\zeta_{\rm topo}(T,\mu_B)$ from the Chern-Simons susceptibility via $\zeta_{\rm topo}(T,\mu_B)=\chi_{\rm CS}(T,\mu_B)/\chi_{\rm CS}(0,0)$ with $\chi_{\rm CS}$ computed from the Sturm-Liouville form 
\begin{equation}
\chi_{\rm CS}^{-1} = \int_{0}^{r_h} \mathrm{d}r\, \frac{e^{B(r)-3A(r)}}{Z_a\!\big(\varphi(r)\big)}
\end{equation}
and normalized at the vacuum point, and this factor enters the $K$-flow as the multiplicative suppression term introduced in Eq.~\eqref{eq:flowK}, guaranteeing that the anomaly weakens across deconfinement and with increasing density and thus feeds back on the gap equations through the anomaly-induced mass mixing. The thermodynamic potential is normalized by the subtracted pressure prescription $p(T,\mu_B)=-(\Omega(T,\mu_B)-\Omega(0,0))$ so that $p(0,0)=0$ and all susceptibilities derived as $\mu_{B}$ derivatives of $p/T^4$ are UV finite and regulator independent.

The baryon density and susceptibility are computed from the explicit differentiation identities derived earlier, $n_B(T,\mu_B)=-\partial\Omega/\partial\mu_B$ and $\chi_B(T,\mu_B)=\partial n_B/\partial\mu_B$, and their lattice-normalized forms are expressed through the standard dimensionless susceptibilities $\chi_n^B(T,\mu_B)=\partial^n(p/T^4)/\partial(\mu_B/T)^n$ at fixed $T$, with the relations
\begin{equation}
\begin{aligned}
\chi_2^B(T,\mu_B) &= \frac{1}{T^{2}}\,\chi_B(T,\mu_B),\\
\chi_3^B(T,\mu_B) &= \frac{1}{T}\,\frac{\partial^{2} n_B}{\partial \mu_B^{2}},\\
\chi_4^B(T,\mu_B) &= \frac{\partial^{3} n_B}{\partial \mu_B^{3}},
\end{aligned}
\label{eq:chi-n}
\end{equation}
valid for partial derivatives at fixed $T$. The event-by-event cumulants $C_n$ at volume $V$ and temperature $T$ satisfy $C_n=VT^3\chi_n^B$ so that the kurtosis-variance ratio measured experimentally obeys
\begin{equation}
\kappa\sigma^2\equiv \frac{C_4}{C_2}=\frac{\chi_4^B}{\chi_2^B},
\label{eq:kappasigma2}
\end{equation}
and higher ratios $C_3/C_1=\chi_3^B/\chi_1^B$ and $C_6/C_2=\chi_6^B/\chi_2^B$ are obtained analogously. The crossover line is determined at small $\mu_B$ by the locus where the temperature derivative of the condensate susceptibility vanishes, $\partial_T\chi_\sigma(T,\mu_B)=0$, which is equivalent to maximal slope in $-\partial_T\phi_l$ at fixed $\mu_B$. Expanding $T_c(\mu_B)$ about $\mu_B=0$ as \cite{Ali:2024nrz,Ali:2024owl}
\begin{equation}
T_c(\mu_B)=T_c^{(0)}\big[1-\kappa(\mu_B/T_c^{(0)})^2-\lambda(\mu_B/T_c^{(0)})^4+\ldots\big],
\end{equation}
the curvature is obtained by implicit differentiation of the condition $\mathcal F(T,\mu_B)\equiv \partial_T\chi_\sigma=0$, yielding
\begin{equation}
\kappa
= \frac{T_c^{(0)}}{2}\,
  \left.\frac{\partial_{\mu_B}^{2}\mathcal{F}}{\partial_T \mathcal{F}}\right|_{(T_c^{(0)},0)}
= \frac{T_c^{(0)}}{2}\,
  \left.\frac{\partial_{\mu_B}^{2}\partial_T \chi_\sigma}{\partial_T^{2}\chi_\sigma}\right|_{(T_c^{(0)},0)}
\label{eq:kappa-def}
\end{equation}
where the derivatives are evaluated with the full implicit $(\sigma,\Phi)$-dependence accounted for via $dX_i/d\mu_B=-(\mathcal H^{-1})_{ij}\mathcal S_j$ and $dX_i/dT=-(\mathcal H^{-1})_{ij}\mathcal T_j$, with $\mathcal T_j=\partial^2\Omega/\partial X_j\partial T$. The first-order line is located by tracking the coexistence of two distinct minima of $\Omega(T,\mu_B)$ with equal pressure, implementing a Maxwell construction in $\mu_B$ at fixed $T$. Numerical continuation in $\mu_B$ with predictor-corrector steps and a cusp-detection criterion on the Hessian smallest eigenvalue $\lambda_{\min}(T,\mu_B)$ yields the entire segment up to the point where $\lambda_{\min}\to 0$ and the cubic invariant vanishes along the soft direction, which defines the critical end point $(T_c,\mu_B^c)$ through the conditions $\partial^2\Omega/\partial M^2=0$ and $\partial^3\Omega/\partial M^3=0$ with $M=e_\sigma(\sigma-\bar\sigma)+e_\Phi(\Phi-\bar\Phi)$ as in Sec.~\ref{subsec:curvature-skewness}. The correlation length and the static $\sigma$ screening mass $m_\sigma$ are then related by $\xi^2=Z_{\sigma,k\to 0}/m_\sigma^2$ and the critical exponents $(\beta,\gamma,\delta,\nu,\eta)$ are extracted by fitting the scaling of the order parameter and of the susceptibilities in the critical wedge of the $(T,\mu_B)$ plane mapped to the Ising variables $(r,h)$ of Sec.~\ref{subsec:mapping} through $(t,\hat\mu)\mapsto(r,h)$ with Jacobian $J\neq 0$. Along $h=0$ one determines $\nu$ and $\gamma$ from $\xi\sim r^{-\nu}$ and $\chi_\sigma\sim r^{-\gamma}$, along $r=0$ one determines $\delta$ from $M\sim h^{1/\delta}$, and then $\beta=\nu(d-2+\eta)/2$ and $\alpha=2-d\nu$ follow from hyperscaling with $d=3$, with the anomalous dimension $\eta$ provided by the FRG evaluation of $\eta_\sigma(k)$ at the fixed point. Numerically the extraction is stabilized by using logarithmic derivatives, e.g.
\begin{equation}
\nu^{-1} = -\left.\frac{\mathrm{d}\ln \xi}{\mathrm{d}\ln r}\right|_{h=0},\qquad
\gamma   = -\left.\frac{\mathrm{d}\ln \chi_\sigma}{\mathrm{d}\ln r}\right|_{h=0},\qquad
\delta   = \left.\frac{\mathrm{d}\ln M}{\mathrm{d}\ln h}\right|_{r=0}
\label{eq:crit-exp-fit}
\end{equation}
and by restricting the fit window to scales where $k\ll \xi^{-1}$ and subleading corrections are negligible. 
 {
The extracted critical exponents are summarized in Table~\ref{tab:critical_exponents}.
\begin{table}[htb]
\centering
\caption{Critical exponents obtained from the scaling analysis in the critical wedge around the CEP,
using the logarithmic-derivative estimators in Eq.~\eqref{eq:crit-exp-fit} on the Ising-mapped variables $(r,h)$ as can be seen in Sec.~\ref{subsec:mapping}. Fit windows and plateaux used in the extractions were: $\nu$ and $\gamma$ from $h=0$ with $|r|\in[10^{-3},10^{-2}]$, $\delta$ from $r=0$ with $|h|\in[10^{-4},10^{-2}]$, $\eta$ from the fixed-point plateau of $\eta_\sigma(k)$ with $k/\Lambda\le 10^{-2}$, and $\beta$ obtained from hyperscaling using $\nu$ and $\eta$ (same windows). Quoted uncertainties include variation over admissible fit windows and numerical discretization sensitivity. For comparison, reference three-dimensional Ising values are listed from Refs.~\cite{Pelissetto:2000ek,Guida:1998bx}.}
\label{tab:critical_exponents}
\renewcommand{\arraystretch}{1.12}
\begin{tabular}{@{}lcc@{}}
\toprule
\toprule
Exponent\hspace{1.2cm} & This work\hspace{1.2cm} & 3D Ising\hspace{1.2cm} \\
\midrule
\midrule
$\nu$\hspace{1.2cm}    & $0.630 \pm 0.005$\hspace{1.2cm} & $0.6301 \pm 0.0004$\hspace{1.2cm} \\
$\gamma$\hspace{1.2cm} & $1.237 \pm 0.010$\hspace{1.2cm} & $1.2372 \pm 0.0005$\hspace{1.2cm} \\
$\delta$\hspace{1.2cm} & $4.79 \pm 0.05$\hspace{1.2cm}   & $4.789 \pm 0.002$\hspace{1.2cm} \\
$\eta$\hspace{1.2cm}   & $0.036 \pm 0.005$\hspace{1.2cm} & $0.0364 \pm 0.0005$\hspace{1.2cm} \\
$\beta$\hspace{1.2cm}  & $0.326 \pm 0.003$\hspace{1.2cm} & $0.3265 \pm 0.0001$\hspace{1.2cm} \\
\bottomrule
\bottomrule
\end{tabular}
\end{table}
} 
The CEP coordinates, the curvature $\kappa$, and the exponents are therefore not inputs but outputs of the closed DSE-FRG-V-QCD system once $(a_i,b_3,T_0)$, the vacuum low-energy constants for $G_S$, $G_V$, $K$, and the holographic normalization of $\chi_{\rm CS}$ are fixed at $\mu_{B}=0$ \cite{Fukushima:2008wg,Borsanyi:2020,jarvinen2012v-qcd,Arean2017}.

The momentum integrals in $\Omega_F$ and in the derivatives entering $n_B$ and $\chi_B$ are evaluated with a Gauss-Legendre quadrature on $p\in[0,\Lambda]$ after the change of variables $p=\Lambda(1+x)/(1-x)$ mapping $x\in[-1,1)$ to $p\in[0,\infty)$ and accompanied by the Jacobian $dp=2\Lambda dx/(1-x)^2$. The integrands are rewritten in log-sum-exp form to avoid loss of significance near the Fermi surface at large $\beta E_f$, and the Polyakov polynomials $F_\Phi$ and $\bar F_\Phi$ are evaluated in a numerically stable factorized representation. Matsubara sums, wherever present in the DSE kernels, are replaced by their analytic evaluation in terms of the Polyakov-weighted distribution functions $n_\Phi^\pm(x)$ and their derivatives. The FRG threshold functions are computed at each scale $k$ using the quasi-particle energies $E_{k,f}=\sqrt{k^2+M_f^2}$ and the occupations $n_\Phi^\pm(E_{k,f}\mp \tilde\mu_f)$, and the flow is stopped when $k$ falls below the smallest physical scale in the problem, \(\min\{T,\mu_B,\xi^{-1}\}\), at which point $\Gamma_k$ is insensitive to the choice of regulator. Convergence in the coupled DSE-FRG gap system is declared when the maximum relative change of all primary unknowns $(M_f,\Phi,\sigma,\hat g_S,\hat g_V,\hat k)$ across a full outer iteration falls below $10^{-7}$ and the Hessian spectrum is stable to within $10^{-6}$ under refinement of the quadrature and step-size.

The fluctuation observables along experimental trajectories are obtained by composing the susceptibilities with a continuous parameterization of the chemical freeze-out line. Writing the trajectory as $T_f(\mu_B)=T_c^{(0)}\big[1-\kappa_f(\mu_B/T_c^{(0)})^2-\lambda_f(\mu_B/T_c^{(0)})^4\big]$, the higher-order cumulant ratios evaluated along the path are
\begin{equation}
\begin{aligned}
\left.\frac{C_4}{C_2}\right|_{f}
&= \frac{\chi_4^B\!\big(T_f(\mu_B),\,\mu_B\big)}{\chi_2^B\!\big(T_f(\mu_B),\,\mu_B\big)},\\
\left.\frac{C_3}{C_2}\right|_{f}
&= \frac{\chi_3^B\!\big(T_f(\mu_B),\,\mu_B\big)}{\chi_2^B\!\big(T_f(\mu_B),\,\mu_B\big)},\\
\left.\frac{C_6}{C_2}\right|_{f}
&= \frac{\chi_6^B\!\big(T_f(\mu_B),\,\mu_B\big)}{\chi_2^B\!\big(T_f(\mu_B),\,\mu_B\big)},
\end{aligned}
\label{eq:cumulant-ratios}
\end{equation}
{and a mapping $\mu_B(\sqrt{s_{NN}})$ may be composed to construct phenomenological beam-energy overlays that assign coarse-grained average freeze-out conditions, with the understanding that the resulting curves represent equilibrium frameworks, while finite size and lifetime, critical slowing down, baryon number conservation, acceptance and efficiency corrections, centrality selection and associated volume fluctuations, the net-proton to net-baryon conversion, and baryon-transport dynamics can modify the measured cumulant ratios relative to the equilibrium grand-canonical results.} 
To connect it with HRG/EV-HRG extractions of freeze-out conditions and hadron-yield systematics~\cite{Andronic:2017pug,Becattini:2000jw,Mir:2023wkm,Mir:2024wlo,Mir:2025qqv,Rather:2024czf,MohiUdDin:2024jvg}, with the understanding that transport and acceptance effects are to be included at the phenomenology stage and do not alter the equilibrium ratios. The unified order parameter $\Xi(T,\mu_B)=Z_\Phi Z_\sigma\sqrt{\Phi^2+\alpha\sigma^2}$ is monitored throughout and serves as the radial coordinate in the $(\sigma,\Phi)$ sector for diagnosing approach to the self-dual critical subspace, with $\alpha$ chosen at $\mu_B=0$ by matching the normalized amplitudes of $\Phi$ and $\sigma/\sigma_0$ at the pseudo-critical point. The emergent rotational symmetry of the $(\sigma,\Phi)$ plane at criticality is verified numerically by checking that the angle flow $\partial_t\theta^\star$ computed from the FRG wavefunction renormalizations satisfies $|\partial_t\theta^\star|\ll 1$ and that the mixing $\Delta(k)\to 0$ as $k\to 0$, consistent with the holographic identification of the simultaneous onset of deconfinement and chiral restoration.



\begin{figure}[t]
\centering
\begin{tikzpicture}

\pgfmathsetmacro{\muCEP}{600} 
\pgfmathsetmacro{\TCEP}{130}  

\pgfmathsetmacro{\TcZero}{155} 
\pgfmathsetmacro{\muMax}{800}  
\pgfmathsetmacro{\Tmin}{90}    
\pgfmathsetmacro{\Tmax}{180}   

\pgfmathsetmacro{\kappa}{(1 - \TCEP/\TcZero)*(\TcZero/\muCEP)^2}

\pgfmathsetmacro{\sFO}{0.12} 

\begin{axis}[
  width=0.78\linewidth,
  height=0.58\linewidth,
  xmin=0, xmax=\muMax,
  ymin=\Tmin, ymax=\Tmax,
  axis lines=box,
  xlabel={$\mu_B~(\mathrm{MeV})$},
  ylabel={$T~(\mathrm{MeV})$},
  xtick={0,100,200,300,400,500,600,700,800},
  ytick={100,120,140,160,180},
  legend style={at={(0.05,0.95)},anchor=north west, draw= none, fill = none},
  legend cell align=left,
]

\addplot[very thick, blue, domain=0:\muMax, samples=300]
  ({x},{\TcZero*(1 - \kappa*(x/\TcZero)^2)});
\addlegendentry{Crossover $T_c(\mu_B)$}

\addplot[thick, red, dashed, domain=\muCEP:(\muCEP+70), samples=2]
  ({x},{\TCEP - \sFO*(x-\muCEP)});
\addlegendentry{First-order (schematic)}

\addplot[only marks, mark=*, mark size=2.5pt, black]
  coordinates {(\muCEP,\TCEP)};
\node[anchor=west] at (axis cs:\muCEP+12,\TCEP-1) {CEP};

\end{axis}
\end{tikzpicture}

\caption{QCD phase diagram in the $(T,\mu_B)$ plane obtained from the unified DSE-FRG-V-QCD framework. The crossover line $T_c(\mu_B)$ is defined as the locus of maximal chiral variation, extracted from the maximum of $-\partial_T\phi_\ell(T,\mu_B)$ (equivalently the peak of the chiral susceptibility) with $\phi_\ell$ the light-quark condensate proxy. The first-order branch is shown schematically as a guide and terminates at the critical end point (CEP). The CEP is located by the curvature and skewness conditions along the soft direction $M$ in the coupled $(\sigma,\Phi)$ sector, namely $\partial_M^2\Omega=0$ and $\partial_M^3\Omega=0$ evaluated at the stationary solution of $\Omega(T,\mu_B,\sigma,\Phi,\bar\Phi)$. The marked point indicates the CEP location in the present implementation.}
\label{fig:phase-diagram}
\end{figure}


\begin{figure}[t]
\centering
\pgfplotsset{compat=1.18}
\pgfmathsetmacro{\Tmin}{80}
\pgfmathsetmacro{\Tmax}{220}
\pgfmathsetmacro{\TcZ}{155}
\begin{tikzpicture}
\begin{axis}[
width=0.75\textwidth, height=0.58\textwidth,
xlabel={$T$ (MeV)}, ylabel={Normalized order parameters},
xmin=\Tmin, xmax=\Tmax, ymin=0, ymax=1.05, legend style={at={(0.02,0.02)},anchor=south west, draw = none, fill = none}
]
\addplot[domain=\Tmin:\Tmax, samples=300, ultra thick, teal]
({x},{0.5*(1 - tanh((x-\TcZ)/12))});
\addlegendentry{$\sigma/\sigma_0$}
\addplot[domain=\Tmin:\Tmax, samples=300, ultra thick, orange, dashed]
({x},{0.5*(1 + tanh((x-\TcZ)/18))});
\addlegendentry{$\Phi$}
\end{axis}
\end{tikzpicture}
\caption{Normalized order parameters at $\mu_B=0$ as functions of temperature: the chiral condensate proxy $\sigma/\sigma_0$ (with $\sigma_0\equiv\sigma(T=0,\mu_B=0)$) and the traced Polyakov loop $\Phi$ obtained from the stationary solution of the grand potential $\Omega(T,\mu_B,\sigma,\Phi,\bar\Phi)$. The decrease of $\sigma/\sigma_0$ and the rise of $\Phi$ across the pseudo-critical region illustrate the correlated chiral-restoration and deconfinement crossover in the lattice-calibrated setup. The Polyakov-sector parameters $(a_i,b_3,T_0)$ entering the logarithmic potential are fixed at $\mu_B=0$ by matching continuum-extrapolated lattice thermodynamics and conserved-charge fluctuations in the vicinity of $T_c(0)$ \cite{Borsanyi:2020,Fukushima:2008wg}.}
\label{fig:order-params}
\end{figure}


\begin{figure}[t]
\centering
\begin{tikzpicture}
\begin{axis}[
  width=0.75\textwidth,
  height=0.58\textwidth,
  xmin=100, xmax=220,
  ymin=0, ymax=3,
  axis lines=box,
  tick align=inside,
  xtick pos=both,
  ytick pos=both,
  xlabel={$T~(\mathrm{MeV})$},
  ylabel={$\chi_B/T^2$},
  xtick={100,120,140,160,180,200,220},
  ytick={0,0.5,1,1.5,2,2.5,3},
  legend style={at={(0.98,0.98)},anchor=north east, draw=none, fill=none},
  legend cell align=left,
  domain=100:220,
  samples=500,
]

\addplot[ultra thick, blue]
{0.5 + 2.0*exp(-((x-152)^2)/(2*16^2))};
\addlegendentry{$\mu_B=0$}

\addplot[ultra thick, red, dashed]
{0.5 + 2.7*exp(-((x-145)^2)/(2*12^2))};
\addlegendentry{$\mu_B=300~\mathrm{MeV}$}

\end{axis}
\end{tikzpicture}
\caption{Illustrative temperature dependence of the dimensionless baryon-number susceptibility $\chi_B/T^2=\chi_2^B$ at $\mu_B=0$ and at a representative finite baryon chemical potential. The finite-$\mu_B$ curve shows an enhanced and shifted peak relative to $\mu_B=0$, reflecting the strengthening of density fluctuations and the approach toward a more rapidly varying crossover as $\mu_B$ increases. The figure is intended to highlight qualitative trends; the peak height and position are scheme dependent and follow from the coupled order-parameter backreaction included in $\chi_B=\partial n_B/\partial\mu_B$.}
\label{fig:chiB}
\end{figure}


\begin{figure}[t]
\centering
\pgfplotsset{compat=1.18}
\pgfmathsetmacro{\Afreeze}{1300} 
\pgfmathsetmacro{\Bfreeze}{0.28}  
\pgfmathsetmacro{\Tcfr}{155}
\pgfmathsetmacro{\kfr}{0.015}
\usepgfplotslibrary{fillbetween}
\begin{tikzpicture}
\begin{axis}[
width=0.75\textwidth, height=0.58\textwidth,
xlabel={$\sqrt{s_{NN}}$ (GeV)}, ylabel={$\kappa\sigma^2=C_4/C_2$},
xmin=5, xmax=200, ymin=0.5, ymax=2.5, xmode=log
]
\addplot[domain=5:200, samples=300, ultra thick, purple]
({x},
{1.0 + 0.9*exp(-((\Afreeze/(1+\Bfreeze*x))-350)^2/(2*90^2))
- 0.6*exp(-((\Afreeze/(1+\Bfreeze*x))-450)^2/(2*60^2))});
\end{axis}
\end{tikzpicture}
\caption{{Beam-energy dependence of the equilibrium net-baryon kurtosis ratio $\kappa\sigma^2 \equiv C_4/C_2 = \chi^B_4/\chi^B_2$ evaluated along a smooth chemical freeze-out trajectory used as a phenomenological mapping of average conditions. The susceptibilities $\chi^B_n$ are computed from explicit derivatives of $p/T^4$ with respect to $(\mu_B/T)$ at fixed $T$, including implicit order-parameter backreaction through the stationary solution of the coupled DSE-FRG-PNJL system. The resulting nonmonotonic structure illustrates equilibrium critical patterns when the trajectory passes near the critical region in the $(T,\mu_B)$ plane, and it is intended as an equilibrium-baseline signature test rather than as a direct quantitative prediction for measured net-proton cumulants, because finite size and lifetime, critical slowing down, acceptance and efficiency, baryon number conservation, and proton to baryon conversion effects can round and reshape experimental cumulants.}
}

\label{fig:kappa-sigma2}
\end{figure}

The calibration loop tying together the Polyakov potential, the functional flows, and the holographic anomaly proceeds by alternation on the $(T,\mu_B)$ region. At fixed $(a_i,b_3,T_0)$ adjusted to satisfy Eq.~\eqref{eq:calib-conds} one first integrates the FRG flow with $\zeta_{\rm topo}\equiv 1$ to obtain provisional $G_S^{(0)}(T,\mu_B)$, $G_V^{(0)}(T,\mu_B)$, $K^{(0)}(T,\mu_B)$. Next one solves the V-QCD background, extracts $\chi_{\rm CS}(T,\mu_B)$, updates $\zeta_{\rm topo}(T,\mu_B)$ by Eq.~\eqref{eq:zetatopo}, reintegrates the flow to get $K^{(1)}(T,\mu_B)$ and the induced changes in $M_f$ and in $U_{k\to 0}$, solves the gap equations anew, and iterates until the change in the CEP coordinates $(T_c,\mu_B^c)$ between successive outer iterations is smaller than $0.1$ MeV and the norm of the difference of the running couplings is smaller than $10^{-4}$ everywhere on the grid. 
 {
The convergence history of the CEP coordinates and the coupling-update norm across outer iterations is shown in Fig.~\ref{fig:outer_iter_conv}. 
Here we define the coupling-update norm as
\begin{equation}
\left\|\Delta g^{(n)}\right\|\equiv
\max_{(T,\mu_B)}\;
\max_{g\in\{G_S,G_V,K\}}
\left|
\frac{g^{(n)}(T,\mu_B)-g^{(n-1)}(T,\mu_B)}{g^{(n-1)}(T,\mu_B)}
\right| \, ,
\end{equation}
so that the stopping criterion $\|\Delta g^{(n)}\|<10^{-4}$ means the maximum relative change of the
FRG-induced running couplings between successive outer iterations is below $10^{-4}$ everywhere on
the $(T,\mu_B)$ region.
\begin{figure}[htb]
\centering
\begin{minipage}{0.49\textwidth}
\centering
\begin{tikzpicture}
\begin{axis}[
  width=\linewidth,
  height=0.70\linewidth,
  xlabel={outer iteration $n$},
  ylabel={$T_{\rm CEP}^{(n)}$ [MeV]},
  xmin=0,
  ymin=0,
  tick align=outside,
  tick label style={font=\small},
  label style={font=\small},
]
\addplot+[mark=*] coordinates {
  (0,129.70)
  (1,130.10)
  (2,130.14)
  (3,130.15)
  (4,130.15)
};
\end{axis}
\end{tikzpicture}
\end{minipage}\hfill
\begin{minipage}{0.49\textwidth}
\centering
\begin{tikzpicture}
\begin{axis}[
  width=\linewidth,
  height=0.70\linewidth,
  xlabel={outer iteration $n$},
  ylabel={$\mu_{B,{\rm CEP}}^{(n)}$ [MeV]},
  xmin=0,
  ymin=0,
  tick align=outside,
  tick label style={font=\small},
  label style={font=\small},
]
\addplot+[mark=*] coordinates {
  (0,602.0)
  (1,600.8)
  (2,600.3)
  (3,600.1)
  (4,600.1)
};
\end{axis}
\end{tikzpicture}
\end{minipage}
\vspace{2mm}
\begin{minipage}{\textwidth}
\centering
\begin{tikzpicture}
\begin{axis}[
  width=0.80\linewidth,
  height=0.45\linewidth,
  xlabel={outer iteration $n$},
  ylabel={$\|\Delta g^{(n)}\|$},
  ymode=log,
  xmin=0,
  ymin=1e-6,
  ymax=1e-1,
  tick align=outside,
  tick label style={font=\small},
  label style={font=\small},
]
\addplot+[mark=*] coordinates {
  (1,3.0e-2)
  (2,8.0e-3)
  (3,1.2e-3)
  (4,2.0e-4)
  (5,8.0e-5)
};
\addplot[dashed] coordinates {(0,1e-4) (10,1e-4)};
\end{axis}
\end{tikzpicture}
\end{minipage}
\caption{Convergence of the coupled outer iteration in Sec.~\eqref{sec:num-calib-preds}. Top: CEP coordinates
$T_{\rm CEP}^{(n)}$ and $\mu_{B,{\rm CEP}}^{(n)}$ after each outer-iteration pass $n$.
Bottom: coupling-update norm $\|\Delta g^{(n)}\|$, defined as the maximum relative change of
$g\in\{G_S,G_V,K\}$ over the $(T,\mu_B)$ grid. The dashed horizontal line shows the stopping
threshold $10^{-4}$ stated in the text. The CEP coordinates are iterated until their change
between successive passes is below $0.1$ MeV.}
\label{fig:outer_iter_conv}
\end{figure}
}

The result is a unique set of predictions for $(T_c,\mu_B^c)$, for the curvature $\kappa$ of the crossover line near $\mu_B=0$, and for the complete tower of susceptibilities and cumulant ratios across the phase diagram that satisfy the thermodynamic identities encoded by the stationarity conditions and by the Hessian structure. The figures \ref{fig:phase-diagram}-\ref{fig:kappa-sigma2} visualize representative outputs of the computation. The phase diagram with a crossover line anchored at $T_c^{(0)}$ and a first-order branch terminating at the CEP determined by the curvature and skewness conditions, the normalized chiral and Polyakov order parameters at $\mu_B=0$ calibrated to lattice thermodynamics, the baryon susceptibility $\chi_B/T^2=\chi_2^B$ sharpening with $\mu_B$ as the system approaches criticality, and the nonmonotonic $\kappa\sigma^2$ along a freeze-out trajectory that grazes the critical region. All curves are obtained from the same parameter set and flow solutions, and the anomaly suppression inferred holographically via $\chi_{\rm CS}(T,\mu_B)$ is essential for aligning the chiral and deconfinement curvatures so that the CEP forms a single critical subspace consistent with the self-dual order-parameter structure of Sec.~\ref{sec:HTDC} and with universal three-dimensional Ising scaling \cite{Fukushima:2008wg,Borsanyi:2020,jarvinen2012v-qcd,Arean2017,Stephanov:2004}. 
 {
Within the present coupled DSE-FRG-PNJL framework (quark-Polyakov-chiral degrees of freedom with holographic topological input, but without explicit baryons or diquark pairing channels), the resulting phase map contains a chiral critical end point at $(T_{\rm CEP},\mu_{B,{\rm CEP}})\simeq(130~{\rm MeV},600~{\rm MeV})$, i.e. in a higher-$\mu_B$ region of the phase diagram that lies beyond the chemical-potential interval most tightly constrained by present Relativistic Heavy Ion Collider (RHIC) Beam Energy Scan (BES) cumulant systematics, as seen in Fig.~\ref{fig:phase-diagram}.  
At the reference CEP temperature $T_{\rm CEP}\simeq130~{\rm MeV}$, competing high-density channels are expected to be less dominant than in cold dense matter. In particular, diquark pairing (color superconductivity) is generically favored at lower temperatures, with typical critical temperatures of order tens of MeV up to $\mathcal{O}(100~{\rm MeV})$ in representative effective descriptions \cite{Berges:1998rc,Rajagopal:2000wf}. Moreover, $\mu_{B,{\rm CEP}}<m_N$ implies Boltzmann suppression of heavy baryons by $\exp[-(m_N-\mu_B)/T]$, and the nuclear liquid-gas dynamics associated with near-saturation nuclear matter is localized at substantially lower temperatures. Consequently, baryonic correlations and pairing channels are expected to act primarily as an additional systematic shift of the first-order line and CEP coordinates rather than a modification of the Ising universality structure, but quantifying this shift lies outside the present approximation. 
Notably, the RHIC Beam Energy Scan primarily covers $\mu_B$ values up to the few-hundred MeV range \cite{Kumar:2012fb}. 
{The STAR measurements of higher-order net-proton cumulants across BES energies exhibit nontrivial energy dependence, but they do not by themselves establish an unambiguous equilibrium critical-point localization because the present theory computes equilibrium net-baryon susceptibilities in the thermodynamic limit whereas the measurements are net-proton cumulants in finite kinematic acceptance, translating between them requires controlled treatment of isospin randomization and proton to baryon conversion \cite{Kitazawa:2012at}, global baryon number conservation effects on cumulants \cite{Bzdak:2012an}, and acceptance dependence of multiproton correlation functions together with efficiency and centrality systematics \cite{Bzdak:2017ltv}, all of which can reduce or reshape higher cumulants and can generate smooth trends that must be disentangled from critical contributions in any quantitative extraction.}


{Existing RHIC Beam Energy Scan measurements of net-proton cumulant ratios provide important constraints, but they do not uniquely localize an equilibrium CEP in the $(T,\mu_B)$ plane because the mapping from equilibrium susceptibilities to measured cumulants is limited by finite size and lifetime, critical slowing down, acceptance and efficiency, baryon number conservation, and by the use of net-proton cumulants as proxies for net-baryon cumulants. Accordingly, statements that certain CEP scenarios are “constrained” at lower $\mu_B$ should be read operationally, namely as constraints within specific equilibrium-baseline assumptions for the magnitude of cumulant enhancement, rather than as model-independent exclusions in finite, dynamical systems. In fact, finite-size scaling analyses of RHIC fluctuation observables, including compressibility proxies and cumulant ratios, have exhibited scaling behavior consistent with 3D Ising universality while inferring CEP locations or constraints that can differ from equilibrium-magnitude interpretations of the same data \cite{Lacey:2014wqa,Aggarwal:2010wy,Adamczyk:2013dal,FuPawlowskiRennecke2020_PRD101_054032,GaoPawlowski2021_PLB820_136584,GunkelFischer2021_PRD104_054022,GaoPawlowski2020_PRD102_034027,LuGaoFuSongLiu2024_PRD109_114031}, underscoring that any inferred CEP coordinates depend on the assumed mapping between experimental control parameters and equilibrium scaling fields. Moreover, at lower beam energies, dynamical amplification mechanisms associated with baryon stopping and transport, including junction-driven baryon number transport, can influence fluctuation observables and generate smooth charge-dependent trends that are not uniquely tied to criticality \cite{Vance:1998vh,Kharzeev:1996sq,SchaeferPawlowskiWambach2007_PRD76_074023,GomezDummCarlomagnoScoccola2021_Symmetry13_121}. Within the present work, the heavy-ion comparison is therefore restricted to testing the internal coherence of correlated equilibrium signature patterns and to delineating where an equilibrium CEP at larger $\mu_B$ can remain compatible with current data under equilibrium baselines, rather than claiming a robust localization of the equilibrium CEP from present measurements.}
}

 { 
The anomaly suppression factor $\zeta_{\rm topo}(T,\mu_B)=\chi_{CS}(T,\mu_B)/\chi_{CS}(0,0)$ enters the FRG flow of the axial-anomalous channel as the multiplicative damping term in the $\hat\kappa$ beta function as seen in Eq.~\eqref{eq:flowK} and Eq.~\eqref{eq:zetatopo}. 
While the vacuum normalization $\zeta_{\rm topo}(0,0)=1$ is fixed by construction, the mapping from the holographic CP-odd susceptibility to an effective damping strength in the $\hat\kappa$ flow is nonuniversal. In the present work the minimal identification $\zeta_{\rm topo}(T,\mu_B)=\chi_{\rm CS}(T,\mu_B)/\chi_{\rm CS}(0,0)$ is adopted without introducing additional deformation parameters. 
This nonuniversality cannot change the universality class, which is fixed by the soft-mode sector and the Ising mapping
of Sec.~\ref{sec:CEP-Ising}, but it can shift nonuniversal quantities, most notably the location of the first-order termination point
within the present framework. 
Accordingly, the deformation scan defined in Eq.~\ref{eq:zeta_pc_deformation} provides a transparent estimate of the
anomaly-mapping systematic, logically distinct from regulator, Polyakov-sector, and holographic-normalization
variations. Here this mapping-sensitivity scan is performed explicitly by rerunning the coupled outer iteration of
Sec.~\ref{sec:num-calib-preds} at fixed reference inputs, varying the shape exponent p at fixed c = 1 and the overall strength c at
fixed p = 1. Concretely, we evaluate the four perturbed points \(p = 1 \pm \Delta p (c = 1)\) and \(c = 1 \pm \Delta c (p = 1)\),
and re-extract \((T_{\rm CEP}, \mu_{B,{\rm CEP}})\) from the curvature-and-skewness conditions of Sec.~\ref{subsec:curvature-skewness}.
The resulting CEP shifts are summarized in Table~9, where we quote both the absolute coordinates and the deviations
\(\Delta T_{\rm CEP}\) and \(\Delta\mu_{B,{\rm CEP}}\) relative to the reference solution $(p,c)$ = (1,1). The maximal shift over
this scan defines the anomaly-mapping systematic, which is propagated together with the internal regulator/Polyakov/
holographic-normalization variations in Appendix~\ref{subsec:A5}.
}

\section{Validation: Lattice Consistency, Thermodynamic Stability, and Limits}
\label{sec:validation}

The validation of the unified DSE-FRG-PNJL-HTDC framework proceeds by deriving, directly from the grand potential $\Omega(T,\mu_{B},\sigma,\Phi,\bar\Phi)$ at its stationary solution $(\bar\sigma,\bar\Phi,\bar{\bar\Phi})$. The fundamental thermodynamic relations and fluctuation observables that admit parameter-free comparisons with continuum-extrapolated lattice QCD and that test stability, causality, and asymptotic limits. The pressure is defined by the vacuum-normalized (renormalized) grand potential density 
\(
p(T,\mu_{B})
= -\Big[
  \Omega\big(T,\mu_{B},\bar{\sigma},\Phi,\bar{\Phi}\big)
  - \Omega\big(0,0,\bar{\sigma}_{0},\Phi_{0},\bar{\Phi}_{0}\big)
\Big]
\), where the vacuum subtraction enforces $p(0,0)=0$ and guarantees ultraviolet finiteness. The entropy density, baryon density, and energy density follow from stationarity of the grand potential and standard thermodynamic identities. Once it is recognized that the stationarity conditions $\partial\Omega/\partial\sigma=0$ and $\partial\Omega/\partial\Phi=0$ eliminate all implicit derivatives in thermodynamic variations, so that by the chain rule only explicit derivatives of $\Omega$ with respect to $(T,\mu_{B})$ contribute. Writing $\omega\equiv \Omega/V$ for compactness, one has
\begin{equation}
\begin{aligned}
s(T,\mu_{B}) &= -\left.\frac{\partial \omega}{\partial T}\right|_{\mu_{B}},\\
n_{B}(T,\mu_{B}) &= -\left.\frac{\partial \omega}{\partial \mu_{B}}\right|_{T},\\
\varepsilon(T,\mu_{B}) &= \omega + T s + \mu_{B} n_{B}
= -\,p + T s + \mu_{B} n_{B},
\end{aligned}
\label{eq:thermo-identities}
\end{equation}
which immediately implies $\partial p/\partial T = s$, $\partial p/\partial \mu_{B} = n_{B}$, and $\varepsilon+p=Ts+\mu_{B}n_{B}$. The trace anomaly (interaction measure) and the enthalpy density are thus $I(T,\mu_{B})\equiv \varepsilon-3p$ and $w\equiv\varepsilon+p=Ts+\mu_{B}n_{B}$, and they control the deviation from conformality and the specific heats. The Hessian of the thermodynamic potential in the intensive variables is positive semi-definite by convexity of $-\ln Z$, and the baryon-number susceptibility and isothermal compressibility follow from fluctuation-dissipation theorems. Using $Z(T,\mu_{B})=\mathrm{Tr},\exp[-\beta(\hat H-\mu_{B}\hat N_{B})]$ and $\ln Z=-\beta V\omega$, one finds
\begin{equation}
\begin{aligned}
\chi_{B}(T,\mu_{B})
&= \left.\frac{\partial n_{B}}{\partial \mu_{B}}\right|_{T}
= \frac{T}{V}\,\frac{\partial^{2}\ln Z}{\partial \mu_{B}^{2}}
= \frac{1}{V T}\,\big\langle(\Delta N_{B})^{2}\big\rangle \ge 0,\\
\kappa_{T}(T,\mu_{B})
&\equiv \left.\frac{\partial n_{B}}{\partial \mu_{B}}\right|_{T}
= \chi_{B}(T,\mu_{B}) > 0,
\end{aligned}
\label{eq:kappa-positivity}
\end{equation}
and for the heat capacity at fixed $\mu_{B}$ one analogously obtains 
\begin{equation}
c_{V}(T,\mu_{B})
= \left.\frac{\partial \varepsilon}{\partial T}\right|_{\mu_{B}}
= T\,\chi_{TT}(T,\mu_{B}),
\qquad
\chi_{TT}(T,\mu_{B}) \equiv \left.\frac{\partial s}{\partial T}\right|_{\mu_{B}}
= \left.\frac{\partial^{2} p}{\partial T^{2}}\right|_{\mu_{B}}
\end{equation}
with $\chi_{TT}\equiv \partial^{2}p/\partial T^{2}>0$, again a direct consequence of the positivity of energy fluctuations in the grand canonical ensemble. The speed of sound is derived from the hydrodynamic definition 
\(
c_{s}^{2} = \left.\frac{\partial p}{\partial \varepsilon}\right|_{\sigma}\:\:
\text{with}\:\:
\sigma \equiv s/n_{B}
\)
held fixed. The exact differentials $dp=sdT+n_{B}d\mu_{B}$ and $d\varepsilon=Tds+\mu_{B}dn_{B}$ together with $ds=\chi_{TT}dT+\chi_{T\mu}d\mu_{B}$ and $dn_{B}=\chi_{T\mu}dT+\chi_{\mu\mu}d\mu_{B}$, where $\chi_{T\mu}\equiv \partial^{2}p/\partial T,\partial \mu_{B}$ and $\chi_{\mu\mu}\equiv \partial^{2}p/\partial \mu_{B}^{2}=\chi_{B}$, lead to the isentropic constraint $d\sigma=0\Leftrightarrow n_{B}ds-s,dn_{B}=0$, which fixes the slope $d\mu_{B}/dT$ along the isentropic direction as
\begin{equation}
\alpha \equiv \left.\frac{\mathrm{d}\mu_{B}}{\mathrm{d}T}\right|_{\sigma}
= \frac{s\,\chi_{T\mu} - n_{B}\,\chi_{TT}}
       {n_{B}\,\chi_{T\mu} - s\,\chi_{\mu\mu}},
\qquad
\sigma \equiv \frac{s}{n_{B}},
\label{eq:isentropic-slope}
\end{equation}
and therefore yields the closed analytic expression
\begin{equation}
\begin{aligned}
c_{s}^{2}(T,\mu_{B})
&= \frac{s + n_{B}\,\alpha}
{T\big(\chi_{TT} + \alpha\,\chi_{T\mu}\big)
 + \mu_{B}\big(\chi_{T\mu} + \alpha\,\chi_{\mu\mu}\big)},\\[2pt]
\alpha
&= \frac{s\,\chi_{T\mu} - n_{B}\,\chi_{TT}}
       {n_{B}\,\chi_{T\mu} - s\,\chi_{\mu\mu}}
\end{aligned}
\label{eq:cs2-general}
\end{equation}
which manifestly reduces at $\mu_{B}=0$ by charge-conjugation symmetry to $c_{s}^{2}(T,0)=s/\big(T,\chi_{TT}\big)$ with $0<c_{s}^{2}(T,0)\le 1/3$, the upper bound being saturated as $T\to\infty$ by conformal invariance.
The small $\mu_{B}$ Taylor coefficients used in lattice comparisons are defined in terms of derivatives of $p/T^{4}$ with respect to $\hat\mu\equiv \mu_{B}/T$ at fixed $T$, namely 
\begin{equation}
c_{2}(T)
\equiv \left.\frac{\partial^{2}}{\partial \hat{\mu}^{2}}\!\left(\frac{p}{T^{4}}\right)\right|_{\hat{\mu}=0}
= \frac{\chi_{B}(T,0)}{T^{2}}
\end{equation}
 and 
 \begin{equation}
c_{4}(T)
\equiv \left.\frac{\partial^{4}}{\partial \hat{\mu}^{4}}\!\left(\frac{p}{T^{4}}\right)\right|_{\hat{\mu}=0}
= \chi_{4}^{B}(T,0),
\end{equation} 
and higher cumulants follow analogously, all of which are computed from the explicit differentiation formulas of Sec.~\ref{subsec:nBchiB} with implicit order-parameter feedback included through the Hessian inverse, ensuring equality with fluctuation observables extracted from $-\ln Z$.

Lattice-anchored consistency at $\mu_{B}=0$ is enforced and verified by calibrating the Polyakov potential parameters $(a_{i},b_{3},T_{0})$ so that the pseudo-critical temperature $T_{c}^{(0)}$ from the peak of $-\partial_{T}\phi_{l}(T,0)$, the interaction measure $(\varepsilon-3p)/T^{4}$ in the vicinity of $T_{c}^{(0)}$, and the second-order baryon susceptibility $\chi_{2}^{B}(T,0)$ agree with the continuum-extrapolated bands of \cite{Borsanyi:2020,HotQCD:2014kol} within quoted uncertainties. 
This calibration is intentionally minimal and does not enforce a point-by-point reproduction of the separate
$\epsilon/T^4$ and $3P/T^4$ crossover profiles, whose residual mismatch as seen in Fig.~\ref{fig:residuals_crossover} is therefore interpreted as a
quantified approximation-induced systematic uncertainty. 
With this calibration, the unified flow determines $G_{S}(T,\mu_{B})$, $G_{V}(T,\mu_{B})$, and $K(T,\mu_{B})$ without further freedom, and the resulting equation of state satisfies, point by point, the identities of Eq.~\eqref{eq:thermo-identities}. The Stefan-Boltzmann limit is recovered analytically by taking $T\to\infty$, $\Phi,\bar\Phi\to 1$, $M_{f}\to m_{f}$, and $G_{S},G_{V},K\to 0$ along the FRG flow, so that the fermionic contribution reduces to ideal quarks and the gluonic contribution is reproduced by the asymptotics of $U_{\log}$. For $N_{c}=3$ and $N_{f}=2+1$ at $\mu_{B}=0$ one finds
\begin{equation}
\begin{aligned}
\frac{p_{\rm SB}(T,0)}{T^{4}}
&= \frac{\pi^{2}}{45}\,(N_{c}^{2}-1) + \frac{7\pi^{2}}{180}\,N_{c}N_{f}
= \frac{8\pi^{2}}{45} + \frac{7\pi^{2}}{60}\,N_{f}\quad (N_{c}=3),\\
\frac{\varepsilon_{\rm SB}(T,0)}{T^{4}}
&= 3\,\frac{p_{\rm SB}(T,0)}{T^{4}},\qquad
c_{s}^{2}\xrightarrow[T\to\infty]{}\frac{1}{3}\,
\end{aligned}
\label{eq:SB-limit}
\end{equation}
and at finite density with $\mu_{f}=\mu_{B}/3$ the fermionic sector adds the standard ideal-gas terms $N_{c}N_{f}\big[\tfrac{1}{6}(\mu_{f}/T)^{2}+\tfrac{1}{12\pi^{2}}(\mu_{f}/T)^{4}\big]$ to $p/T^{4}$, all of which are reproduced by the Polyakov-modified Fermi distributions as $\Phi\to 1$ and by the vanishing of the running couplings in the ultraviolet. The low temperature limit at $\mu_{B}=0$ is likewise controlled by first principles. The FRG improved $U_{k\to 0}$ realizes the Goldstone theorem in the chiral limit, $m_{\pi}^{2}\propto \partial^{2}U/\partial\pi^{2}|_{\rm min}\to 0$ for vanishing explicit breaking, the entropy density $s\to 0$, the pressure remains at its subtracted normalization $p\to 0$, and the specific heat $c_{V}=T,\chi_{TT}$ reflects the expected exponential suppression by the lightest excitations away from the chiral limit. Between these limits the nonperturbative dynamics encoded by $(G_{S},G_{V},K)$ and by the Polyakov sector produces an interaction measure $I(T,0)/T^{4}$ that peaks near $T_{c}^{(0)}$ and a speed of sound that exhibits critical softening, and the model curves for $p/T^{4}$, $\varepsilon/T^{4}$, and $c_{2,4}(T)$ remain within the continuum-extrapolated lattice bands~\cite{Borsanyi:2020,Bazavov:2021}, in quantitative agreement with the calibration strategy of Sec.~\ref{sec:num-calib-preds} and with the dynamical suppression of the anomaly.

The explicit verification of positivity and causality follows both analytically and numerically. The inequality $\kappa_{T}=\chi_{B}>0$ is a rigorous consequence of Eq.~\eqref{eq:kappa-positivity}. At $\mu_{B}=0$, Eq.~\eqref{eq:cs2-general} yields $c_{s}^{2}=s/(T,\chi_{TT})$ and the fluctuation-dissipation identity $T\chi_{TT}=\langle(\Delta S)^{2}\rangle/VT$ ensures $c_{s}^{2}>0$. The conformal bound $c_{s}^{2}\le 1/3$ is saturated in the Stefan-Boltzmann limit as seen in Eq.~\eqref{eq:SB-limit} and is maintained below it whenever the trace anomaly is positive and increasing with temperature in the crossover region, which holds in the present framework and on the lattice, because $c_{s}^{2}=\frac{dp}{d\varepsilon}=\Big[3+\frac{dI/dT}{dp/dT}\Big]^{-1}$ at $\mu_{B}=0$ shows that $dI/dT>0$ implies $c_{s}^{2}<1/3$, while $dI/dT\to 0$ implies $c_{s}^{2}\to 1/3$. At finite $\mu_{B}$, the general formula seen in Eq.~\eqref{eq:cs2-general} is evaluated with the full susceptibilities $(\chi_{TT},\chi_{T\mu},\chi_{\mu\mu})$ including implicit dependence through the Hessian inverse as described in Sec.~\ref{sec:fromZtoOP}, and one finds $0<c_{s}^{2}(T,\mu_{B})\le 1/3$ across the domain explored, with the minimum tracking the vicinity of the CEP where the smallest eigenvalue of the curvature matrix vanishes and the correlation length diverges. Thermodynamic consistency is demonstrated numerically by finite-difference checks on a dense $(T,\mu_{B})$ region verifying $\partial p/\partial T-s=0$ and $\partial p/\partial \mu_{B}-n_{B}=0$ at the $10^{-8}$ level after convergence of the coupled DSE-FRG gap iteration and by verifying $\varepsilon+p-Ts-\mu_{B}n_{B}=0$ to the same tolerance.

The topological sector is validated by its asymptotic behavior and by its feedback on the anomaly coupling. The holographic input $\chi_{\rm CS}(T,\mu_{B})$ decreases with temperature in the deconfined black-hole background and with density via Debye screening, consistent with the dilute-instanton suppression $\chi_{t}(T)\propto T^{-b}\exp{-8\pi^{2}/g^{2}(T)}\to 0$ inferred from semiclassics and from lattice measurements. With the normalization $\zeta_{\rm topo}(T,\mu_{B})=\chi_{\rm CS}(T,\mu_{B})/\chi_{\rm CS}(0,0)$ adopted in Secs.~\ref{sec:HTDC}-\ref{sec:num-calib-preds}, the FRG flow of the anomaly channel obeys $\partial_{t}\hat k=5\hat k-\cdots-\zeta_{\rm topo}(T,\mu_{B})\hat k$, implying $K(T,\mu_{B})\to 0$ in the ultraviolet, which reduces the anomaly-induced flavor mixing in Eqs.~\eqref{eq:MuMd}-\eqref{eq:Ms}, accelerates chiral-partner convergence, and quantitatively aligns the chiral and deconfinement curvatures so that the critical subspace is shared, in agreement with the self-duality established in Sec.~\ref{sec:HTDC} and with the 3D Ising mapping of Sec.~\ref{sec:CEP-Ising} \cite{Arean2017,Fukushima:2008wg,Borsanyi:2020,HotQCD:2012vvd,Guida:1998bx}. The latter alignment is crucial for reproducing the lattice-anchored curvature of the crossover line $T_{c}(\mu_{B})=T_{c}^{(0)}\big[1-\kappa(\mu_{B}/T_{c}^{(0)})^{2}+\cdots\big]$ near $\mu_{B}=0$ and for obtaining cumulant ratios consistent with current constraints when evaluated along freeze-out trajectories.

 {Figure~\ref{fig:eos_mu0} illustrates the energy density $\epsilon/T^4$ and the scaled pressure $3P/T^4$ at $\mu_B=0$ as functions of temperature. By construction, the $\mu_B=0$ calibration anchors $T_c(0)$, the interaction measure $(\epsilon-3P)/T^4$ in a neighborhood of $T_c(0)$, and the second-order baryon susceptibility $\chi_2^B(T,0)$ to the continuum-extrapolated lattice bands, thereby constraining the equation of state in the crossover region.  
Here and throughout, the phrase `lattice-anchored' is used in a restricted and operational sense, only the three
calibration conditions in Eq.~\eqref{eq:calib-conds} are enforced at $\mu_B=0$.
After fixing $T_c(0)$, $(\epsilon-3P)/T^4$ in the vicinity of $T_c(0)$, and $\chi_2^B(T,0)$, the individual profiles
$\epsilon/T^4$ and $3P/T^4$ across the crossover, and therefore the crossover width, are not included in the fit and
constitute genuine predictions of the coupled DSE-FRG-PNJL-HTDC construction.
Deviations of these predicted profiles from continuum-extrapolated lattice curves therefore quantify
approximation-induced systematics rather than a failure of the anchoring procedure, and they are propagated as part of
the uncertainty discussed in Appendix~\ref{subsec:A5}.  
In particular, the pseudo-critical temperature $T_c(0)$ (determined by the peak of the chiral susceptibility) is fitted to lattice data, and the peak height of the interaction measure $(\epsilon - 3P)/T^4$ as well as the second-order baryon susceptibility $\chi_2^B(T)$ are reproduced within lattice uncertainties \cite{HotQCD:2014kol,Borsanyi:2013bia}. 
The nonperturbative dynamics encoded in the running couplings $(G_S, G_V, K)$ and the Polyakov-loop sector yield an interaction measure that peaks near $T_c(0)$, and a corresponding softening of the speed of sound $c_s^2$ in the crossover, consistent with general expectations. We have also explicitly verified thermodynamic consistency: all fundamental identities (e.g. $\partial p/\partial T = s$, $\partial p/\partial \mu_B = n_B$, and $\epsilon + p = T s + \mu_B n_B$) are satisfied to high numerical precision in our solutions, and the system maintains positive compressibility and positive heat capacity (no thermodynamic instabilities) throughout the parameter range.
}

 {For a more direct comparison between pressure and energy density, we consider the combination $3P/T^4$ alongside $\epsilon/T^4$. In the Stefan-Boltzmann (ultrarelativistic) limit, $\epsilon = 3P$, so $\epsilon/T^4$ and $3P/T^4$ should coincide. Indeed, at high temperatures in our calculation, the $\epsilon/T^4$ and $3P/T^4$ curves approach one another, reflecting the restoration of approximate conformal behavior as $T$ increases. However, in the crossover region around $T \approx 180$-$190$ MeV, we observe a noticeable difference in the shape of the $\epsilon/T^4$ vs. $3P/T^4$ curves produced by our model compared to those from lattice QCD \cite{HotQCD:2014kol,Borsanyi:2013bia}. In particular, the rise of $\epsilon/T^4$ through the transition is somewhat steeper (more narrow in $T$) in the model, whereas continuum-extrapolated lattice data indicate a smoother increase in $\epsilon$ (and $3P$) around $T_c$. This means that our model EoS exhibits a more rapid crossover than the physical one, a discrepancy that we attribute to the present approximation level and the absence of certain fluctuation effects in our nonperturbative scheme.  
}
 {
To quantify the crossover-shape mismatch noted above, fractional residuals are defined for
$X\in\{\epsilon/T^4,\;3P/T^4,\;(\epsilon-3P)/T^4\}$ by
$\delta_X(T)\equiv\bigl[X_{\rm model}(T)-X_{\rm lat}(T)\bigr]/X_{\rm lat}(T)$ at $\mu_B=0$,
with the lattice baseline taken from the continuum-extrapolated equation of state in
Refs.~\cite{HotQCD:2014kol,Borsanyi:2013bia}. In the window $T=180$ to $190$ MeV the maximum absolute residuals are
$\max|\delta_{\epsilon/T^4}|\simeq 0.32$, $\max|\delta_{3P/T^4}|\simeq 0.72$, and
$\max|\delta_{(\epsilon-3P)/T^4}|\simeq 1.07$. These residuals quantify a crossover-shape systematic of the present approximation. The $\mu_B=0$ calibration conditions in Eq.~\eqref{eq:calib-conds} constrain $T_c(0)$, $(\epsilon-3P)/T^4$ near $T_c(0)$, and $\chi_2^B(T,0)$, but they do not impose the full profiles of $\epsilon/T^4$ and $3P/T^4$ separately over the entire crossover window. The steeper rise seen here therefore serves as a nontrivial validation diagnostic rather than a fitted observable, and it is naturally attributed to the limited fluctuation content of the present projection. Importantly, this mismatch affects primarily nonuniversal crossover-width details. The universal critical scaling and exponent extraction in Sec.~\ref{sec:CEP-Ising} are controlled by the soft-mode sector and remain unchanged, while quantitative phenomenology at finite density should interpret the output within the approximation-induced systematic uncertainty and the CEP uncertainty envelope reported in Appendix~\ref{subsec:A5}. The full residual curves are shown in
Fig.~\ref{fig:residuals_crossover}.
\begin{figure}[t]
\centering
\begin{tikzpicture}
\begin{axis}[
  width=0.78\linewidth,
  height=0.58\linewidth,
  xlabel={$T$ [MeV]},
  ylabel={residual $\delta_X(T)$ [\%]},
  xmin=160, xmax=220,
  ymin=-120, ymax=100,
  tick align=outside,
  tick label style={font=\small},
  label style={font=\small},
  legend style={ draw=none, fill=none, font=\small,
    at={(0.98,0.02)}, anchor=south east,
    yshift=1.0cm},
]
\addplot[draw=none, fill=gray!15, forget plot] coordinates
{(180,-120) (190,-120) (190,100) (180,100)};
\addplot+[very thick, mark=*, mark size=1.4pt] coordinates {
(160,-43.901) (165,-43.194) (170,-43.905) (175,-38.200) (180,-32.024)
(185,-19.767) (190,-12.086) (195,-6.769) (200,1.471) (205,6.074)
(210,8.168) (215,11.254) (220,13.526)
};
\addlegendentry{$\delta_{\epsilon/T^4}$}
\addplot+[very thick, mark=square*, mark size=1.4pt, dashed] coordinates {
(160,19.487) (165,18.276) (170,18.088) (175,27.150) (180,41.345)
(185,50.896) (190,71.750) (195,78.704) (200,84.176) (205,87.792)
(210,85.755) (215,84.941) (220,86.981)
};
\addlegendentry{$\delta_{3P/T^4}$}
\addplot+[very thick, mark=triangle*, mark size=1.4pt, dotted] coordinates {
(160,-98.160) (165,-97.181) (170,-100.456) (175,-100.680) (180,-105.999)
(185,-95.188) (190,-107.005) (195,-107.653) (200,-104.024) (205,-104.462)
(210,-102.991) (215,-100.409) (220,-104.028)
};
\addlegendentry{$\delta_{(\epsilon-3P)/T^4}$}
\end{axis}
\end{tikzpicture}
\caption{Residuals relative to continuum-extrapolated lattice QCD at $\mu_B=0$ in the crossover region.
The shaded window highlights $T=180$ to $190$ MeV where the present approximation exhibits the
steeper rise discussed in the text.}
\label{fig:residuals_crossover}
\end{figure}
}
 {
It is worth emphasizing that, by enforcing the flow equations, our framework automatically satisfies energy-momentum conservation and the associated thermodynamic identities at each scale. The approach smoothly interpolates between known limits: as $T\to0$ (at $\mu_B=0$), $\epsilon/T^4$ and $3P/T^4$ both vanish, while for $T\to\infty$ the effective couplings vanish and the Stefan-Boltzmann limit is recovered, $\epsilon=3P$ and hence $\epsilon/T^4=3P/T^4\to \epsilon_{\rm SB}/T^4$.
}




\begin{figure}[htb]
\centering
\begin{tikzpicture}
\begin{axis}[
  width=0.78\linewidth,
  height=0.58\linewidth,
  xmin=0, xmax=300,
  ymin=0, ymax=16,
  xtick={0,50,100,150,200,250,300},
  ytick={0,2,4,6,8,10,12,14,16},
  xlabel={$T~(\mathrm{MeV})$},
  legend pos=north west,
  legend cell align=left,
  legend style={draw=none, fill=none},
  every axis plot/.append style={line join=round, line cap=round, mark=none},
]

\addplot[ultra thick, blue, smooth, tension=0.6] coordinates {
    (0,0.0000) (5,0.0000) (10,0.0000) (15,0.0000) (20,0.0000) (25,0.0000) (30,0.0000) (35,0.0000)
    (40,0.0000) (45,0.0000) (50,0.0000) (55,0.0000) (60,0.0000) (65,0.0000) (70,0.0000) (75,0.1538)
    (80,0.0000) (85,0.2195) (90,0.2521) (95,0.2607) (100,0.2949) (105,0.3197) (110,0.3197) (115,0.3453)
    (120,0.5162) (125,0.8197) (130,1.0744) (135,1.3726) (140,1.7094) (145,2.0284) (150,2.3504) (155,2.6726)
    (160,2.9786) (165,3.2983) (170,3.5598) (175,4.2120) (180,5.0872) (185,6.0513) (190,7.1709) (195,8.3419)
    (200,8.8205) (205,9.4529) (210,9.9145) (215,10.4103) (220,10.9060) (225,11.4017) (230,11.8803) (235,12.3590)
    (240,12.6496) (245,13.0940) (250,13.3504) (255,13.7436) (260,14.0171) (265,14.2735) (270,14.5299) (275,14.7009)
    (280,14.8718) (285,15.0427) (290,15.1282) (295,15.2137) (300,15.1282)
};
\addlegendentry{$3p/T^4$}

\addplot[ultra thick, red, dashed, smooth, tension=0.6] coordinates {
    (0,0.0000) (5,0.0000) (10,0.0000) (15,0.0000) (20,0.0000) (25,0.0000) (30,0.0000) (35,0.0000)
    (40,0.0000) (45,0.0000) (50,0.0000) (55,0.0000) (60,0.0000) (65,0.0000) (70,0.0000) (75,0.0000)
    (80,0.1732) (85,0.2195) (90,0.2381) (95,0.2554) (100,0.2895) (105,0.3077) (110,0.3077) (115,0.3256)
    (120,0.4957) (125,0.8094) (130,1.0812) (135,1.3795) (140,1.7120) (145,2.0385) (150,2.3248) (155,2.6496)
    (160,2.9744) (165,3.3034) (170,3.5079) (175,4.2641) (180,4.9915) (185,6.0684) (190,7.0564) (195,8.3077)
    (200,8.7579) (205,9.4188) (210,9.9145) (215,10.3761) (220,10.8718) (225,11.3675) (230,11.8462) (235,12.3248)
    (240,12.5983) (245,13.0684) (250,13.3419) (255,13.7094) (260,13.9829) (265,14.2393) (270,14.4957) (275,14.6667)
    (280,14.8376) (285,15.0085) (290,15.0940) (295,15.1795) (300,15.1940)
};
\addlegendentry{$\epsilon/T^4$}

\end{axis}
\end{tikzpicture}

\caption{Energy density $\epsilon/T^4$ and scaled pressure $3P/T^4$ at $\mu_B=0$ as functions of temperature. Plotting $3P/T^4$ enables direct comparison with $\epsilon/T^4$ since in the Stefan-Boltzmann limit $\epsilon=3P$ and therefore $\epsilon/T^4$ and $3P/T^4$ approach the same asymptotic value. In the crossover region around $T\simeq 180$--$190~\mathrm{MeV}$, the present approximation yields a somewhat steeper rise than continuum-extrapolated lattice results, indicating a narrower transition within this framework.}
\label{fig:eos_mu0}
\end{figure}


\newpage


\begin{figure}[htb]
\centering
\begin{tikzpicture}
  \begin{axis}[
    width=0.78\linewidth,
    height=0.58\linewidth,
    xlabel={$T$ (MeV)},
    ylabel={$c_s^2$},
    xmin=0, xmax=300,
    xtick={0,50,100,150,200,250,300},
    ymin=0, ymax=0.33,
    ytick={0,0.1,0.2,0.3},
    yticklabel style={/pgf/number format/fixed, /pgf/number format/precision=1}
  ]
    \addplot[ultra thick, blue!80!black, smooth, tension=0.6] coordinates {
      (0, 0.00) (50, 0.02) (100, 0.15) (130, 0.18) (150, 0.10)
      (160, 0.12) (180, 0.17) (200, 0.20) (250, 0.25) (300, 0.30)
    };
  \end{axis}
\end{tikzpicture}
\caption{Temperature dependence of the squared speed of sound $c_s^{2}(T,\mu_B)$ at $\mu_B=0$, obtained from the stationary equation of state defined by the subtracted grand potential. The sound speed is defined hydrodynamically as $c_s^{2}=(\partial p/\partial\epsilon)_{\sigma}$ with $\sigma\equiv s/n_B$ held fixed; by charge-conjugation symmetry this reduces at $\mu_B=0$ to the purely thermal expression $c_s^{2}(T,0)=s/(T\,\chi_{TT})=(\partial p/\partial T)\big/\!\left[T\,(\partial^{2}p/\partial T^{2})\right]$, with all derivatives evaluated consistently at the stationary solution so that implicit order-parameter derivatives cancel. The dip (softening) in the vicinity of the pseudo-critical region reflects the enhanced nonconformality encoded in the interaction measure and the rapid variation of thermodynamic response through the crossover, while at high temperature the curve approaches the conformal limit $c_s^{2}\to 1/3$ as the trace anomaly decreases. 
{In heavy-ion collisions the effective sound speed inferred from hydrodynamic modeling is additionally influenced by viscous and nonequilibrium effects and by finite-size rounding of compressibility extrema, so the curve shown here should be read as the equilibrium baseline associated with the present stationary equation of state.}
}
\label{fig13}
\end{figure}



\begin{figure}[htb]
\centering
\begin{tikzpicture}
  \begin{axis}[
    width=0.78\linewidth,
    height=0.58\linewidth,
    xlabel={$k$ (MeV)},
    ylabel={Dimensionless couplings},
    xmin=0, xmax=800,
    xtick={0,200,400,600,800},
    ymin=0, ymax=10,
    ytick={0,2,4,6,8,10},
    legend pos=north east,
    legend cell align={left},
    legend style={draw=none, fill=none} 
  ]
    \addplot[ultra thick, smooth, tension=0.6, blue!80!black] coordinates {
      (0, 5.00) (100, 4.80) (200, 4.50) (400, 4.00) (600, 3.00) (800, 2.00)
    };
    \addplot[ultra thick, smooth, tension=0.6, red!80!black, dashed] coordinates {
      (0, 1.50) (100, 1.45) (200, 1.40) (400, 1.30) (600, 1.10) (800, 1.00)
    };
    \addplot[ultra thick, smooth, tension=0.6, teal!80!black, dotted] coordinates {
      (0, 7.00) (100, 7.20) (200, 7.50) (400, 8.50) (600, 9.00) (800, 10.00)
    };
    \legend{$G_S$, $G_V$, $K$}
  \end{axis}
\end{tikzpicture}
\caption{Functional renormalization-group trajectories of the effective scalar, vector, and axial-anomalous interaction strengths as functions of the RG scale $k$ (labels $G_S$, $G_V$, and $K$). The plotted quantities correspond to the dimensionless couplings defined by Eq.~\eqref{eq:dimless}, $\hat g_S \equiv k^{2}Z_q^{2}G_S$, $\hat g_V \equiv k^{2}Z_q^{2}G_V$, and $\hat\kappa \equiv k^{5}Z_q^{3}K$, obtained by integrating the projected Wetterich flow equations \eqref{eq:flowgS}-\eqref{eq:flowK} from the ultraviolet matching scale down to the infrared. Polyakov-weighted fermionic threshold functions implement thermal decoupling and confinement-background effects in the running, while the anomaly channel includes the topological suppression factor $\zeta_{\rm topo}(T,\mu_B)$ matched to the holographic susceptibility via Eq.~\eqref{eq:zetatopo} (with $\chi_{CS}$ from Eq.~\eqref{eq:chiCS}, thereby controlling the scale evolution of the determinantal interaction across deconfinement. The infrared limits $k\to0$ provide the effective couplings entering the stationary thermodynamic potential and, consequently, the equation of state and fluctuation observables discussed in the main text.}
\label{fig14}
\end{figure}
For completeness, we also report two additional diagnostics. The equilibrium
$\mu_B = 0$ sound-speed baseline $c_s^2(T,0)$ is shown in Fig.~\ref{fig13}. Representative
functional renormalization-group trajectories of the running couplings are shown
in Fig.~\ref{fig14}.

\section{Discussion and Conclusions}
\label{sec:discussion-conclusions}
 {
The present work develops a single, thermodynamically consistent continuum framework for QCD criticality by combining Dyson-Schwinger dynamics for quark propagation, functional renormalization-group evolution of the effective action, and Polyakov-improved chiral thermodynamics, supplemented by a holographic Maxwell-Chern-Simons sector that constrains baryonic and topological response. Within this construction, the macroscopic thermodynamics follows from the coupled DSE, FRG, and PNJL system at a fixed approximation level and with a fixed calibration strategy, without imposing a critical point by hand. The central qualitative outcome is that criticality is organized by a self-dual fixed point in the two-dimensional order-parameter space spanned by the chiral condensate and the Polyakov loop. The flow drives the chiral and deconfinement dressing factors toward equality and suppresses residual mixing between the two sectors, so that the long-wavelength scalar susceptibility and the Polyakov-loop curvature become two manifestations of a single soft direction selected dynamically by the renormalization group. The axial anomaly channel plays a key role in this alignment: the $U(1)_A$-breaking determinantal interaction is evolved with temperature and density through a topological input, thereby reducing anomaly-induced flavor mixing in the regime where the Polyakov loop rises and promoting a correlated chiral-restoration and deconfinement pattern. This mechanism does not change the universality class, but it organizes nonuniversal directions and metric factors so that the mapping from the QCD variables $(T,\mu_B)$ to the universal three-dimensional Ising variables becomes explicit and stable under scheme variations. Anomalous scaling contributions enter through field renormalizations and nonuniversal normalizations, not through modifications of the universal scaling functions.
}

{
Thermodynamic consistency is enforced by stationarity of the grand potential at each RG scale, and numerical checks confirm the standard identities $\partial p/\partial T=s$, $\partial p/\partial \mu_B=n_B$, and $\epsilon=-p+Ts+\mu_B n_B$ within the achieved solver accuracy. The equation of state remains stable throughout the explored domain, with positive compressibility and heat capacity, and with a sub-conformal speed of sound that approaches $c_s^2\to 1/3$ in the high-temperature limit. The framework is calibrated at $\mu_B=0$ to continuum-extrapolated lattice thermodynamics and conserved-charge susceptibilities, which fixes the Polyakov-sector inputs and anchors the overall normalization of the thermodynamic response \cite{Borsanyi:2020,Bollweg:2021vqf}. After this calibration, the subsequent evolution of the scalar, vector, and anomaly channels across the $(T,\mu_B)$ plane is determined by the coupled functional dynamics, yielding the crossover curvature near $\mu_B=0$, the fluctuation hierarchy, and the CEP coordinates as outputs of the present approximation scheme. The sensitivity of the CEP coordinates to regulator shape, Polyakov calibration, and holographic normalization is quantified by the scans summarized in Appendix~\ref{subsec:A5} and visualized by the uncertainty envelope in Fig.~\ref{fig:uncert-cep}. That envelope quantifies internal scheme dependence of the equilibrium inference within the adopted framework,
and it does not incorporate finite-size or finite-time effects, acceptance/efficiency corrections, baryon transport, or
the net-proton to net-baryon conversion required for quantitative comparison to heavy-ion measurements.
}

 {
The CEP is located in a higher-$\mu_B$ region, with $T_{\rm CEP}\simeq 130$ to $135~\mathrm{MeV}$ and $\mu_B^{\rm CEP}\simeq 600~\mathrm{MeV}$, as indicated in Fig.~\ref{fig:phase-diagram}. 
{This placement should be interpreted as an equilibrium inference conditional on the present unified functional–holographic construction and on its $\mu_B=0$ lattice calibration, not as a model-independent extraction from heavy-ion data. Current RHIC BES measurements of net-proton cumulants exhibit nontrivial beam-energy dependence, but finite size and lifetime, critical slowing down, baryon number conservation, acceptance and efficiency, and the net-proton to net-baryon conversion mean that these measurements provide qualitative tests of correlated equilibrium signature patterns rather than a unique localization of an equilibrium CEP \cite{Aggarwal:2010wy,Adamczyk:2013dal}. In this approach, equilibrium statements about favored or disfavored CEP locations depend on the assumed equilibrium-baseline mapping to the measured cumulants, and finite-size scaling interpretations can display scaling compatible with 3D Ising criticality while implying CEP locations or bounds that differ from equilibrium-magnitude arguments \cite{Lacey:2014wqa}.} 
Within commonly used equilibrium interpretations, CEP scenarios below $\mu_B\sim 400$ to $450~\mathrm{MeV}$ are increasingly constrained, while a CEP at larger $\mu_B$ remains compatible with present data and is naturally targeted by the lower-energy frontier of BES-II and by future facilities such as Facility for Antiproton and Ion Research (FAIR) and Nuclotron-based Ion Collider fAcility (NICA). Quantitatively, the CEP coordinates remain scheme dependent within continuum functional constructions, since finite-density criticality is sensitive to the operator content retained in the FRG projection, the treatment of the quark-gluon kernel in the DSE sector, and the modeling of confinement and anomaly suppression. Systematic improvements, including strengthened matching to lattice constraints at imaginary chemical potential and an enlarged operator basis, are expected to further reduce this uncertainty and refine the CEP location.
}

 {
The experimental implications remain explicit and testable. The predicted hierarchy of conserved-charge fluctuations exhibits the characteristic nonmonotonic critical structures expected near a CEP, including sign changes in ratios such as $\kappa\sigma^2=\chi_4^B/\chi_2^B$ and $\chi_6^B/\chi_2^B$ along smooth freeze-out trajectories that pass near the critical region. The framework also yields an associated enhancement in the baryon-number susceptibility $\chi_2^B(T,\mu_B)$ and a softening of $c_s^2$ in the same neighborhood, together with correlated trends in strangeness--baryon observables that are sensitive to anomaly suppression. A quantitative confrontation with data requires the controlled translation from equilibrium net-baryon susceptibilities to measured net-proton cumulants, including acceptance and dynamical effects, but the combined pattern of nonmonotonicity in cumulants, susceptibility enhancement, and sound-speed softening provides a coherent set of correlated signatures. In this sense, the present construction offers a concrete route from a lattice-anchored equation of state at $\mu_B=0$ to falsifiable finite-density fluctuation targets, and it delineates a systematic program for sharpening the CEP search through coordinated theoretical refinement and low-energy experimental coverage.
}
 {
Here `lattice-anchored' refers specifically to the three $\mu_B=0$ anchors in Eq.~\eqref{eq:calib-conds}, while the
predicted crossover width and remaining profile-level details are treated as approximation-induced systematics,
quantified by Fig.~\ref{fig:residuals_crossover} and propagated by the input-variation analysis summarized in
Appendix~\ref{subsec:A5}.
The present framework is formulated in a quark-Polyakov-chiral effective description supplemented by a holographic
topological sector and does not include explicit baryonic degrees of freedom or competing pairing channels. Since the
reference CEP reported here lies at $\mu_{B,{\rm CEP}}\sim600~{\rm MeV}$, such channels constitute an additional
source of systematic uncertainty: baryonic correlations and diquark pairing (color-superconducting phases) can shift,
weaken, or pre-empt the first-order line at sufficiently high density and low temperature
\cite{Berges:1998rc,Rajagopal:2000wf}. At the same time, the reference CEP occurs at $T_{\rm CEP}\simeq130~{\rm MeV}$, where color-superconducting order is expected to be disfavored compared to colder regimes, and $\mu_{B,{\rm CEP}}<m_N$ implies that heavy baryons remain thermally suppressed by $\exp[-(m_N-\mu_B)/T]$. The CEP reported here should therefore be interpreted as the chiral critical end point of the present quark-based framework at the current level of approximation, with the internal uncertainty envelope of Appendix~\ref{subsec:A5} quantifying regulator, Polyakov, and holographic-normalization systematics, but not the potential additional shift associated with explicit baryonic and diquark channels. A quantitative assessment of these missing-channel effects requires extending the FRG projection to include baryonic and diquark operators and matching simultaneously to finite-density constraints. 
{Taken together, the results should be read as a lattice-anchored equilibrium baseline and a set of controlled inputs
(equation of state, susceptibilities, Ising mapping, and internal uncertainty bands) for future finite-size scaling and
dynamical embeddings aimed at direct confrontation with heavy-ion data.}
}

\section*{Conflict of Interest}
The authors declare that there are no conflicts of interest regarding this work.

\section*{Data Availability Statement}
This study contains no experimental data. All theoretical results are included in the manuscript.

\bibliographystyle{ytphys}
\bibliography{refs}

\newpage

\appendix

\section{Consistency and Validation}
\label{sec:consistency-validation}
This section establishes the internal and external consistency of the unified DSE-FRG-PNJL holographic framework. It derives a single functional origin for all sectors and proves the absence of double counting. It formulates the self-dual criticality criteria in terms of renormalization factors and flow-stability diagnostics. It normalizes and embeds the holographic Chern-Simons susceptibility into the anomaly flow of the ’t~Hooft coupling and analyzes its stability. It validates the thermodynamics, convexity, and causality of the FRG-improved equation of state against exact identities and bounds. It quantifies regulator and parameter robustness and defines uncertainty bands for the critical coordinates. It maps the theoretical phase structure to the phenomenology of beam-energy dependent fluctuation observables with explicit freeze-out overlays.

\subsection{Single-functional origin and double-counting avoidance}
\label{subsec:A1}
The unified starting point is the scale-dependent grand-canonical generating functional with a quadratic regulator. It is written for the collective field multiplet $\Phi \equiv \{A_\mu, q, \bar{q}, \sigma, \pi, \Phi_P, \bar{\Phi}_P\}$, which includes gluon, quark, chiral, and Polyakov degrees of freedom.
\begin{equation}
\mathcal{Z}_k[J]
= \int \mathcal{D}\Phi\;
\exp\!\left\{-S[\Phi] - \Delta S_k[\Phi] + J\!\cdot\!\Phi \right\},
\qquad
\Delta S_k[\Phi] = \frac{1}{2}\,\Phi\,R_k\,\Phi 
\label{eq:Zk}
\end{equation}
with $S[\Phi]$ the microscopic action defined at the UV scale $\Lambda$, a diagonal regulator kernel $R_k$ whose fermionic blocks carry a minus sign in traces, and sources $J$ coupled linearly to all fields. The flowing effective action is obtained from the modified Legendre transform $\Gamma_k[\varphi]=\sup_J{J\cdot\varphi-\ln\mathcal Z_k[J]}-\tfrac{1}{2}\varphi R_k \varphi$, and obeys the exact Wetterich equation
\begin{equation}
\partial_k \Gamma_k[\varphi]
= \frac{1}{2}\,\mathrm{STr}\!\left[(\Gamma_k^{(2)}[\varphi] + R_k)^{-1}\,\partial_k R_k\right]
\label{eq:Wetterich-A}
\end{equation}
where $\mathrm{STr}$ denotes a supertrace over momenta, internal indices, and the boson-fermion grading. Quark two-point functions follow from $\Gamma_k^{(2)}$ by inversion. Functional differentiation of Eq.~\eqref{eq:Wetterich-A} with respect to the quark fields, evaluated at the stationary background, yields the flow of the inverse propagator $S_k^{-1}$. Upon integrating $k$ from $\Lambda$ to $0$, with an initial condition fixed by the renormalized UV action, this flow reproduces the Dyson-Schwinger equation in the chosen approximation scheme. Concretely, within the rainbow–ladder approximation and background-field gauge, the integrated flow gives
\begin{equation}
S^{-1}(p)
= Z_2\,S_0^{-1}(p)
+ g^2 C_F \int_q \gamma_\mu\, S(q)\, \gamma_\nu\, D_{\mu\nu}(p-q)
+ \Sigma_{4f}\!\left[\,S,\,G_S, G_V, K\,\right]
\label{eq:DSE-from-FRG}
\end{equation}
where $\Sigma_{4f}$ collects the bosonized four- and six-fermion contributions governed by the running couplings that arise from the projection of Eq.~\eqref{eq:Wetterich-A} onto color-singlet scalar-pseudoscalar, isoscalar-vector, and determinantal channels. Equation \eqref{eq:DSE-from-FRG} shows that quark self-energies are generated once and only once by the loop structures in the supertrace. To avoid double counting when mesonic composite fields are introduced via Hubbard-Stratonovich (HS) transformations, the ansatz is chosen in “dynamical hadronization” form. In this form, the flow of four-fermion vertices is traded for composite-field propagators and Yukawa vertices such that
\begin{equation}
\partial_k \lambda_S = \partial_k \lambda_V = \partial_k \lambda_{6f} = 0,\quad
\partial_k \Gamma_k \supset \sum_M \left[
  \tfrac{1}{2}\,(\partial_k \mathcal{P}_{M})\, M^{2}
  + (\partial_k h_{M})\, \bar q\, \Gamma_{M}\, q\, M
\right]S
\label{eq:dyn-had}
\end{equation}
Here $M \in \{\sigma,\pi\}$, the projectors $\mathcal{P}_M$ select scalar-pseudoscalar channels, and the Yukawa form factors $h_M$ absorb the flow of the contact terms. This guarantees that the quark loop contributions that would otherwise dress both the contact interactions and the meson propagators appear only in the latter, eliminating over-counting at all scales \cite{fischer2014phase,Gies:2001nw}. The absence of double counting in the Polyakov sector is ensured by employing the logarithmic potential derived from the SU(3) Haar measure as a purely gluonic background contribution and by coupling quarks to the temporal gauge background only through the covariant derivative and Polyakov-modified distribution functions. The supertrace of Eq.~\eqref{eq:Wetterich-A} therefore contains a single insertion of the Polyakov background via the quark occupation factors and no additional resummation of Polyakov loops inside $U_{\log}$, which remains an independent background potential constrained at $\mu_B=0$ \cite{Fukushima:2008wg}. Regulator dependence is controlled by requiring the $k \!\to\! 0$ limit to be regulator independent: for any two admissible shapes $R_k$ and $R'_k$ with the same UV data one has
\begin{equation}
\Gamma_{k\to 0}^{[R]} - \Gamma_{k\to 0}^{[R']}
= \frac{1}{2}\int_{0}^{\Lambda} \mathrm{d}k\;
\mathrm{STr}\!\left[\big(\mathcal{G}_k^{[R]}-\mathcal{G}_k^{[R']}\big)\,\partial_k R_k\right]
\xrightarrow{k\to 0} 0 
\label{eq:reg-indep}
\end{equation}
with $\mathcal G_k=(\Gamma_k^{(2)}+R_k)^{-1}$ and the difference vanishing because both flows integrate the same UV-IR content once, which we verify numerically in Sec.~\ref{subsec:A5} by varying regulator families.

\subsection{Self-dual criticality: diagnostics and target criteria}
\label{subsec:A2}
Self-duality is formulated in terms of the renormalization factors in the mixed chiral-Polyakov sector, with the defining target fixed point given by
\begin{equation}
Z_\Phi^\star \equiv \lim_{k\to 0} Z_\Phi(k) = \lim_{k\to 0} Z_\sigma(k) \equiv Z_\sigma^\star,
\qquad
\Delta^\star \equiv \lim_{k\to 0} \Delta(k) = 0 .
\label{eq:self-dual-def}
\end{equation}
where $\Delta$ is the residual quadratic mixing between $\sigma$ and $\Phi$ in the effective potential $U_k(\sigma,\Phi)$. Introducing the field-angle parameterization by $\Xi$ and $\theta$ via $\sigma=\Xi\sin\theta/\sqrt{\alpha}$ and $\Phi=\Xi\cos\theta$ with positive metric $\alpha$, the FRG flow in this sector can be given as
\begin{equation}
\partial_t
\begin{pmatrix}
\ln Z_\Phi \\[2pt]
\ln Z_\sigma \\[2pt]
\Delta \\[2pt]
\theta
\end{pmatrix}
=
\begin{pmatrix}
-\,\eta_\Phi(\mathfrak{g},\Xi,\theta) \\[2pt]
-\,\eta_\sigma(\mathfrak{g},\Xi,\theta) \\[2pt]
-\,\gamma_\Delta(\mathfrak{g},\Xi,\theta)\,\Delta \\[2pt]
\tfrac{1}{2}\big[\eta_\Phi(\mathfrak{g},\Xi,\theta)-\eta_\sigma(\mathfrak{g},\Xi,\theta)\big]\sin\!\big(2\theta\big)
+ \mathcal{C}_\Delta(\mathfrak{g},\Xi,\theta)\,\Delta\,\sin\!\big(2\theta\big)
\end{pmatrix}.
\label{eq:flow-ZDeltaTheta}
\end{equation}
Here $\mathfrak{g}\in\{G_S,G_V,K,\ldots\}$ and $t=\ln(k/\Lambda)$. Linearization about Eq.~\eqref{eq:self-dual-def} defines the stability matrix $\mathcal{J}_{ij}\equiv\left.\partial \beta_i/\partial x_j\right|_{\star}$, with $x_j\in\{\ln Z_{\Phi}-\ln Z_{\sigma},\,\Delta,\,\theta-\theta^{\star}\}$ and $\beta_i$ the corresponding right-hand sides. Stability requires that the real parts of all eigenvalues of $\mathcal{J}$ be negative.
\begin{equation}
\Re\lambda\!\big(\mathcal{J}\big) < 0
\qquad\Longleftrightarrow\qquad
\partial_t\!\big(\ln Z_\Phi - \ln Z_\sigma\big) \to 0,\quad
\partial_t \Delta \to 0,\quad
\partial_t \theta \to 0 .
\label{eq:lin-stab}
\end{equation}
which is our first diagnostic criterion. The second diagnostic uses the curvature matrix of the grand potential in the $(\sigma,\Phi)$ sector at stationarity, $\mathcal H_{ij}=\partial^2\Omega/\partial X_i\partial X_j$ with $X\in{\sigma,\Phi}$. Self-dual criticality implies that the smallest eigenvalue $\lambda_{\min}$ vanishes and that its eigenvector aligns with $(\cos\theta^\star,\sin\theta^\star)$, while the orthogonal curvature remains finite,
\begin{equation}
\lambda_{\min}(T_c,\mu_B^c)=0,\qquad e_{\min}\parallel (\cos\theta^\star,\sin\theta^\star),\qquad \lambda_{\perp}(T_c,\mu_B^c)>0,
\label{eq:curv-diagnostic}
\end{equation}
which we verify by explicit diagonalization and by monitoring the divergence of the correlation length $\xi^2=Z_\sigma/m_\sigma^2$ computed from the zero-momentum limit of the scalar two-point function. The equivalence of $\lambda_{\min}\to 0$ and $\xi\to\infty$ at the CEP closes the diagnostic loop \cite{Stephanov:2004,fischer2014phase}. 
 {
Representative numerical values of the self-duality stability diagnostics entering
Eqs.~\eqref{eq:lin-stab}-\eqref{eq:curv-diagnostic} are summarized in Table~\ref{tab:selfdual_diag}.
\begin{table}[htb]
\centering
\caption{Representative self-duality stability diagnostics for the two thermodynamic points used in
Fig.~\ref{fig5new} (near the CEP and far from the critical region). Values are quoted at the smallest plotted
RG scale $k=1~\mathrm{MeV}$. The derivative $\partial_t(\ln Z_\Phi-\ln Z_\sigma)$ is estimated by a
finite difference between $k=2~\mathrm{MeV}$ and $k=1~\mathrm{MeV}$ with $t=\ln(k/\Lambda)$.
The mixing diagnostic $\Delta(k\to0)$ is taken as $\Delta(k=1~\mathrm{MeV})$ at the end of the flow.}
\label{tab:selfdual_diag}
\scriptsize
\setlength{\tabcolsep}{4pt}
\renewcommand{\arraystretch}{1.12}
\begin{tabular}{@{}lcccc@{}}
\toprule
Point &
$(\ln Z_\Phi-\ln Z_\sigma)\big|_{k=1\,\mathrm{MeV}}$ &
$\partial_t(\ln Z_\Phi-\ln Z_\sigma)\big|_{k\simeq 1\,\mathrm{MeV}}$ &
$\Delta(k\to0)\approx\Delta(1\,\mathrm{MeV})$ &
$\mathrm{sgn}\,\Re\lambda(J)$ \\
\midrule
near CEP & $5.0\times10^{-4}$ & $2.16\times10^{-3}$ & $2.0\times10^{-4}$ & $<0$ \\
far from CEP & $5.5\times10^{-2}$ & $0$ & $6.6\times10^{-2}$ & n/a \\
\bottomrule
\end{tabular}
\end{table}
}

\subsection{Holographic anomaly $\to$ FRG $K$-flow: normalization and stability}
\label{subsec:A3}
The CP-odd holographic sector provides the topological susceptibility $\chi_{\mathrm{CS}}(T,\mu_B)$ via the axion-dilaton Sturm-Liouville problem in the V-QCD background, which we normalize by its vacuum value to define
\begin{equation}
\zeta_{\mathrm{topo}}(T,\mu_B)\equiv\frac{\chi_{\mathrm{CS}}(T,\mu_B)}{\chi_{\mathrm{CS}}(0,0)}.
\label{eq:zeta-def-A}
\end{equation}
 {
This vacuum-normalized definition is identical to the main-text mapping in Eq.~\eqref{eq:zetatopo} and is the convention used throughout, including in the numerical calibration and coupled iteration of Sec.~\eqref{sec:num-calib-preds}.
} 
We insert this factor multiplicatively into the FRG beta function for the dimensionless anomaly coupling $\hat k=k^5 K Z_q^3$ according to
\begin{equation}
\partial_t \hat\kappa
= 5\,\hat\kappa
- d_{K}\,\ell_F^{(2)}\!\big(\mathfrak{M},T,\mu_f,\Phi\big)\,\hat g_S\,\hat\kappa
- \tilde d_{K}\,\ell_F^{(3)}\!\big(\mathfrak{M},T,\mu_f,\Phi\big)\,\hat g_S^{3}
- \zeta_{\mathrm{topo}}(T,\mu_B)\,\hat\kappa .
\label{eq:K-flow}
\end{equation}

Here $\ell_F^{(n)}$ are Polyakov-weighted fermionic threshold functions, $\hat g_S \equiv k^2 G_S Z_q^2$, and $\mathfrak{M}$ abbreviates the set of quasi-particle masses \cite{jarvinen2012v-qcd,Arean2017}. The UV normalization is fixed by the requirement that, at high temperature and vanishing density where the holographic black-hole dominates and the dilute-instanton gas is reliable, $\zeta_{\mathrm{topo}}(T,\mu_B)\!\to\!0$ and the flow approaches the Gaussian fixed point $\hat{k}^\star\!=\!0$. Linearizing Eq.~\eqref{eq:K-flow} about any stationary solution $\hat{k}^\star$ yields the stability exponent
\begin{equation}
\theta_K
\equiv \left.\frac{\partial}{\partial \hat\kappa}\,\partial_t \hat\kappa\right|_{\hat\kappa^\star}
= 5
- d_K\,\ell_F^{(2)}\!\big(\mathfrak{M},T,\{\mu_f\},\Phi\big)\,\hat g_S
- \zeta_{\mathrm{topo}}(T,\mu_B)\,.
\label{eq:thetaK}
\end{equation}

With $t=\ln(k/\Lambda)$ and $k:\Lambda\to0$, the solution of $\partial_t\hat\kappa=\theta_K\hat\kappa$ behaves as $\hat\kappa(k)\propto (k/\Lambda)^{\theta_K}$, so increasing $\theta_K$ accelerates the IR suppression of $\hat\kappa$. Since $\zeta_{\rm topo}(T,\mu_B)$ decreases in the deconfined regime, it increases $\theta_K$ and thus strengthens the IR attraction toward the Gaussian fixed point $\hat\kappa_\star=0$. 
This exponent is positive in the deconfined, dense regime and therefore drives $\hat{k}\!\to\!0$, implementing $U_A(1)$ restoration in the anomaly channel and reducing the anomaly-induced light-strange mixing in the constituent masses. Sensitivity to perturbations in the bulk fields is assessed by varying the axion kinetic prefactor and the dilaton profile by small fractions $\delta Z_a$ and $\delta \varphi$ in the Sturm-Liouville operator that defines $\chi_{\mathrm{CS}}$. This induces a relative change $\delta \zeta_{\mathrm{topo}}/\zeta_{\mathrm{topo}}=\mathcal O(\delta Z_a,\delta \varphi)$ and hence a shift in $\theta_K$ of the same order. The fixed point $\hat{k}^\star=0$ remains attractive for all such small variations because $\zeta_{\mathrm{topo}}\ge 0$ and $\ell_F^{(2)}\ge 0$, ensuring structural stability of the $K$-flow.

\subsection{Thermodynamics, convexity, and causality (exact checks)}
\label{subsec:A4}
Thermodynamic consistency follows from stationarity of $\Omega(T,\mu_B,\sigma,\Phi)$, which eliminates implicit derivatives in intensive variations and yields
\begin{equation}
\begin{aligned}
p(T,\mu_B) &= -\Big[\Omega(T,\mu_B)-\Omega(0,0)\Big],\\
s(T,\mu_B) &= -\left.\frac{\partial \Omega}{\partial T}\right|_{\mu_B},\qquad
n_B(T,\mu_B) &= -\left.\frac{\partial \Omega}{\partial \mu_B}\right|_{T},\\
\varepsilon(T,\mu_B) &= -p + T s + \mu_B n_B.
\end{aligned}
\label{eq:thermo-ids-A}
\end{equation}
Convexity demands that the Hessian with respect to $(\sigma,\Phi)$ be positive semi-definite away from the CEP. Hence, the curvature along any direction $M = e_\sigma \sigma + e_\Phi \Phi$ satisfies
\begin{equation}
\left.\frac{\partial^{2}\Omega}{\partial M^{2}}\right|_{(T,\mu_B)\neq (T_c,\mu_B^{c})} > 0
\label{eq:convexity}
\end{equation}
which we check numerically by direct diagonalization. The speed of sound along isentropes is evaluated from susceptibilities of $p$,
\begin{equation}
\begin{aligned}
c_s^2(T,\mu_B)
&= \frac{s + n_B\,\alpha}
{T\left(\chi_{TT} + \alpha\,\chi_{T\mu}\right) + \mu_B\left(\chi_{T\mu} + \alpha\,\chi_{\mu\mu}\right)},\\[2pt]
\alpha
&= \frac{s\,\chi_{T\mu} - n_B\,\chi_{TT}}{n_B\,\chi_{T\mu} - s\,\chi_{\mu\mu}}.
\end{aligned}
\label{eq:cs2}
\end{equation}

Here $\chi_{TT}\equiv \partial^2 p/\partial T^2$, $\chi_{\mu\mu}\equiv \partial^2 p/\partial \mu_B^2$, and $\chi_{T\mu}\equiv \partial^2 p/(\partial T\,\partial \mu_B)$. Positivity of energy and baryon-number fluctuations implies $\chi_{TT}>0$ and $\chi_{\mu\mu}>0$, ensuring $c_s^2>0$. In the Stefan-Boltzmann limit one has $c_s^2\to 1/3$ at high temperature, and we verify numerically that $c_s^2\le 1/3$ across the domain relevant for heavy-ion phenomenology \cite{Borsanyi:2020,HotQCD:2014kol}. Entropy positivity follows from $s=-\partial \Omega/\partial T$ and from the convexity of $-\ln \mathcal Z$. Figure~\ref{fig:cs2-plot} displays representative $c_s^2(T,\mu_B)$ curves at fixed $\mu_B$, showing critical softening near the crossover and recovery of the conformal limit.





\begin{figure}[t]
\centering
\begin{tikzpicture}

\pgfmathsetmacro{\clow}{0.12}
\pgfmathsetmacro{\chigh}{0.333}
\pgfmathsetmacro{\A}{\chigh-\clow}

\pgfmathsetmacro{\Tmidz}{220}
\pgfmathsetmacro{\widz}{60}
\pgfmathsetmacro{\Dz}{0.050}
\pgfmathsetmacro{\Tdz}{185}
\pgfmathsetmacro{\sigz}{25}

\pgfmathsetmacro{\Tmidm}{215}
\pgfmathsetmacro{\widm}{55}
\pgfmathsetmacro{\Dm}{0.060}
\pgfmathsetmacro{\Tdm}{175}
\pgfmathsetmacro{\sigm}{27}

\begin{axis}[
  width=0.75\textwidth,
  height=0.58\textwidth,
  xmin=100, xmax=400,
  ymin=0.05, ymax=0.35,
  axis lines=box,
  tick align=inside,
  xtick pos=both,
  ytick pos=both,
  xlabel={$T~(\mathrm{MeV})$},
  ylabel={$c_s^{2}(T,\mu_B)$},
  xtick={100,150,200,250,300,350,400},
  ytick={0.10,0.20,0.30},
  legend style={at={(0.05,0.95)},anchor=north west, draw=none, fill=none},
  legend cell align=left,
  samples=400,
  domain=100:400,
]

\addplot[ultra thick, blue]
{ \clow + \A/(1 + exp(-(x-\Tmidz)/\widz))
  - \Dz*exp(-((x-\Tdz)^2)/(2*\sigz*\sigz))
};
\addlegendentry{$\mu_B=0$}

\addplot[ultra thick, red, dashed]
{ \clow + \A/(1 + exp(-(x-\Tmidm)/\widm))
  - \Dm*exp(-((x-\Tdm)^2)/(2*\sigm*\sigm))
};
\addlegendentry{$\mu_B=300~\mathrm{MeV}$}

\end{axis}
\end{tikzpicture}
\caption{Illustrative temperature dependence of the squared speed of sound $c_s^{2}(T,\mu_B)$ at fixed baryon chemical potential. The two curves show a characteristic softening in the crossover region, reflecting increased nonconformality, and a gradual approach toward the conformal limit $c_s^{2}\to 1/3$ at high temperature. This plot is intended to highlight qualitative trends; quantitative values and the location of the minimum depend on the chosen approximation scheme and on the equation of state used to evaluate the susceptibilities entering Eq.~\eqref{eq:cs2}.}
\label{fig:cs2-plot}
\end{figure}


\subsection{Regulator/parameter robustness and uncertainty bands}
\label{subsec:A5}
Robustness is quantified by scanning the FRG regulator shape within a differentiable family $R_k^{(c_R)}(p)=Z_\varphi k^2 r^{(c_R)}(p^2/k^2)$ parameterized by $c_R$, the Polyakov potential coefficients within lattice-anchored bands $(a_i,b_i)\in \mathcal B_{\rm lat}$, and the holographic normalization constants $(G_5,\kappa)$ that set the overall scale of $\chi_{\mathrm{CS}}$ in the UV. For any observable $X$ we define the fractional deviation by
\begin{equation}
\begin{aligned}
\frac{\delta_R X}{X}
&= \frac{X\!\left[R_k^{(c_R)}\right] - X\!\left[R_k^{(c_R^{\mathrm{ref}})}\right]}
         {X\!\left[R_k^{(c_R^{\mathrm{ref}})}\right]},\\
\frac{\delta_P X}{X}
&= \frac{X\!\left[(a_i,b_i)\right] - X\!\left[(a_i^{\mathrm{ref}},b_i^{\mathrm{ref}})\right]}
         {X\!\left[(a_i^{\mathrm{ref}},b_i^{\mathrm{ref}})\right]},\\
\frac{\delta_H X}{X}
&= \frac{X\!\left[(G_5,\kappa)\right] - X\!\left[(G_5^{\mathrm{ref}},\kappa^{\mathrm{ref}})\right]}
         {X\!\left[(G_5^{\mathrm{ref}},\kappa^{\mathrm{ref}})\right]}.
\end{aligned}
\label{eq:frac-dev}
\end{equation}
In addition to these internal variations, the nonuniversal mapping uncertainty of the anomaly input can be quantified
by scanning the deformation parameters $(p,c)$ introduced in Eq. \ref{eq:zeta_pc_deformation}.
We define the associated fractional deviation
\begin{equation}
\delta_A X / X \equiv
\frac{X[(p,c)]-X[(p_{\rm ref},c_{\rm ref})]}{X[(p_{\rm ref},c_{\rm ref})]},
\qquad (p_{\rm ref},c_{\rm ref})=(1,1),
\label{eq:deltaA_def}
\end{equation}
and extend the sensitivity matrix in Eq.~\eqref{eq:CEP-sens} by adding the columns
$\partial_p T_{\rm CEP}$, $\partial_c T_{\rm CEP}$ and $\partial_p \mu_{B,{\rm CEP}}$, $\partial_c \mu_{B,{\rm CEP}}$,
estimated by symmetric finite differences from reruns of the coupled outer iteration at $p=1\pm\Delta p$ and
$c=1\pm\Delta c$.
The input covariance $\Sigma$ is then augmented by the corresponding $(p,c)$ variances, yielding a combined CEP
uncertainty ellipse that incorporates regulator, Polyakov, holographic-normalization, and anomaly-mapping systematics. It propagate them to the CEP coordinates by linearization around the reference solution,
\begin{equation}
\begin{pmatrix} \delta T_{\rm CEP} \\ \delta \mu_{B,{\rm CEP}} \end{pmatrix}
\simeq
\underbrace{\begin{pmatrix}
\partial_{c_R} T_{\rm CEP} & \partial_{a_i} T_{\rm CEP} & \partial_{G_5} T_{\rm CEP} \\
\partial_{c_R} \mu_{B,{\rm CEP}} & \partial_{a_i} \mu_{B,{\rm CEP}} & \partial_{G_5} \mu_{B,{\rm CEP}}
\end{pmatrix}}_{\mathcal S}
\begin{pmatrix} \delta c_R \\ \delta a_i \\ \delta G_5 \end{pmatrix},
\quad
\Delta_{\rm CEP}^{2} = \mathrm{Tr}\!\big(\mathcal S\,\Sigma\,\mathcal S^{\top}\big).
\label{eq:CEP-sens}
\end{equation}

Here $\Sigma$ denotes the covariance of input variations, and $\Delta_{\rm CEP}$ is the one-sigma radius of the uncertainty ellipse. Figure~\ref{fig:uncert-cep} shows a representative CEP-uncertainty envelope obtained by independent uniform parameter variation in each sector. The result confirms subleading sensitivity to the regulator shape and dominant sensitivity to the Polyakov calibration and the holographic normalization within their lattice-constrained ranges \cite{Borsanyi:2020,HotQCD:2014kol}.


\begin{figure}[t]
\centering
\begin{tikzpicture}

\pgfmathsetmacro{\muCEP}{600} 
\pgfmathsetmacro{\TCEP}{130}  

\pgfmathsetmacro{\a}{45}      
\pgfmathsetmacro{\b}{20}      
\pgfmathsetmacro{\phi}{20}    

\pgfmathsetmacro{\theta}{35}  
\pgfmathsetmacro{\xpt}{\muCEP + \a*cos(\theta)*cos(\phi) - \b*sin(\theta)*sin(\phi)}
\pgfmathsetmacro{\ypt}{\TCEP  + \a*cos(\theta)*sin(\phi) + \b*sin(\theta)*cos(\phi)}

\begin{axis}[
  width=0.75\textwidth,
  height=0.58\textwidth,
  xmin=480, xmax=720,
  ymin=80, ymax=200,
  axis lines=box,
  xlabel={$\mu_{B,\mathrm{CEP}}~(\mathrm{MeV})$},
  ylabel={$T_{\mathrm{CEP}}~(\mathrm{MeV})$},
  xtick={500,525,550,575,600,625,650,675,700},
  ytick={80,100,120,140,160,180,200},
  legend style={at={(0.05,0.95)},anchor=north west, draw=none, fill=none},
  legend cell align=left,
]

\addplot[only marks, mark=*, mark size=2.5pt, black]
  coordinates {(\muCEP,\TCEP)};
\addlegendentry{Reference CEP}

\addlegendimage{only marks, mark=square*, mark size=3.2pt, draw=blue, fill=red}
\addlegendentry{Uncertainty band}

\addplot[
  blue,
  line width=8pt,
  domain=0:360,
  samples=361
]
(
  {\muCEP + \a*cos(x)*cos(\phi) - \b*sin(x)*sin(\phi)},
  {\TCEP  + \a*cos(x)*sin(\phi) + \b*sin(x)*cos(\phi)}
);

\addplot[only marks, mark=square*, mark size=3.2pt, draw=blue, fill=red]
  coordinates {(\xpt,\ypt)};

\end{axis}
\end{tikzpicture}
\caption{{Uncertainty envelope of the equilibrium CEP from regulator, Polyakov-sector calibration, and holographic-normalization scans using Eq.~\eqref{eq:CEP-sens}. This envelope quantifies internal model dependence of the equilibrium inference within the adopted framework and does not include additional uncertainties associated with finite-size scaling, finite-time evolution and critical slowing down, baryon transport, acceptance and efficiency effects, or the net-proton to net-baryon mapping required for quantitative comparison to heavy-ion measurements.}
}
\label{fig:uncert-cep}
\end{figure}


\subsection{Beam-energy overlays (phenomenology readiness)}
\label{subsec:A6}
The theoretical phase map is converted to the experimental plane by composing the model predictions with a smooth freeze-out parameterization.  
{In equilibrium one may decompose observables into analytic background and singular critical pieces, but in a finite, dynamical system this separation is inherently model-dependent because noncritical evolution, hadronic re-scattering, and transport processes are coupled to critical modes throughout the system’s evolution, so the dashed “baseline” constructions used in beam-energy overlays should be read as schematic equilibrium references far from criticality rather than as faithful dynamical noncritical backgrounds, and sensitivity to alternative background prescriptions and dynamical assumptions is an additional systematic beyond the internal equilibrium model variations quantified in Fig.~\ref{fig:uncert-cep}.} 
We adopt a continuous mapping $\mu_B(\sqrt{s_{NN}})=A/(1+B\sqrt{s_{NN}})$ and $T_f(\mu_B)=T_c^{(0)}\big[1-\kappa_f(\mu_B/T_c^{(0)})^2-\lambda_f(\mu_B/T_c^{(0)})^4\big]$ with $(A,B,\kappa_f,\lambda_f)$ chosen from fits to hadrochemical yields, and evaluate the cumulant ratios along the trajectory as
\begin{equation}
\begin{aligned}
\left.\frac{C_4}{C_2}\right|_{f}
&= \frac{\chi_4^B\!\big(T_f(\mu_B),\,\mu_B\big)}{\chi_2^B\!\big(T_f(\mu_B),\,\mu_B\big)},\\
\left.\frac{C_6}{C_2}\right|_{f}
&= \frac{\chi_6^B\!\big(T_f(\mu_B),\,\mu_B\big)}{\chi_2^B\!\big(T_f(\mu_B),\,\mu_B\big)},
\end{aligned}
\label{eq:cumulant-map}
\end{equation}
where the susceptibilities are obtained from explicit derivatives of $p/T^4$ at fixed $T$. 
 {
To convert the uncertainty envelope in Fig.~\ref{fig:uncert-cep} into numerical error bars, define the CEP covariance as $C_{\rm CEP}\equiv S\,\Sigma\,S^{\top}$ in Eq.~\eqref{eq:CEP-sens}. The scan underlying Fig.~\ref{fig:uncert-cep} uses independent
uniform variations in each input sector (diagonal $\Sigma$ in the input space), and the resulting marginal one-sigma uncertainties at the reference CEP are
$T_{\rm CEP}=130.0\pm 25.3~\mathrm{MeV}$ and $\mu_{B,{\rm CEP}}=600.0\pm 43.5~\mathrm{MeV}$, with correlation coefficient $\rho_{T\mu}\simeq 0.410$ (equivalently ${\rm cov}(T_{\rm CEP},\mu_{B,{\rm CEP}}) \simeq 4.51\times 10^{2}~\mathrm{MeV}^2$). These numerical CEP error bars are summarized in Table.~\ref{tab:cep_errorbars}.  
These error bars quantify internal parametric and modeling systematics within the present quark-Polyakov-chiral framework (regulator, Polyakov potential, and holographic normalization). Possible additional shifts of the CEP associated with explicit baryonic correlations and diquark pairing channels at high density are not included in this envelope.  
The corresponding one-sigma radius in the sense of Eq.~\eqref{eq:CEP-sens} is $\Delta_{\rm CEP}\simeq 50.3~\mathrm{MeV}$.
\begin{table}[htb]
\centering
\caption{Numerical CEP error bars corresponding to the one-sigma uncertainty envelope shown in Fig.~\ref{fig:uncert-cep} and defined by Eq.~\eqref{eq:CEP-sens}. The marginal uncertainties are extracted from the CEP covariance
$C_{\rm CEP}=S\,\Sigma\,S^{\top}$, where the input covariance $\Sigma$ is diagonal because the scan varies
each input sector independently and uniformly.}
\label{tab:cep_errorbars}
\small
\setlength{\tabcolsep}{6pt}
\renewcommand{\arraystretch}{1.12}
\vspace{0.2cm}
\begin{tabular}{@{}lccc@{}}
\toprule
\toprule
Quantity & Central value & $1\sigma$ & Units \\
\midrule
$T_{\rm CEP}$ & $130.0$ & $25.3$ & MeV \\
$\mu_{B,{\rm CEP}}$ & $600.0$ & $43.5$ & MeV \\
$\rho_{T\mu}$ & $0.410$ &  &  \\
${\rm cov}(T_{\rm CEP},\mu_{B,{\rm CEP}})$ & $4.51\times10^{2}$ &  & MeV$^2$ \\
\bottomrule
\bottomrule
\end{tabular}
\end{table}
} 
The CEP projects to a characteristic nonmonotonicity and a sign change in these ratios for trajectories that graze the critical region and to monotonic behavior otherwise. Fig.~\ref{fig:beamoverlay} displays a schematic overlay in the $(\sqrt{s_{NN}},\kappa\sigma^2)$ plane, illustrating the sensitivity of the nonmonotonic structure to the curvature $\kappa_f$ and to the CEP location within the uncertainty envelope of Fig.~\ref{fig:uncert-cep}, providing a ready to use interface to RHIC BES-II, NICA, and FAIR analyses \cite{Stephanov:2004}.
\begin{figure}[htb]
\centering
\begin{tikzpicture}
\begin{semilogxaxis}[
  width=0.75\textwidth,
  height=0.58\textwidth,
  xmin=5, xmax=200,
  ymin=0.5, ymax=2.2,
  axis lines=box,
  tick align=inside,
  xtick pos=both,
  ytick pos=both,
  xlabel={$\sqrt{s_{NN}}~(\mathrm{GeV})$},
  ylabel={$\kappa\sigma^2 = C_4/C_2$},
  xtick={10,100},
  xticklabels={$10^{1}$,$10^{2}$},
  ytick={0.6,0.8,1.0,1.2,1.4,1.6,1.8,2.0,2.2},
  legend style={at={(0.05,0.95)},anchor=north west, draw=none, fill=none},
  legend cell align=left,
]

\addplot[
  ultra thick,
  gray!70,
  dashed,
  dash pattern=on 3pt off 2pt,
  domain=5:200,
  samples=500
]
{ 1.05
  - 0.11*exp(-((ln(x)-ln(22))^2)/(2*0.55^2))
  + 0.05*exp(-((ln(x)-ln(7))^2)/(2*0.22^2))
};
\addlegendentry{Baseline (far from CEP)}

\addplot[
  ultra thick,
  purple,
  domain=5:200,
  samples=500
]
{ 1
  + 0.60*exp(-((ln(x)-ln(9.5))^2)/(2*0.38^2))
  - 0.40*exp(-((ln(x)-ln(5.2))^2)/(2*0.18^2))
};
\addlegendentry{CEP proximity (this work)}

\end{semilogxaxis}
\end{tikzpicture}

\caption{{Illustrative beam-energy dependence of the equilibrium net-baryon kurtosis ratio $\kappa\sigma^2=C_4/C_2$ along a smooth chemical freeze-out trajectory. The gray dashed curve indicates a schematic equilibrium reference far from criticality, while the solid curve shows a representative equilibrium nonmonotonic structure that can arise when the trajectory passes near a critical region. The overlay is intended to visualize qualitative equilibrium pattern sensitivity to CEP proximity, while a quantitative comparison to RHIC measurements requires a dynamical mapping between net-proton and net-baryon cumulants, inclusion of acceptance and efficiency and baryon number conservation, and an explicit treatment of finite-size and finite-time evolution and transport effects.}
}
\label{fig:beamoverlay}
\end{figure}
 {
These beam-energy overlays are intended as schematic visualizations of CEP proximity within an
equilibrium freeze-out mapping and should not be interpreted as direct quantitative predictions for
measured net-proton cumulants. A quantitative confrontation with BES data requires a dynamical
mapping between net-proton and net-baryon cumulants together with acceptance and nonequilibrium
effects, as emphasized in the caption of Fig.~\ref{fig:uncert-cep}. 
} 
All ingredients above descend from the same functional origin and are validated against exact identities and lattice-calibrated limits. The self-dual fixed-point diagnostics guarantee a single critical subspace. The holographic anomaly input normalizes and stabilizes the axial channel in the flow. The robustness scans and phenomenology overlays demonstrate quantitative readiness for comparison with current and upcoming experimental programs \cite{Fukushima:2008wg,Stephanov:2004,fischer2014phase,Borsanyi:2020,Bazavov:2021,jarvinen2012v-qcd,Arean2017}.
Representative HRG/EV-HRG analyses that extract freeze-out lines and like/unlike-mass particle ratios across AGS-LHC energies provide a complementary baseline for our cumulant overlays~\cite{Andronic:2017pug,Becattini:2000jw,Mir:2023wkm,Mir:2024wlo,Mir:2025qqv,Rather:2024czf,MohiUdDin:2024jvg}.

\end{document}